\documentclass[12pt]{article}
\usepackage{amsfonts}
\usepackage{amsthm}
\usepackage{amsmath}
\usepackage{epsfig}
\usepackage{latexsym}
\usepackage{amssymb}
\usepackage{float}

\usepackage{listings}
\usepackage[english]{babel}
\usepackage{natbib}
\usepackage{appendix}
\usepackage{graphicx}
\usepackage{subcaption}
\usepackage{enumerate}

\usepackage{multirow}

\usepackage{hyperref}

\newcommand{\E}  {\mbox{E}}
\newcommand{\var}  {\mbox{var}}

\def\beqn{\begin{eqnarray*}}
\def\eeqn{\end{eqnarray*}}
\def\beq{\begin{eqnarray}}
\def\eeq{\end{eqnarray}}

\def\cond{\;\vert\;}

\def\suma[#1][#2]{\sum\limits_{#1}^#2}

\DeclareMathOperator*{\argmin}{arg\,min}

\usepackage[dvipsnames]{color}

\newtheorem{theorem}{Theorem}[section]
\newtheorem{lemma}{Lemma}[section]
\newtheorem{corollary}{Corollary}[section]

\newcommand{\blind}{1}

\begin{document}
	
	\def\spacingset#1{\renewcommand{\baselinestretch}%
		{#1}\small\normalsize} \spacingset{1}

%\title{Bagging cross-validated bandwidths in regression with application to Big Data}
%\author{D. Barreiro-Ures \and R. Cao \and M. Francisco-Fern\'andez}
%\date{}
%
%\maketitle

\if1\blind
{
	\title{\bf Bagging cross-validated {bandwidth selection in nonparametric regression estimation} with applications to {large-sized samples}}
	\author{D. Barreiro-Ures\\
		Department of Mathematics, {CITIC,} University of A Coru\~na\\
		and \\
		R. Cao \\
		Department of Mathematics, {CITIC,} University of A Coru\~na and {ITMATI}\\
		and \\
		M. Francisco-Fern\'andez \thanks{\textit{Corresponding author: E-mail: mariofr@udc.es}}\hspace{.2cm}\\
		Department of Mathematics, {CITIC,} University of A Coru\~na and {ITMATI}}
	\maketitle
} \fi

\if0\blind
{
	\bigskip
	\bigskip
	\bigskip
	\begin{center}
		{\LARGE\bf Bagging cross-validated {bandwidth selection in nonparametric regression estimation} with applications to {large-sized samples}}
	\end{center}
	\medskip
} \fi

\bigskip
\begin{abstract}
%	Cross-validation is a popular method of selecting the bandwidth in nonparametric regression estimation. However, this technique has two important drawbacks: (i) the large variability of the selected bandwidths, and (ii) the inability to provide results in a reasonable time for very large sample sizes due to its high computational complexity. To overcome these problems, the use of bagging in conjunction with cross-validation is considered in this paper. The approach studied consists in computing cross-validation bandwidths for a finite number of subsamples and then rescaling the averaged bandwidths to the original sample size. The asymptotic properties of the bagging cross-validation selector are derived and its behavior is shown through several simulation studies and an application to a real dataset related to the current COVID-19 pandemic.
	{Cross-validation is a well-known and widely used bandwidth selection method in nonparametric regression estimation. However, this technique has two remarkable drawbacks: (i) the large variability of the selected bandwidths, and (ii) the inability to provide results in a reasonable time for very large sample sizes. To overcome these problems, bagging cross-validation bandwidths are analyzed in this paper. 
	This approach consists in computing the cross-validation bandwidths for a finite number of subsamples and then rescaling the averaged smoothing parameters to the original sample size. Under a random-design regression model, asymptotic expressions up to a second-order for the bias and variance of the leave-one-out cross-validation bandwidth for the Nadaraya--Watson estimator are obtained. Subsequently, the asymptotic bias and variance and the limit distribution are derived for the bagged cross-validation selector. Suitable choices of the number of subsamples and the subsample size lead to an $n^{-1/2}$ rate for the convergence in distribution of the bagging cross-validation selector, outperforming the rate $n^{-3/10}$ of leave-one-out cross-validation. Several simulations and an illustration on a real dataset related to the COVID-19 pandemic show the behavior of our proposal and its better performance, in terms of statistical efficiency and computing time, when compared to leave-one-out cross-validation.}
\end{abstract}

\noindent%
{\it Keywords:}  cross-validation, subsampling, regression, Nadaraya--Watson, bagging%3 to 6 keywords, that do not appear in the title
\vfill

\newpage
\spacingset{1.5} % DON'T change the spacing!

\section{Introduction}
\label{sec:intro}
{The study of a variable of interest depending on other variable(s) is a common problem that appears in many disciplines. To deal with this issue, an appropriate regression
model setting up the possible functional relationship between the variables is
usually formulated. As part of this analysis, the unknown regression function, describing
the general relationship between the variable of interest and the explanatory variable(s), has to be estimated. 
This task can be carried out using nonparametric methods that do not assume any parametric form for the regression function, providing flexible procedures and avoiding misspecification problems. Among the available nonparametric approaches, kernel-type regression estimators  \citep{wand95} are perhaps the most popular. To compute this type of estimators the user has to select a kernel function (typically a density function)
and a bandwidth or smoothing parameter that regulates the amount of smoothing to
be used, which in turn determines the trade-off between the bias and the variance of the estimator. Although the choice of the kernel function is of secondary importance, the
smoothing parameter plays a crucial role. In this regard,
numerous contributions have been made over the last decades, providing methods to
select the {bandwidth}. These approaches
include, among others,  cross-validation
methods \citep{Hardle1988} {and} plug-in selectors  \citep{Ruppert1995}.
%, and bootstrap bandwidth selectors  \citep{Cao1993}.
In \cite{Sperlich2014}, a complete review and an extensive simulation study of different data-drive{n} bandwidth selectors for kernel regression are presented.
Due to {their} wide applicability and the good performance obtained in this complete comparison,
in the present paper, we focus on analyzing cross-validation bandwidth selection  techniques.}

Cross-validation is a popular method of model selection that precedes an early discussion of the method by \cite{Stone}. In its simplest form, cross-validation consists of splitting {the dataset {under} study} into two parts, using one part to fit one or more models, and then predicting the data in the second part with the models so-built. In this way, by not using the same data to fit and validate the models, it is possible to objectively compare the predictive capacity of different models. The leave-one-out version of cross-validation, which will be of interest in the present paper, is somewhat more involved. It excludes one datum from the {dataset}, fits a model from the remaining observations, uses this model to predict the datum left out, and then repeats this process for all the data.

The present paper studies {the leave-one-out cross-va\-li\-da\-tion bandwidth selection method and} the application of bagging {\citep{Breiman1996}} to {this procedure. We derive some asymptotic properties of the corresponding selectors when considering a random-design regression model and the Nadaraya--Watson kernel-type estimator is used. The Nadaraya--Watson estimator can be seen as a particular case of a wider class of nonparametric estimators, the so-called local polynomial estimators \citep{Stone1977, Cleveland1979, Fan1992}, when performing a local constant fit.} Given a random sample of size $n$, bagging {cross-validation} consists of selecting $N$ subsamples of size $r < n$, each without replacement, from the $n$ observations. One then computes a cross-validation bandwidth from each of the $N$ subsets, averages them, and then scales the average down appropriately to account for the fact that $r < n$. It is well-known that the use of bagging can lead to substantial reductions in the variability of an estimator that is nonlinear in the observations \citep[see][]{FriedHall}, as occurs in the case of {the cross-validation criterion function}.
% bandwidth selection.
 The use of bagging in conjunction with cross-validation for bandwidth selection has already been studied in the case of kernel density estimation by several authors \citep[see, for example][]{PaperBiometrika, HallRobinson2009}. In addition to the potential improvement in statistical precision, {even in the case of small sample sizes,} the use of bagging {(with appropriate elections of $r$ and $N$)} can drastically reduce computation times{, especially for very large sample sizes}. {Note that} the complexity of cross-validation is $O(n^2)${, while the complexity of bagging cross-validation is $O(N r^2)$. Larger reductions in computation time can also be additionally achieved with the application of binning techniques in the bagging procedure.} 

{Apart from the theoretical analysis of the cross-validation bandwidth selection methods, another goal of this study is} to apply the techniques studied in the present paper to a dataset related to the current COVID-19 pandemic. In particular, {using} a moderately large sample{, provided by the Spanish Center for Coordinating Sanitary Alerts and Emergencies,} 
{consisting }
% with observations 
of the age and {the time in hospital}
% hospitalization time 
of people infected with COVID-19 in Spain, {we are interested in studying} the relationship between those two variables by means of the Nadaraya--Watson estimator. 
% {The study of these confidential data is part of a much more complete research arisen from the collaboration of one of the authors of the present paper with the Spanish health authorities, with the aim of analyzing several databases related to the pandemic using different mathematical and statistical techniques.}
Apart from its purely epidemiological interest and due to the considerable size of the sample, this dataset {is also} useful to put into practice the techniques analyzed in the present paper. 

The {remainder of the} paper {is} as follows. In Section \ref{sec:model}, {the regression model considered, the Nadaraya--Watson regression estimator
% , making mention of 
{and} the important {problem of bandwidth selection} {are presented}. In Section \ref{sec:cvbw}, the leave-one-out cross-validation bandwidth selection method} {is described} {and}  {some} asymptotic properties of {the corresponding} selector {are provided} {when} the Nadaraya--Watson estimator {is used}. {Section} \ref{sec:baggedcv} {considers} the use of bagging {for} {cross-validation} in {bandwidth selection for} the Nadaraya--Watson estimator. {A}symptotic expressions for the bias and the variance of the proposed bandwidth selector, as well as for its limit distribution{, are also derived in this section}. In Section \ref{sec:optsubsize}, an {algorithm {is proposed} to automatically choose} the {subsample size} for the bandwidth selector studied in Section \ref{sec:baggedcv}. The techniques proposed are empirically tested through several simulation studies in Section \ref{sec:sims} and applied to the previously mentioned COVID-19 dataset in Section \ref{sec:realdata}. Finally, concluding remarks are given in Section \ref{sec:discussion}. {The detailed} proofs {and some additional plots concerning the simulation study are} included in {the accompanying supplementary materials}.

\section{{Regression model and Nadaraya--Watson estimator}}
\label{sec:model}

Let $\mathcal{X} = \{(X_1, Y_1),\dots,(X_n, Y_n)\}$ be {an independent and identically distributed (i.i.d.) sample of size $n$ of} the two-dimensional random variable $(X,Y)$, drawn from the nonparametric regression model:
\beq \label{eq:model}
Y = m(X)+\varepsilon,
\eeq
where $m(x) = \E(Y\cond X=x)$ denotes the regression function, and $\varepsilon$ {is} the error term, {satisfying that} $\E\left(\varepsilon\cond X=x\right)=0$ and $\E\left(\varepsilon^2\cond X=x\right) = \sigma^2(x)$.
%\beqn
%\E\left(\varepsilon\cond X=x\right) &=& 0,\\
%\E\left(\varepsilon^2\cond X=x\right) &=& \sigma^2(x).
%\eeqn

The Nadaraya--Watson estimator or local constant estimator \citep{Nadaraya1964, Watson1964} offers a {nonparametric} way to estimate the unknown regression function, $m$. {It}  is given by:
\beq\label{eq:nw}
\hat m_h(x) = \frac{\sum\limits_{i=1}^n K_h\left(x-X_i\right)Y_i}{\sum\limits_{i=1}^n K_h\left(x-X_i\right)},
\eeq
where $h>0$ denotes the bandwidth {or smoothing parameter} and $K$ the kernel function. {As pointed out in the introduction,} the value of the bandwidth is of great importance since it determines the amount of smoothing performed by the estimator and, therefore, heavily influences its behavior. 
{Thus, in practice, data-driven bandwidth selection methods are needed.}
% {Then, in practice, it is useful to have available an automated method based on sample observations for the selection of this parameter.}
{In the present paper, we deal} with {this problem, analyzing cross-validation bandwidth selection techniques when using the estimator (\ref{eq:nw}) and considering the {general} regression model (\ref{eq:model})}.

{A possible way to select a (global) optimal bandwidth for \eqref{eq:nw} consists in minimizing a (global) {optimality criterion}, for instance,} the mean integrated squared error (MISE), defined as:
\beq\label{eq:mise}
M_n(h) = \E\left\{\int \left[\hat m_h(x)-m(x)\right]^2f(x)\,dx\right\},
\eeq
where $f$ denotes the marginal density function of $X$. The bandwidth that minimizes \eqref{eq:mise} is called the MISE bandwidth and {it will be denoted} by $h_{n0}$, that is,
\beq\label{eq:hn0}
h_{n0} = \argmin\limits_{h>0} M_n(h).
\eeq

The MISE bandwidth depends on $m$ and $f$ and, since in practice {both functions} are {often}  unknown, $h_{n0}$ cannot be directly calculated. However, it can be estimated{, for example, using {the cross-validation method}.}

{In the following section, we present the leave-one-out cross-validation bandwidth selection criterion and provide the asymptotic properties of the corresponding selector when using the Nadaraya--Watson  estimator \eqref{eq:nw} and considering the regression model \eqref{eq:model}.}

\section{Cross-validation bandwidth}\label{sec:cvbw}

Cross-validation is a method that offers a criterion for optimality which works as an empirical analogue of the MISE and so it allows us to estimate $h_{n0}$. The cross-validation function is defined as:
\beq\label{eq:cv}
CV_n(h) = \frac{1}{n}\sum\limits_{i=1}^n \left[\hat m_h^{(-i)}(X_i)-Y_i\right]^2,
\eeq
where { $\hat m_h^{(-i)}$ denotes} the Nadaraya--Watson estimator constructed using $\mathcal{X}\setminus\{(X_i, Y_i)\}$, that is, leaving out the $i$-th observation,
\beq\label{eq:nwloo}
\hat m_h^{(-i)}(x) = \frac{\sum\limits_{\substack{j=1\\j\neq i}}^n K_h\left(x-X_j\right)Y_j}{\sum\limits_{\substack{j=1\\j\neq i}}^n K_h\left(x-X_j\right)}.
\eeq

Hence, the cross-validation bandwidth, $\hat h_{CV,n}$, can be defined as 
% the bandwidth that minimizes \eqref{eq:cv}, that is,
\beq\label{eq:hcv}
\hat h_{CV,n} = \argmin\limits_{h>0} CV_n(h).
\eeq

In order to obtain the asymptotic properties of \eqref{eq:hcv} as an estimator of \eqref{eq:hn0}, {it is necessary to study} certain moments of \eqref{eq:cv} and its derivatives. However, the fact that the Nadaraya--Watson estimator has a random denominator makes this a very difficult task. To overcome this problem, the following  {unobservable} modified version of the Nadaraya--Watson estimator was proposed in \cite{tesisInes}:
\beq\label{eq:nwtram}
\tilde m_h(x) = m(x)+\frac{1}{nf(x)}\sum\limits_{i=1}^n K_h\left(x-X_i\right)\left[Y_i-m(x)\right].
\eeq

Expression \eqref{eq:nwtram} does not define an estimator, but a theoretical approximation of \eqref{eq:nw}. {Of course, \eqref{eq:nwtram} can be used} to define a modified version of the cross-validation criterion,
\beq\label{eq:cvtram}
\widetilde{CV}_n(h) = \frac{1}{n}\sum\limits_{i=1}^n \left[\tilde m_h^{(-i)}(X_i)-Y_i\right]^2,
\eeq
where $\tilde m_h^{(-i)}$ denotes the leave-one-out version of \eqref{eq:nwtram} {without} the $i$-th observation, that is,
\beq\label{eq:nwlootram}
\tilde m_h^{(-i)}(x) = m(x)+\frac{1}{(n-1)f(x)}\sum\limits_{\substack{j=1\\j\neq i}}^n K_h\left(x-X_j\right)\left[Y_j-m(x)\right].
\eeq

For {the sake of} simplicity and convenience, from now on, we will denote by $CV_n(h)$ and $\hat h_{CV,n}$ the modified version of the cross-validation criterion defined in \eqref{eq:cvtram} and the bandwidth that minimizes \eqref{eq:cvtram}, respectively. {Using} Taylor expansions, we can obtain the following approximation:
\beq\label{eq:hcvtaylor}
\hat h_{CV,n}-h_{n0} &\approx& -\frac{CV_n'(h_{n0})-M_n'(h_{n0})}{M_n''(h_{n0})}\nonumber\\
&+&\frac{[CV_n'(h_{n0})-M_n'(h_{n0})][CV_n''(h_{n0})-M_n''(h_{n0})]}{M_n''(h_{n0})^2},
\eeq
where the second term of \eqref{eq:hcvtaylor} is negligible with respect to the first one when it comes to the bias and the variance of $\hat h_{CV,n}$. Since the first-order terms of $\E[CV_n^{k)}(h)]$ and $M_n^{k)}(h)$ coincide for every $k \geq 1$, we need to calculate the second-order terms of both $\E[CV_n'(h_{n0})]$ and $M_n'(h_{n0})$ in order to analyze the bias of the modified cross-validation bandwidth. As {for} the variance of the modified cross-validation bandwidth, it suffices to calculate the first-order term of $\var[CV_n'(h_{n0})]$. It is well-known that
{under suitable regularity conditions}, up to first order,
\beqn
M_n(h) = B_1 h^4 + V_1 n^{-1}h^{-1}+O\left(h^6+n^{-1}h\right),
\eeqn
where
\beqn
B_1 &=& \frac{1}{4}\mu_2(K)^2\int \left[m''(x)+2\frac{m'(x)f'(x)}{f(x)}\right]^2 f(x)\,dx,\\
V_1 &=& R(K)\int \sigma^2(x)\, dx,
\eeqn
{with $R(g)=\int g^2(x)\,{\rm d}x$ and $\mu_j(g) = \int x^jg(x)\, {\rm d}x$, $j=0,1, \ldots$, 
provided that these integrals exist finite. Then,} the first-order term of the MISE bandwidth, $h_n$, has the following expression:
\beqn
h_n = C_0n^{-1/5},
\eeqn
where
\beqn
C_0 = \left(\frac{V_1}{4B_1}\right)^{1/5}.
\eeqn

However, {there are no results in the  literature that obtain} 
% it seems that the literature has never delved into 
the {higher-order} terms of the MISE {of the Nadaraya--Watson estimator}. {To undertake this task}, let us start by defining
\beqn
a &=& m(x)f(x),\\
e &=& f(x),\\
\hat a &=& \frac{1}{n}\sum\limits_{i=1}^n K_h\left(x-X_i\right)Y_i,\\
\hat e &=& \frac{1}{n}\sum\limits_{i=1}^n K_h\left(x-X_i\right).
\eeqn

Then, we have that
\beq\label{eq:mhtram_roto}
\hat m_h(x) = A+B+C+D+E+F,
\eeq
where
\beqn
A &=& \frac{\hat a}{e},\\
B &=& \frac{a(e-\hat e)}{e^2},\\
C &=& \frac{(\hat a-a)(e-\hat e)}{e^2},\\
D &=& \frac{a}{e}\frac{(e-\hat e)^2}{e^2},\\
E &=& \frac{\hat a-a}{e}\frac{(e-\hat e)^2}{e^2},\\
F &=& \frac{\hat a}{\hat e}\frac{(e-\hat e)^3}{e^3}.
\eeqn

{Expression (\ref{eq:mhtram_roto}) splits} $\hat m_h(x)$ as a sum of five terms with no random denominator plus an additional term, $F$, which has a random denominator. {This} {last term is} negligible with respect to the others. This approach will help us study the {higher-order} terms of the MISE {of \eqref{eq:nw}}. {Let us define} $S = A+B+C+D+E$, {then} we have {that}
\beqn
M_n(h) = \int\left[\E(S)-m(x)\right]^2f(x)\,dx+\int\var(S)f(x)\,dx+O\left(h^8+n^{-1}h^2\right).
\eeqn

\subsection{Asymptotic results}
{The asymptotic bias and variance of the cross-validation bandwidth minimizing \eqref{eq:cvtram} are derived {in this section}. For this, some previous {lemmas} are proved. The detailed proof of these results can be found in the supplementary material.
The following assumptions are needed:
\begin{itemize}
	\item[A1.] $K$ is a symmetric and differentiable kernel function.
	\item[A2.] For every {$j = 0, \dots, 6$,} the integrals $\mu_j(K)$, $\mu_j(K')$ and $\mu_j(K^2)$ exist and are finite.
	\item[A3.] The functions $m$ and $f$ are {eight times differentiable}.
	\item[A4.] The function $\sigma^2$ is {four times differentiable}.
\end{itemize}}

Lemma \ref{th:bias_var_mh} provides expressions for the first and second order terms of both the bias and the variance of $S$.

\begin{lemma}\label{th:bias_var_mh}
	{Under assumptions {\rm A1}--{\rm A4},} the bias and the variance of $S = A+B+C+D+E$ satisfy:
	\beqn
	{\rm E}(S)-m(x) &=& \mu_2(K)\left[\frac{1}{2}m''(x)+\frac{m'(x)f'(x)}{f(x)}\right]h^2\\\nonumber
	&+&\left\{\mu_4(K)\left[\frac{1}{24}m^{4)}(x)+\frac{1}{6}\frac{m'''(x)f'(x)}{f(x)}+\frac{1}{4}\frac{m''(x)f''(x)}{f(x)}\right.\right.\\\nonumber
	&+&\left.\left.\frac{1}{6}\frac{m'(x)f'''(x)}{f(x)}\right]-\mu_2(K)^2\frac{f''(x)}{f(x)}\left[\frac{1}{4}m''(x)+\frac{m'(x)f'(x)}{f(x)}\right]\right\}h^4\\\nonumber
	&+&O\left(h^6+n^{-1}h\right)
	\eeqn
and
	\beqn
	{\rm var}(S) &=& R(K)\sigma^2(x)f(x)^{-1}n^{-1}h^{-1}\\
	&+&\left\{\mu_2(K^2)f(x)^{-2}\left[\varphi_3(x)+\frac{1}{2}m(x)^2f''(x)-2\varphi_1(x)m(x)f(x)\right]\right.\\
	&-&\left.R(K)\mu_2(K)\sigma^2(x)f(x)^{-2}f''(x)\right\}n^{-1}h+O(n^{-1}h^2).
	\eeqn
	%	\beqn
	%	\var(S) &=& R(K)\sigma^2(x)f(x)^{-1}n^{-1}h^{-1}+\left[\mu_2(K^2)f(x)^{-2}\left(\frac{1}{2}f''(x)\sigma^2(x)\right.\right.\\
	%	&+&\left.\left.m'(x)^2f(x)+\frac{1}{2}{\sigma^2}''(x)f(x)+f'(x){\sigma^2}'(x)\right)\right.\\
	%	&-&\left.R(K)\mu_2(K)\sigma^2(x)f(x)^{-2}f''(x)\right]n^{-1}h+O\left(n^{-1}h^2\right).
	%	\eeqn
\end{lemma}

It follows from Lemma \ref{th:bias_var_mh} that
\beqn
M_n(h) = B_1 h^4 + V_1 n^{-1}h^{-1} + B_2 h^6 + V_2 n^{-1}h + O\left(h^8+n^{-1}h^2\right),
\eeqn
where
\beqn
B_2 &=& 2\mu_2(K)\int\left[\frac{1}{2}m''(x)+\frac{m'(x)f'(x)}{f(x)}\right]\left\{\mu_4(K)\left[\frac{1}{24}m^{4)}(x)+\frac{1}{6}\frac{m'''(x)f'(x)}{f(x)}\right.\right.\\
&+&\frac{1}{4}\frac{m''(x)f''(x)}{f(x)}+\left.\left.\frac{1}{6}\frac{m'(x)f'''(x)}{f(x)}\right]-\mu_2(K)^2\frac{f''(x)}{f(x)}\left[\frac{1}{4}m''(x)+\frac{m'(x)f'(x)}{f(x)}\right]\right\}f(x)\,dx,\\
V_2 &=& \int \left\{\mu_2(K^2)f(x)^{-2}\left[\frac{1}{2}f''(x)\sigma^2(x)+m'(x)^2f(x)+\frac{1}{2}{\sigma^2}''(x)f(x)+f'(x){\sigma^2}'(x)\right]\right.\\
&-&\left.R(K)\mu_2(K)\sigma^2(x)f(x)^{-2}f''(x)\right\}f(x)\,dx.
\eeqn

Lemma \ref{th:bias_var_cv1} provides expressions for the first and second order terms of both the expectation and variance of $CV_n'(h)$.
\begin{lemma}\label{th:bias_var_cv1}
Let us define
\beqn
A_1 &=& 12\mu_2(K)\mu_4(K)\int f(x)^{-1}\left[\frac{1}{2}m''(x)f(x)+m'(x)f'(x)\right]\left[\frac{1}{24}m^{4)}(x)f(x)\right.\\
&+&\left.\frac{1}{6}m'''(x)f'(x)+\frac{1}{4}m''(x)f''(x)\right]\,dx,\\
A_2 &=& \mu_2(K^2)\int f(x)^{-1}\left[\frac{1}{2}{\sigma^2}''(x)f(x)+{\sigma^2}'(x)f'(x)+\frac{1}{2}\sigma^2(x)f''(x)+m'(x)^2f(x)\right]\,dx,\\
R_1 &=& 32R(K)^2\mu_2(K)^2\int \sigma^2(x)f(x)^{-1}\left[\frac{1}{4}m''(x)^2f(x)^2+m'(x)m''(x)f(x)f'(x)\right.\\
&+&\left.m'(x)^2f'(x)^2\right]\,dx,\\
R_2 &=& 4\mu_2\left[(K')^2\right]\int \sigma^2(x)^2\,dx.
\eeqn

Then, {under assumptions {\rm A1}--{\rm A4},}
\beq
{\rm E}\left[CV_n'(h)\right] &=& 4B_1 h^3-V_1 n^{-1}h^{-2}+A_1h^5+A_2n^{-1}+O\left(h^7+n^{-1}h^2\right),\label{eq:espCV1_th}\\
%\var\left(CV_n'(h)\right) &=& 32R(K)^2\mu_2(K)^2\int \sigma^2(x)f(x)^{-1}\left(\frac{1}{4}m''(x)^2f(x)^2+m'(x)m''(x)f(x)f'(x)\right.\\\nonumber
%&+&\left.m'(x)^2f'(x)^2\right)\,dx\,n^{-1}h^2
%+4\mu_2\left((K')^2\right)\int \sigma^2(x)^2\,dx\,n^{-2}h^{-3}+O\left(n^{-1}h^4+n^{-2}h\right).\label{eq:varCV1_th}
{\rm var}\left[CV_n'(h)\right] &=& R_1n^{-1}h^2+R_2n^{-2}h^{-3}+O\left(n^{-1}h^4+n^{-2}h^{-1}\right).\label{eq:varCV1_th}
\eeq
\end{lemma}

Finally, Theorem \ref{th:dist_asint_h_tramposo}, which can be derived from \eqref{eq:hcvtaylor}, Lemma \ref{th:bias_var_mh} and Lemma \ref{th:bias_var_cv1}, {provides the asymptotic bias and variance of the cross-validation bandwidth that minimizes \eqref{eq:cvtram}.}

\begin{theorem}\label{th:dist_asint_h_tramposo}
{Under assumptions {\rm A1}--{\rm A4},} the {asymptotic} bias {and variance} of the bandwidth that minimizes \eqref{eq:cvtram} {are:}
\beqn
{\rm E}\left(\hat h_{CV,n}\right)-h_{n0} &=& Bn^{-3/5}+o\left(n^{-3/5}\right),\\
{\rm var}\left(\hat h_{CV,n}\right) &=& Vn^{-3/5}+o\left(n^{-3/5}\right),
\eeqn
where
\beqn
B &=& \frac{6B_2C_0^5+V_2-A_1C_0^5-A_2}{12B_1C_0^2+2V_1C_0^{-3}},\\
V &=& \frac{R_1C_0^2+R_2C_0^{-3}}{\left(12B_1C_0^2+2V_1C_0^{-3}\right)^2}.
\eeqn
\end{theorem}

\begin{corollary}\label{cor:limdist_hcv}
{Under {the} assumptions of Theorem \ref{th:dist_asint_h_tramposo}}, the asymptotic distribution of the bandwidth that minimizes \eqref{eq:cvtram} satisfies
\beqn
n^{3/10}\left(\hat h_{CV,n}-h_{n0}\right) \xrightarrow[]{d} N(0,V),
\eeqn
where the constant $V$ was defined in Theorem \ref{th:dist_asint_h_tramposo}.
\end{corollary}

\section{Bagged cross-validation bandwidth}\label{sec:baggedcv}
{While the cross-validation method 
% studied in the previous section 
is very useful to select reliable bandwidths in nonparametric regression, it also has the handicap of requiring a high computing time 
% to obtain a selector 
if the sample size is very large. This problem can be partially circumvented by using bagging \citep{Breiman1996},  a statistical technique belonging to the family of ensemble methods \citep{Opitz1999}, in the bandwidth selection procedure. In this section, we explain how bagging may be applied in the cross-validation context. Additionally, the asymptotic properties of the corresponding selector {are obtained}. 
% We will see that, 
Apart from the obvious reductions in computing time, the bagging cross-validation selector also presents better theoretical properties than the leave-one-out cross-validation bandwidth. This will be corroborated in the numerical studies presented in Sections \ref{sec:sims} and \ref{sec:realdata}.}

{Let $\mathcal{X}^* = \{(X_1^*,Y_1^*),\dots,(X_r^*,Y_r^*)\}$ be a random sample of size $r < n$ drawn without replacement from the i.i.d sample $\mathcal{X}$ defined in Section \ref{sec:model}.} This subsample is used to calculate a cross-validation bandwidth, $\hat h_r$. A rescaled version of $\hat h_r$, $\tilde h_r = (r/n)^{1/5}\hat h_r$, can be viewed as a feasible estimator of the optimal MISE bandwidth, $h_{n0}$, for $\hat m_h$. Bagging consists of repeating this resampling procedure independently $N$ times, leading to $N$ rescaled bandwidths, $\tilde h_{r,1},\dots,\tilde h_{r,N}$. The bagging bandwidth is then defined as:
\beq\label{eq:hcvbag}
\hat h(r,N) = \frac{1}{N}\suma[i=1][N] \tilde h_{r,i}.
\eeq

In the case of kernel density estimation, both the asymptotic properties and the empirical behavior of this type of bandwidth selector have been studied in \cite{HallRobinson2009} for $N = \infty$ and generalized in \cite{PaperBiometrika}, where the asymptotic properties of the bandwidth selector are  derived for the more practical case of a finite $N$.

{In the next section, the asymptotic bias and variance of the bagging bandwidth \eqref{eq:hcvbag} when using the Nadaraya--Watson  estimator \eqref{eq:nw} and considering the regression model \eqref{eq:model} {are obtained}. Moreover, its asymptotic distribution {is also derived}.}

\subsection{Asymptotic results}
Expressions for the {asymptotic} bias and the variance of \eqref{eq:hcvbag} are given in Theorem \ref{th:hcvbag_moments}.
{The following additional assumption is needed:
\begin{itemize}
\item[A5.] As $r,n \to \infty$, $r = o(n)$ and $N$ tends to a
        positive constant or $\infty$.
\end{itemize}}

\begin{theorem}\label{th:hcvbag_moments}
{Under assumptions {\rm A1}--{\rm A5},} the {asymptotic bias and the variance of the} bagged cross-validation bandwidth defined in \eqref{eq:hcvbag} {are:}
\beqn
{\rm E}\left[\hat h(r,N)\right]-h_{n0} &=& (B+C_1)r^{-2/5}n^{-1/5}+o\left(r^{-2/5}n^{-1/5}\right),\\
{\rm var}\left[\hat h(r,N)\right] &=& Vr^{-1/5}n^{-2/5}\left[\frac{1}{N}+\left(\frac{r}{n}\right)^2\right]+o\left(\frac{r^{-1/5}n^{-2/5}}{N}+r^{9/5}n^{-12/5}\right),
\eeqn
where the constants $B$ and $V$ were defined in Theorem \ref{th:dist_asint_h_tramposo} and the constant $C_1$ is defined in {expression {\rm (124)} in the supplementary material}.
\end{theorem}

\begin{corollary}\label{cor:limdist_hcvbag}
{Under the assumptions of {Theorem} \ref{th:hcvbag_moments},} the asymptotic distribution of the bagged cross-validation bandwidth defined in \eqref{eq:hcvbag} satisfies:
\beqn
\frac{r^{1/10}n^{1/5}}{\sqrt{\frac{1}{N}+\left(\frac{r}{n}\right)^2}}\left[\hat h(r,N)-h_{n0}\right] \xrightarrow[]{d} N(0, V),
\eeqn
where the constant $V$ was defined in Theorem \ref{th:dist_asint_h_tramposo}. In particular, {assuming} that $r = o\left(n/\sqrt{N}\right)$, then,
\beqn
r^{1/10}n^{1/5}\sqrt{N}\left[\hat h(r,N)-h_{n0}\right] \xrightarrow[]{d} N(0,V).
\eeqn
\end{corollary}

It should be noted that, while $\hat h_{CV,n}-h_{n0}$ converges in distribution at the rate $n^{-3/10}$, this result can be improved {with} the use of bagging and letting $r$ and $N$ tend to infinity at adequate rates. For example, if both $r$ and $N$ {tended} to infinity at the rate $\sqrt{n}$, then $\hat h(r,N)-h_{n0}$ would converge in distribution at the rate $n^{-1/2}$, which is {indeed} a faster rate of convergence than $n^{-3/10}$.

\section{Choosing an optimal subsample size}\label{sec:optsubsize}
In practice, an important step of our approach is, for fixed {values of} $n$ and $N$, choosing the \emph{optimal} subsample size, $r_0$. A possible optimality criterion {could} be to consider the value of $r$ that minimizes the {main} term of the variance of $\hat h(r,N)${. In this} case, we would get:
\beqn
r_0^{(1)} = \frac{n}{3\sqrt{N}}
\eeqn
and the variance of the bagging bandwidth would converge to zero at the rate
\beqn
\var\left[\hat h\left(r_0^{(1)},N\right)\right] \sim n^{-3/5}N^{-9/10},
\eeqn
which is a faster rate of convergence than that of the standard cross-validation bandwidth. In particular,
\beqn
\frac{\var\left[\hat h\left(r_0^{(1)},N\right)\right]}{\var\left(\hat h_{CV,n}\right)} \sim N^{-9/10}.
\eeqn

The obvious drawback of this criterion is that it would not allow any improvement in terms of computational {efficiency,} since the complexity of the algorithm would be the same as in the case of standard cross-validation, $O(n^2)$. {This} makes this choice of $r_0$ {inappropriate for} very large sample sizes. Another possible criterion for {selecting} $r_0$ would be to minimize, as a function of $r$,
 the asymptotic mean squared error (AMSE) of $\hat h(r, N)$, {given by:}
 %The obvious drawback of this first criterion lies in the fact that, unless $N$ tends to infinity at a suitable rate, the value of $r_0$ would be too close to the original sample size and so the computational gain would be minimal or even nonexistent. However, if $N$ were to tend to infinity at an adequate rate, an advantage of this criterion is the fact that no time would need to be wasted in estimating $r_0$. If, say, $N = \sqrt{n}$, then the computational complexity associated to bagged cross-validation would be $O(n^{5/4})$, while that of standard cross-validation would be $O(n^2)$. Another possible criterion for the selection of $r_0$ would be to minimize the asymptotic mean squared error (AMSE) of $\hat h(r, N)$, as a function of $r$,
\beq\label{eq:amse}
\mbox{AMSE}\left[\hat h(r,N)\right] = (B+C_1)^2r^{-4/5}n^{-2/5}+Vr^{-1/5}n^{-2/5}\left[\frac{1}{N}+\left(\frac{r}{n}\right)^2\right].
\eeq

Since $B$, $C_1$ and $V$ are unknown, we propose the following method to estimate
\beqn
r_0 = \argmin\limits_{r>1}\mbox{AMSE}\left[\hat h(r,N)\right].
\eeqn

\begin{enumerate}
	\item[\emph{Step 1.}] Consider $s$ subsamples of size $p<n$, drawn without replacement from the original sample of size $n$.
	\item[\emph{Step 2.}] For each of these subsamples, obtain an estimate, $\hat f$, of the marginal density function of the explanatory variable ({using} kernel density estimation, for example) and an estimate, $\hat m$, of the regression function (for instance, by fitting a polynomial of a certain degree). Do the same for the required derivatives of both $f$ and $m$.
	\item[\emph{Step 3.}] Use the estimates obtained in the previous step to compute the constants $B^{[i]}$, $C_1^{[i]}$ and $V^{[i]}$ for each subsample, where $i$ {($i=1, \ldots, s$)} denotes the {subsample index}.
	\item[\emph{Step 4.}] Compute the bagged estimates of the unknown constants, that is,
	\beqn
	\hat B &=& \frac{1}{s}\sum\limits_{i=1}^s B^{[i]},\\
	\hat C_1 &=& \frac{1}{s}\sum\limits_{i=1}^s C_1^{[i]},\\
	\hat V &=& \frac{1}{s}\sum\limits_{i=1}^s V^{[i]},
	\eeqn
	and obtain $\widehat{\mbox{AMSE}}\left[\hat h(r,N)\right]$ by plugging these bagged estimates into \eqref{eq:amse}.
	\item[\emph{Step 5.}] Finally, estimate $r_0$ by:
	\beqn
	\hat r_0 = \argmin\limits_{r>1}\widehat{\mbox{AMSE}}\left[\hat h(r,N)\right].
	\eeqn
\end{enumerate}

Additionally, {assuming} that $r = o\left(n/\sqrt{N}\right)$, {then}
\beqn
r_0^{(2)} = \left[-\frac{4(B+C_1)^2}{V}N\right]^{5/3}
\eeqn
and the rate of convergence to zero of the AMSE of the bagging bandwidth would be:
\beqn
\mbox{AMSE}\left[\hat h\left(r_0^{(2)},N\right)\right] \sim n^{-2/5}N^{-4/3}.
\eeqn

Hence,
\beqn
\frac{\mbox{AMSE}\left[\hat h\left(r_0^{(2)},N\right)\right]}{\mbox{AMSE}\left(\hat h_{CV,n}\right)} \sim n^{1/5}N^{-4/3},
\eeqn
and this ratio would tend to zero if $N$ {tended} to infinity at a rate faster than $n^{3/20}$. Furthermore, if we let $N = n^{3/20}$ and $r = r_0^{(2)}$, then the computational complexity of the algorithm would be $O\left(n^{13/20}\right)$, much lower than that of standard cross-validation. In fact, by selecting $r_0$ {in} this way, the complexity of the algorithm will only equal {to} that of standard cross-validation when $N$ tends to infinity at the rate $n^{6/13}$.

\section{Simulation studies}\label{sec:sims}
\begin{figure}[htb]
\begin{centering}
	\includegraphics[scale=0.35]{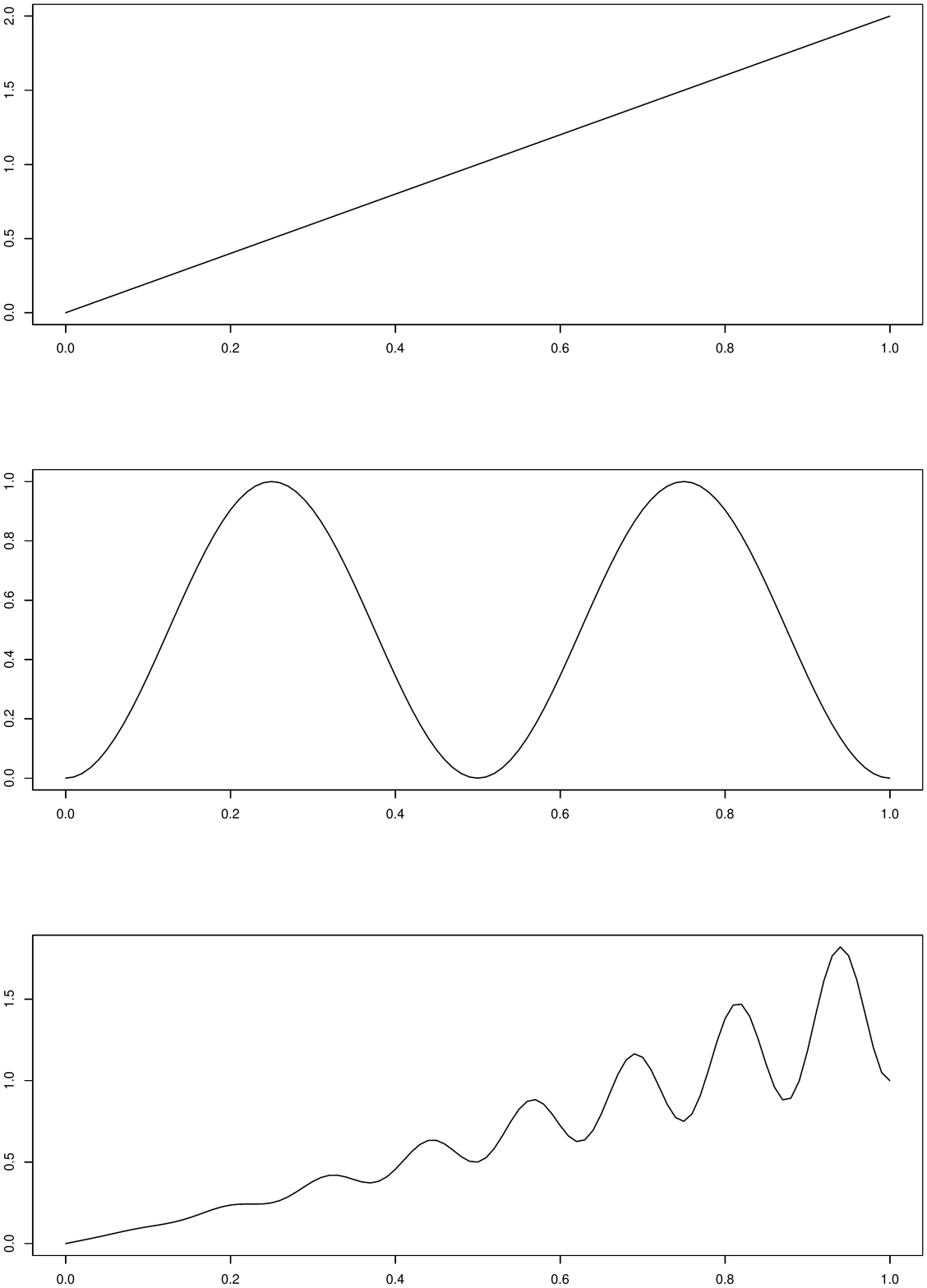}
	\par\end{centering}
	\caption{Regression function of models M1 (top), M2 (middle) and M3 (bottom).\label{fig:regmodels}}
\end{figure}

{The behavior of the leave-one-out and bagged cross-validation bandwidths is evaluated by simulation in this section. We considered} the following regression models:
\begin{enumerate}
	\item[M1:] $Y = m(X)+\varepsilon$, $m(x) = 2x$, $X \sim \mbox{Beta}(3,3)$, $\varepsilon \sim \mbox{N}(0,0.1^2)$,
	\item[M2:] $Y = m(X)+\varepsilon$, $m(x) = \sin(2\pi x)^2$, $X \sim \mbox{Beta}(3,3)$, $\varepsilon \sim \mbox{N}(0,0.1^2)$,
	\item[M3:] $Y = m(X)+\varepsilon$, $m(x) = x+x^2\sin(8\pi x)^2$, $X \sim \mbox{Beta}(3,3)$, $\varepsilon \sim \mbox{N}(0,0.1^2)$,
\end{enumerate}
whose regression functions are plotted in Figure \ref{fig:regmodels}. {The Gaussian kernel was used {for computing} the Nadaraya--Watson estimator throughout this section. Moreover, to reduce computing time in the simulations, we used binning to select the ordinary and the bagged cross-validation bandwidths.}

{In a first step, we empirically checked how close the bandwidths that minimize the MISE of \eqref{eq:nw} and \eqref{eq:nwtram} are. For this, we simulated $100$ samples of different sizes ($100$, $1000$ and $5000$) from models M1, M2 and M3 and compute the corresponding MISE curves for the standard Nadaraya--Watson estimator and for its modified version, given in \eqref{eq:nwtram}. For {the sake of brevity}, the plot containing these curves {is included} in the accompanying supplementary materials. {That plot shows} that the bandwidth that minimizes the MISE of \eqref{eq:nwtram} and the MISE of the standard Nadaraya--Watson estimator appear to be quite close, even for {moderate} sample sizes. {Of course}, the distance between the minima of both curves tends to zero as the sample size increases. Moreover, {the} standard cross-validation bandwidths {and} the modified cross-validation selectors (using the standard and the modified version of the Nadaraya--Watson estimator, respectively) {are} obtained for samples of sizes ranging from $600$ to $5000$ drawn from model M2. The corresponding figure is also included in the supplementary material. {It shows} that both bandwidth selectors provide similar results, which in turn get closer as $n$ increases.}

\begin{figure}[htb]
	\begin{centering}
		\includegraphics[width=\linewidth]{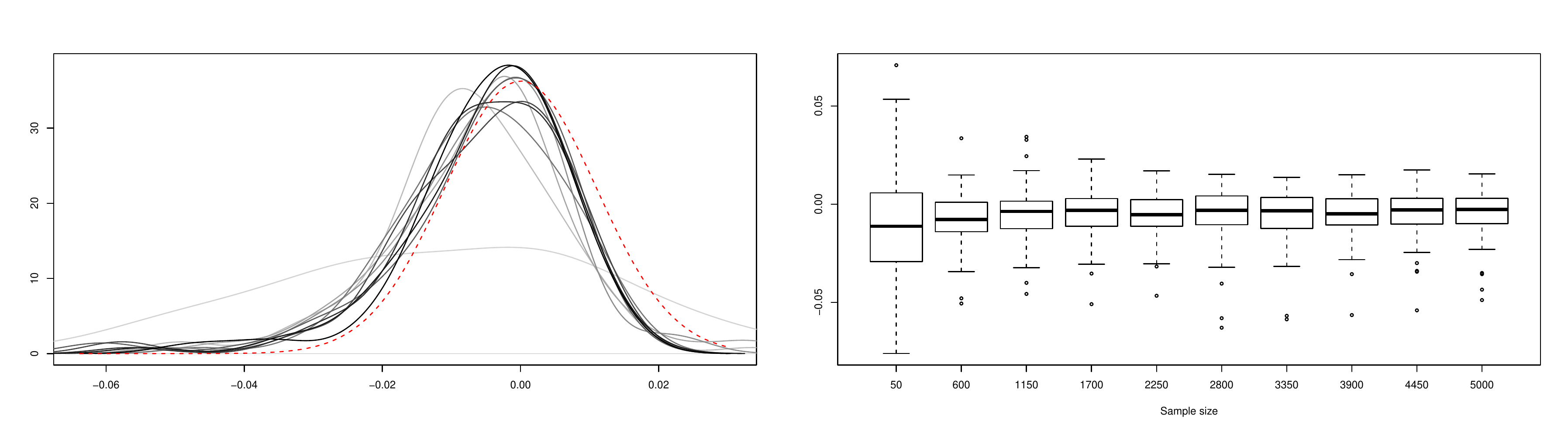}
		\par\end{centering}
	\caption{{Sampling distribution of $S_n = n^{3/10}(\hat h_{CV,n}-h_{n0})$: kernel density estimates {(left panel)} and boxplots {(right panel)}, for samples drawn from model M2 and considering values of $n$ between $50$ and $5000$. In the {left panel}, sharper lines correspond to larger values of $n$. {T}he limit distribution of $S_n$ is {also} shown (dashed line).}\label{fig:sampdist_hcv}}
\end{figure}

{In a second step, we checked how fast the statistic $S_n = n^{3/10}\left(\hat h_{CV,n}-h_{n0}\right)$ approaches its limit distribution. For this, 1000 samples of size $n$ were simulated from model M2 (with values of $n$ ranging from $50$ to $5000$) and the corresponding values of $S_n$ were computed. Figure \ref{fig:sampdist_hcv} shows the kernel density estimates and boxplots constructed using these samples of $S_n$. The empirical behavior observed in Figure \ref{fig:sampdist_hcv} is in agreement with the result reflected in Corollary \ref{cor:limdist_hcv}, since the sampling distribution of $S_n$ seems to tend to a normal distribution with zero mean and constant variance.}
{Similar plots were obtained when considering models M1 and M3. They are not shown here for {the sake of brevity}.}

%\begin{figure}[htb]
%\begin{centering}
%	\includegraphics[scale=0.40]{plots/boxplots_hcv}
%	\par\end{centering}
%	\caption{Sampling distribution of the standard cross-validation bandwidth (left), $\hat h_{CV, n}$, and sampling distribution of $n^{3/10}(\hat h_{CV,n}-h_{n0})$ (right), for samples drawn from model M2 and considering values of $n$ between $50$ and $5000$.\label{fig:boxplots_hcv}}
%\end{figure}
%
%\begin{figure}[htb]
%\begin{centering}
%	\includegraphics[scale=0.34]{plots/densidades_hcv-1}
%	\par\end{centering}
%	\caption{Kernel density estimates of $S_n = n^{3/10}(\hat h_{CV,n}-h_{n0})$, for $n$ ranging from $50$ to $5000$ and samples drawn from model M2. {The larger $n$, the sharper the lines.} The limit distribution of $S_n$ is shown {with a {dashed} line}.\label{fig:densidades_hcv}}
%\end{figure}

%Finally, Figures \ref{fig:boxplots_hbag_m1}, \ref{fig:boxplots_hbag_m2} and \ref{fig:boxplots_hbag_m3} show the sampling distribution of $\hat h(r,N)/h_{n0}$, for models M1, M2 and M3, respectively, and considering $n = 10^5$, $r \in \{100, 500, 1000, 5000, 10000\}$ and $N = 25$. For each model and subsample size, $1000$ samples were simulated to approximate the sampling distribution of $\hat h(r,N)/h_{n0}$.
{In the next part of the study, we focused on empirically analyzing the performance of the bagged cross-validation bandwidth $\hat h(r,N)$, given in \eqref{eq:hcvbag}, for different values of $n$, $r$ and $N$.}
Figure \ref{fig:boxplots_hbag_models} shows the sampling distribution of {$\hat h_n/h_{n0}$, where $\hat h_n$ denotes either the ordinary or bagged cross-validation bandwidth. For this, 1000 samples of size $n = 10^5$ from models M1, M2 and M3 were generated, considering in the case of $\hat h(r,N)$ the values} $r \in \{100, 500, 1000, 5000, 10000\}$ and $N = 25$. {For all three models, {it is observed} how the bias and variance of the bagging bandwidth decrease as the subsample size increases
and how its mean squared error seems to stabilize for values of $r$ close to 5000. Moreover, the behavior of the bagging selector
turns out to be quite positive even when considering subsample sizes as small as $r=100$, perhaps excluding the case of model M3 for which
the variance of the bagging bandwidth is still relatively high for $r=100$, although it undergoes a rapid reduction as the subsample size
increases slightly. 

\begin{figure}[htb]
\begin{center}
	\begin{subfigure}{0.3\textwidth}
		\includegraphics[width=\textwidth, height=\textwidth]{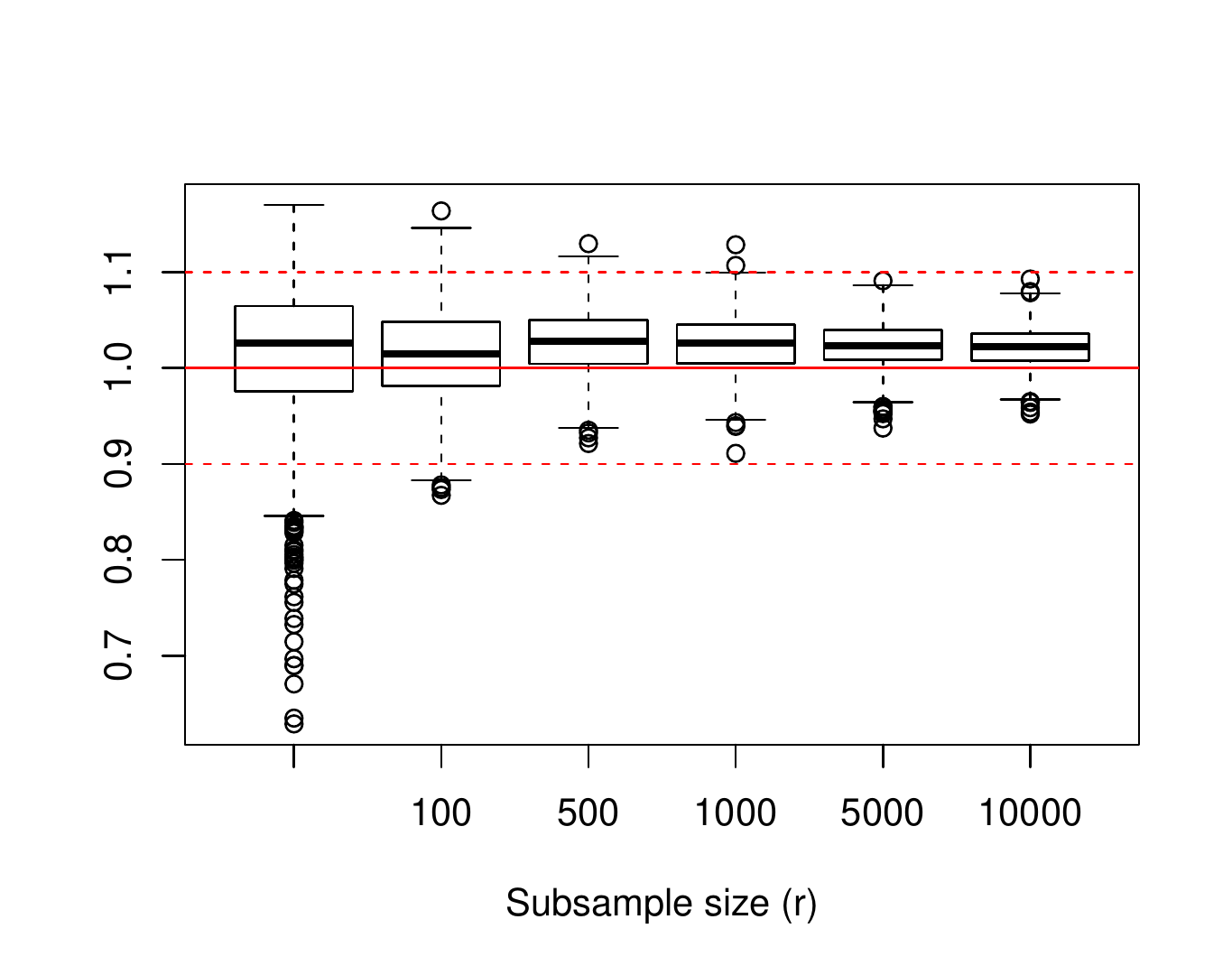}
	\end{subfigure}
	\begin{subfigure}{0.3\textwidth}
		\includegraphics[width=\textwidth, height=\textwidth]{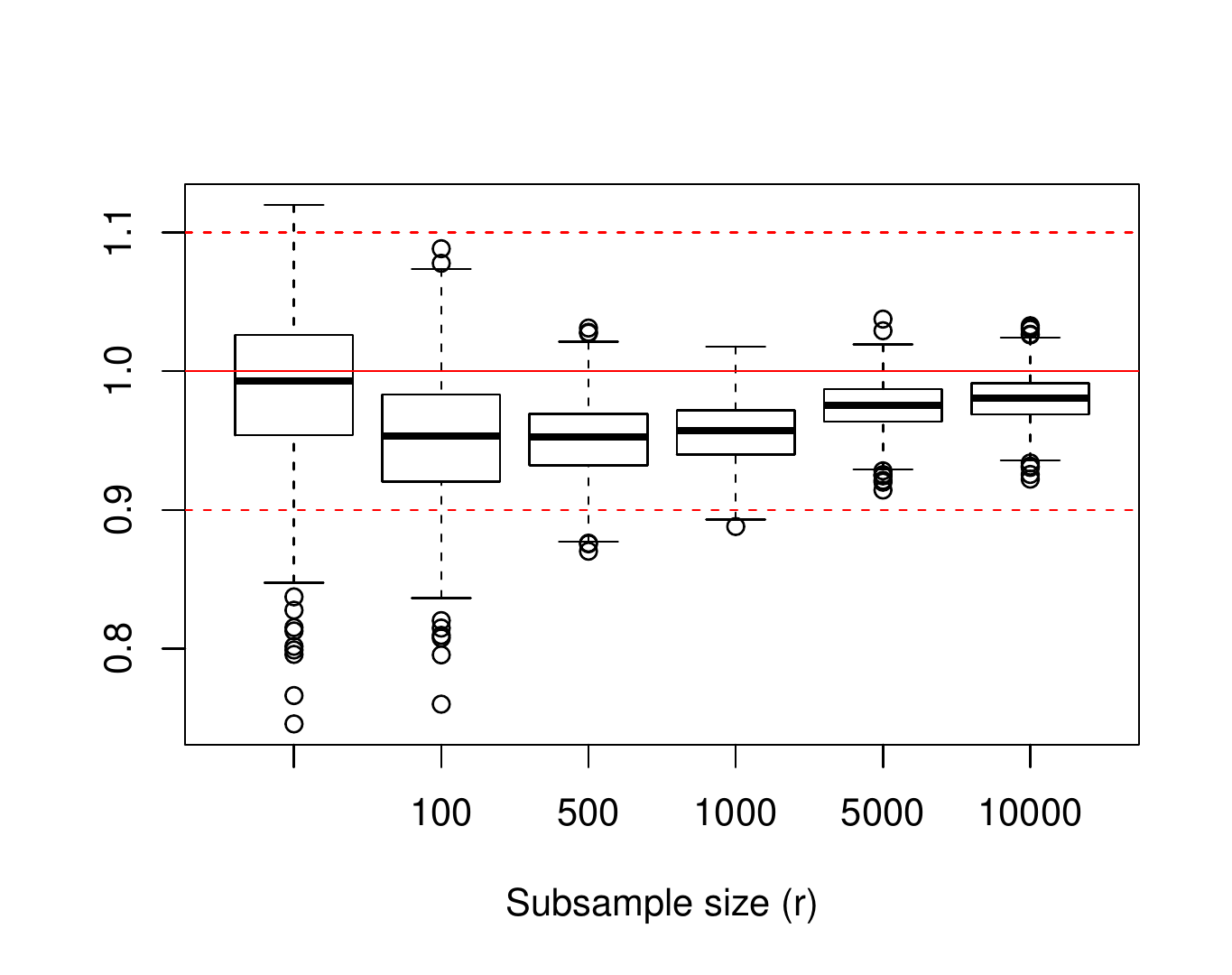}
	\end{subfigure}
	\begin{subfigure}{0.3\textwidth}
		\includegraphics[width=\textwidth, height=\textwidth]{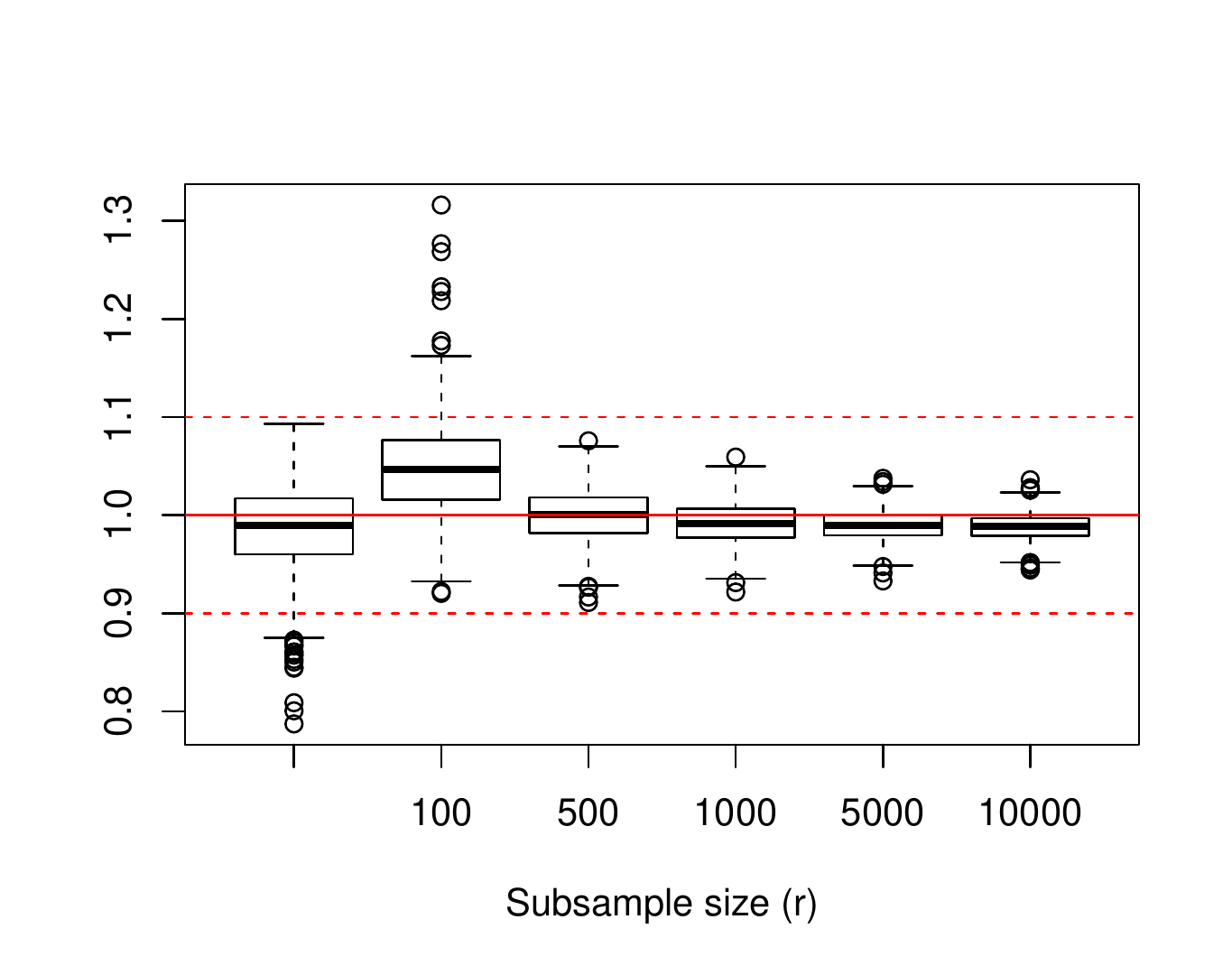}
	\end{subfigure}
	\end{center}
	\caption{Sampling distribution {of $\hat h_{CV,n}/h_{n0}$ (first boxplot on each panel) and $\hat h(r,N)/h_{n0}$ (second to sixth boxplots on each panel)} for models M1 ({left panel}), M2 ({central panel}) and M3 ({right panel}), where the considered subsample sizes are $r \in \{100,500,1000,5000,10^4\}$ and the number of subsamples is $N = 25$. {The original sample size is $n = 10^5$.}
	{{Dashed} lines are plotted
at values 0.9 and 1.1 for reference.}\label{fig:boxplots_hbag_models}}
\end{figure}

{The effect that $r$ has on the mean squared error of the bagged bandwidth is also illustrated in Table \ref{tab_relmse}, which shows the ratio of the mean squared errors of the bagged bandwidth and the ordinary cross-validation bandwidth, MSE$[\hat h(r,N)]/$MSE$(\hat h_{CV,n})$, for the three models.}}

\begin{table}[htb]
	\begin{center}
		%\scalebox{0.8}{
		{
			\begin{tabular}{c|ccc}
				\multicolumn{1}{c}{} & \multicolumn{3}{c}{Model}\\
				\cline{2-4}
				\multicolumn{1}{c}{} & M1 & M2 & M3\\
				\hline
				Subsample size ($r$) & \multicolumn{3}{c}{MSE ratio}\\
				\hline
				$100$ & $0.47$ & $1.47$ & $2.16$ \\
				$500$ & $0.32$ & $1.06$ & $0.33$\\
				$1,000$ & $0.26$ & $0.80$ & $0.23$\\
				$5,000$ & $0.19$ & $0.30$ & $0.17$\\
				$10,000$ & $0.16$ & $0.22$ & $0.16$\\
				\hline
			\end{tabular}
		%}
		}
	\end{center}

	\label{tab_relmse}
	
	\caption{ \label{tab_relmse}
		{Ratio of the mean squared errors of the bagged and the ordinary cross-validation bandwidths for models M1--M3. Different values of $r$ and $N = 25$ were considered.} }
\end{table}

\begin{figure}[htb]
\begin{centering}
	\includegraphics[scale=0.75,width=\linewidth]{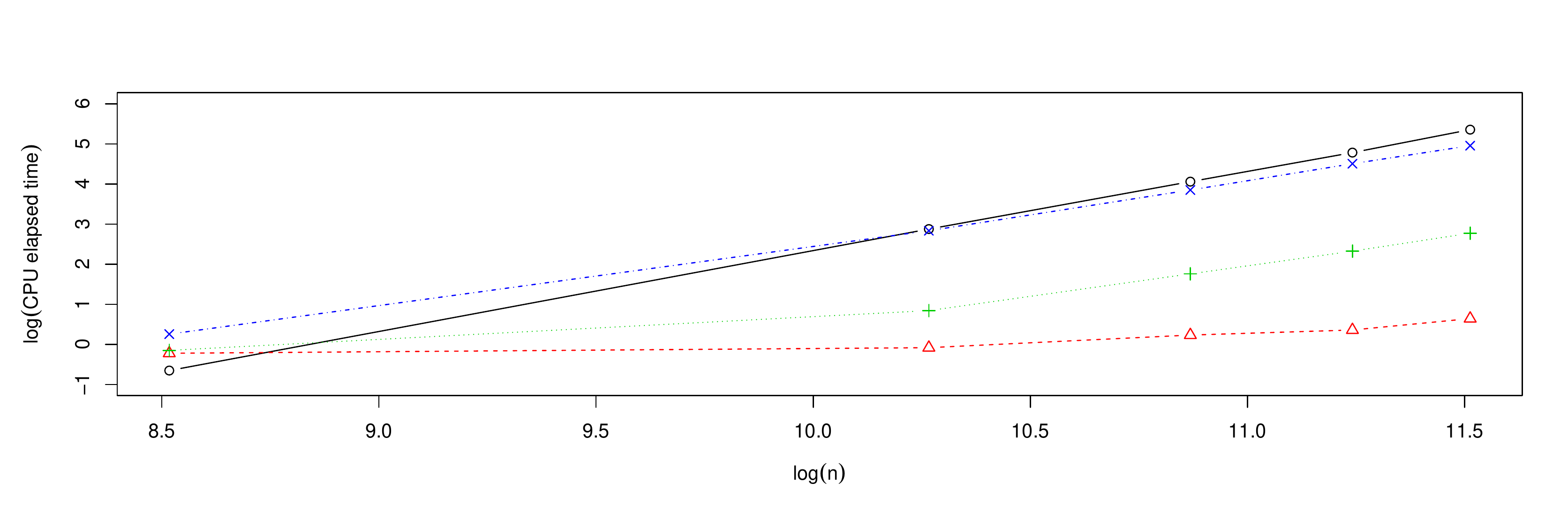}
	\par\end{centering}
	\caption{{CPU elapsed time (seconds) as a function of the sample size of standard cross-validation {(solid line-circles)} and bagged cross-validation. Both variables are shown on a logarithmic scale. A fixed number of subsamples was used, $N = 25$. Three growth rates for $r$ were considered, namely, $r=n^{0.7}$ {(dashed line-triangles)}, $r=n^{0.8}$ {(dotted line-pluses)} and $r=n^{0.9}$ {(dashed-dotted line-crosses)}.}\label{fig:cpu_times}}
\end{figure}

{Apart from a better statistical precision of the cross-validation bandwidths selected using bagging, another potential advantage of employing this approach is the reduction of computing times, especially with large sample sizes. To analyze this issue,}
Figure \ref{fig:cpu_times} shows{, as a function of the sample size, $n$,} the CPU elapsed times for {computing} the {standard and the bagged cross-validation bandwidths. {Both variables are shown on a logarithmic scale.} In the case of the bagging selector, three different subsample size values ($r$) depending on $n$ {were considered: $r=n^{0.7}$, $r=n^{0.8}$ and $r=n^{0.9}$}. C}alculations were performed in parallel using an Intel Core i5-8600K 3.6GHz CPU. {The sample size}  $n \in \{5000,28750,52500,76250,10^5\}$ {and} a fixed number of subsamples, $N = 25$, {were used}. {In this experiment, binning techniques were employed using a} number of bins {of} $0.1n$ {for standard cross-validation and} $0.1r$ in the case of bagged cross-validation. The time required to compute the bagged cross-validation bandwidth was measured considering {the} three possible growth rates for $r$, {mentioned above}. 

%{Fitting an appropriate model, these CPU elapsed times could be used to predict the computing times of the different selectors for larger sample sizes. Considering Figure \ref{fig:cpu_times}, the following log-linear model was used: 
%\beq
%\label{log_lin}
%T(n) = \alpha n^{\beta},
%\eeq
%where $T(n)$ denotes the CPU elapsed time as a function of the original sample size ($n$). In the case of the bagged cross-validation bandwidths we realized that, because they were computed in parallel, there was a ``residual time", $\gamma$, corresponding to the time required for the setting up of the parallel socket cluster that would have to be taken into account when fitting the log-linear model. The value of $\gamma$, which does not depend on $n$, $r$ or $N$, but only on the CPU and the number of cores used in the parallelization, was estimated to be $0.79$. Using this value, the corrected CPU elapsed times obtained for the bagged bandwidths, $T-\gamma$, were employed to fit the log-linear model \eqref{log_lin} estimating $\alpha, \beta > 0$ by least squares  and, subsequently, to make predictions. Table \ref{tab_pred_times} shows the predicted CPU elapsed time for ordinary and bagged cross-validation for large sample sizes. Although we should take these predictions with caution, the results in Table \ref{tab_pred_times} serve to illustrate the important reductions in computing time that bagging can provide for certain choices of $r$ and $N$, especially when working with very large sample sizes.} 
{Fitting an appropriate model, these CPU elapsed times could be used to predict the computing times of the different selectors for larger sample sizes. Considering Figure \ref{fig:cpu_times}, the following log-linear model was used: 
	\beq
	\label{log_lin}
	T(n) = \alpha n^{\beta},
	\eeq
	where $T(n)$ denotes the CPU elapsed time as a function of the original sample size, $n$. In the case of the bagged cross-validation bandwidths, there is a fixed time corresponding to the one required for the setting up of the parallel socket cluster. This time, which does not depend on $n$, $r$ or $N$, but only on the CPU and the number of cores used in the parallelization, was estimated to be $0.79$. Using this value, the corrected CPU elapsed times obtained for the bagged bandwidths, $T- 0.79$, were employed to fit the log-linear model \eqref{log_lin} estimating $\alpha, \beta > 0$ by least squares  and, subsequently, to make predictions. Table \ref{tab_pred_times} shows the predicted CPU elapsed time for ordinary and bagged cross-validation for large sample sizes. Although we should take these predictions with caution, the results in Table \ref{tab_pred_times} serve to illustrate the important reductions in computing time that bagging can provide for certain choices of $r$ and $N$, especially for very large sample sizes.} 
%Their values are collected in Table \ref{estimated_parameters}}.
 %are unknown. For each of the scenarios considered, we obtained the following estimates for the parameters of the previous model:
%\beqn
%\mbox{Standard CV:}&& \hat \alpha = 2.08\cdot 10^{-8},\, \hat \beta = 2.0,\\
%\mbox{Bagged CV ($r = n^{0.7}$):}&& \hat \alpha = 7.90\cdot 10^{-2},\, \hat \beta = 0.26,\\
%\mbox{Bagged CV ($r = n^{0.8}$):}&& \hat \alpha = 2.10\cdot 10^{-4},\, \hat \beta = 0.95,\\
%\mbox{Bagged CV ($r = n^{0.9}$):}&& \hat \alpha = 1.99\cdot 10^{-6},\, \hat \beta = 1.57.
%\eeqn
%
%

\begin{table}[h]
% \def~{\hphantom{0}}
% \tbl{Estimated parameters for the CPU elapsed time model.}{%
\begin{center}
{
	%\scalebox{0.8}{
\begin{tabular}{l|ccc}
\multicolumn{1}{c}{} & \multicolumn{3}{c}{Sample size ($n$)}\\
%\multicolumn{4}{r}{Sample size ($n$)}\\
\cline{2-4}
\multicolumn{1}{c}{}& $10^6$ & $10^7$ & $10^8$\\
\hline
Method & \multicolumn{3}{c}{Computing time}\\
\hline
Standard CV & $6$ hours & $24$ days & $7$ years \\
Bagged CV ($r = n^{0.7}, N = 25$) & $40$ seconds & $25$ minutes & $16$ hours \\
Bagged CV ($r = n^{0.8}, N = 25$) & $16$ minutes & $17$ hours & $45$ days \\
Bagged CV ($r = n^{0.9}, N = 25$) & $3$ hours & $11$ days & $2$ years\\
\hline
\end{tabular}
%}
}
\end{center}
% }
\label{tab_pred_times}

\caption{ \label{tab_pred_times}
{Predicted CPU elapsed time for the standard and the bagging cross-validation method using three different choices for the subsample size.} }
\end{table}

{Next, the influence of the number of subsamples, $N$, in the computing times of the bagged badwidths was studied.} {Similarly to Figure \ref{fig:cpu_times}}, Figure \ref{fig:cpu_times_varN} shows the CPU elapsed times for computing the cross-validation {bandwidths (standard and bagged). {For} the bagging method, {the} number of subsamples, $N$, {was selected depending on} the original sample size ($n$) {by} $N = \sqrt{n}$. The growth rates {used} for $r$ are the same as in the case of Figure \ref{fig:cpu_times}.

\begin{figure}[htb]
\begin{centering}
	\includegraphics[scale=0.75,width=\linewidth]{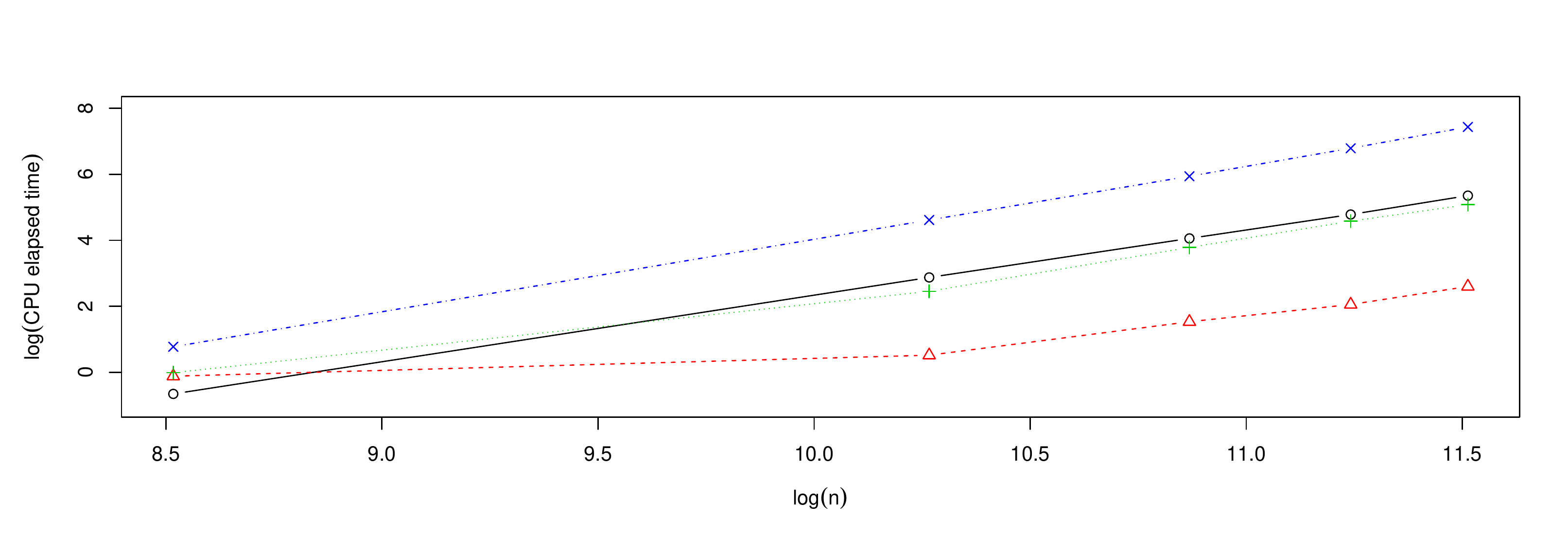}
	\par\end{centering}
	\caption{CPU elapsed time (seconds) as a function of the sample size of standard cross-validation (solid line-circles) and bagged cross-validation. Both variables are shown on a logarithmic scale. The number of subsamples grows with $n$ at the rate $N = \sqrt{n}$. Three growth rates for $r$ were considered, namely, $r=n^{0.7}$ {(dashed line-triangles)}, $r=n^{0.8}$ {(dotted line-pluses)} and $r=n^{0.9}$ {(dashed-dotted line-crosses)}.}
	\label{fig:cpu_times_varN}
\end{figure}

{It should also be stressed that although the quadratic complexity of the cross-validation algorithm is not so critical in terms of computing time for small sample sizes, even in these cases, the use of bagging can still lead to substantial reductions in mean squared error of the corresponding bandwidth selector with respect to the one selected by ordinary cross-validation.
In order to show this, $1000$ samples from model M1 of sizes $n \in \{50,500,5000\}$ were simulated and the ordinary and bagged cross-validation bandwidths for each of these samples were computed. In the case of the bagged cross-validation bandwidth, both the size of the subsamples and the number of subsamples were selected depending on $n$, choosing $r=N=4\sqrt{n}$. Figure \ref{fig:sampdist_hcv_hcvbag_smalln_M1} shows the sampling distribution of $\hat h_n/h_{n0}$, where $\hat h_n$ denotes either the ordinary or bagged cross-validation bandwidth. In the three scenarios, it can be observed the considerable reductions in variance produced by bagging more than offset the slight increases in bias, thus obtaining significant reductions in mean squared error with respect to the ordinary cross-validation bandwidth selector. Specifically, the relative reductions in mean squared error achieved by the bagged bandwidth turned out to be $69.3\%$, $90.1\%$ and $93.8\%$ for $n = 50$, $n = 500$ and $n = 5000$, respectively. This experiment was repeated with models M2 and M3, obtaining similar results.}

\begin{figure}[htb]
	\begin{centering}
		\includegraphics[width=\linewidth]{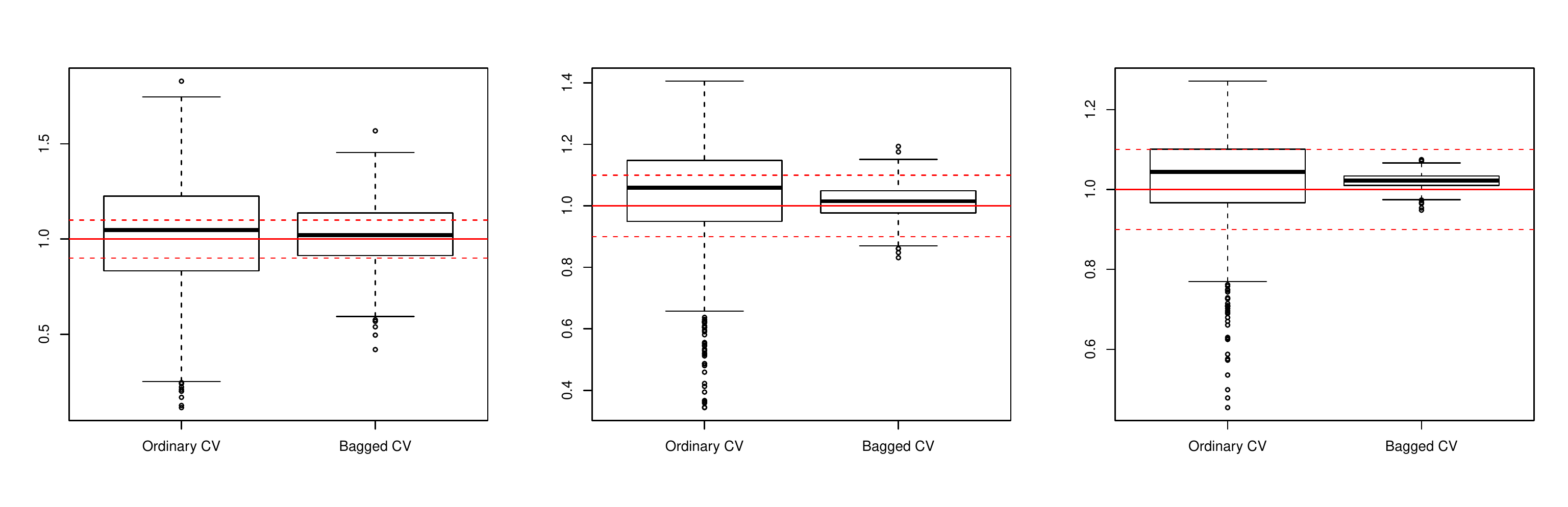}
		\par\end{centering}
	\caption{{Sampling distribution of $\hat h_n/h_{n0}$, where $\hat h_n$ denotes either the ordinary or bagged cross-validation bandwidth, for samples of size $n = 50$ (left panel), $n = 500$ (central panel) and $n = 5,000$ (right panel) drawn from model M1. The values of $r$ and $N$ were chosen as $r=N=4\sqrt{n}$.} {{Dashed} lines are plotted
at values 0.9 and 1.1 for reference.}}
	\label{fig:sampdist_hcv_hcvbag_smalln_M1}
\end{figure}

\section{Application to COVID-19 data}\label{sec:realdata}
{In order} to illustrate the performance of the techniques studied in the previous sections, {the COVID-19 dataset briefly mentioned in the introduction is considered}. 
{It consists of} a sample of size $n = 105,235$ which contains the age ({the} explanatory variable) and {the time in hospital} ({the} response variable) of people infected with COVID-19 in Spain from January 1, 2020 to December 20, 2020. Due to the high number of ties in {this dataset} and in order to avoid problems when performing cross-validation, we decided to remove the ties by jittering the data. In particular, three independent random samples of size $n$, $U_1$, $U_2$ and $U_3$, drawn from a continuous uniform distribution defined on the interval $(0,1)$, {were generated. Then,} $U_1$ {was added} to the original explanatory variable and {$U_2-U_3$} to the original response variable. Figure \ref{fig:covid_muestra} shows {scatterplots} for the {complete} sample as well as for three randomly chosen subsamples of size $1,000$. 

\begin{figure}[htb]
\begin{centering}
	\includegraphics[scale=0.37]{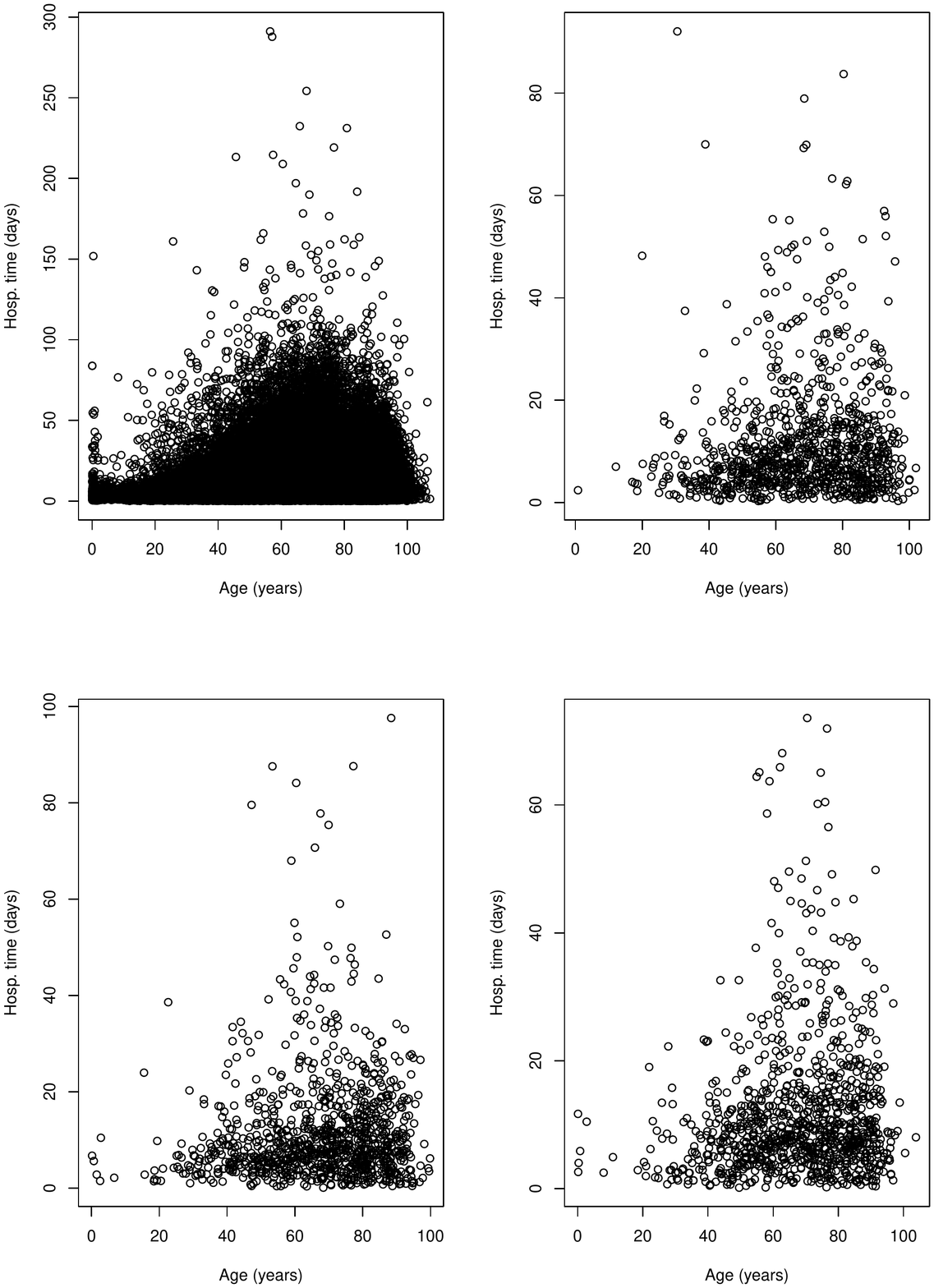}
	\par\end{centering}
	\caption{Whole COVID-19 sample (top left {panel}) as well as three randomly chosen subsamples of size $1000$.\label{fig:covid_muestra}}
\end{figure}

To compute the standard cross-validation bandwidth {using binning}, the number of bins was set to {$10,000$}, that is, roughly $10\%$ of the sample size. The value of the bandwidth thus obtained was $1.84$ and {computing it} took $72$ seconds. {For }the bagged bandwidth, $10$ subsamples of size $30,000$ {were considered. Binning was used again for} {each subsample, fixing} the number of bins to $3,000$. {The} calculations associated with each subsample were performed in parallel using $5$ cores. The value of the bagged bandwidth was $1.52$ and its {computing time was} $33$ seconds. Figure \ref{fig:covid_est} shows {the} Nadaraya--Watson {estimates} with both standard and bagged cross-validation bandwidths. {For comparative purposes,} the local linear {regression estimate} with direct plug-in bandwidth {\citep{Ruppert1995}} was also computed. 

\begin{figure}[htb]
\begin{centering}
	\includegraphics[scale=0.75]{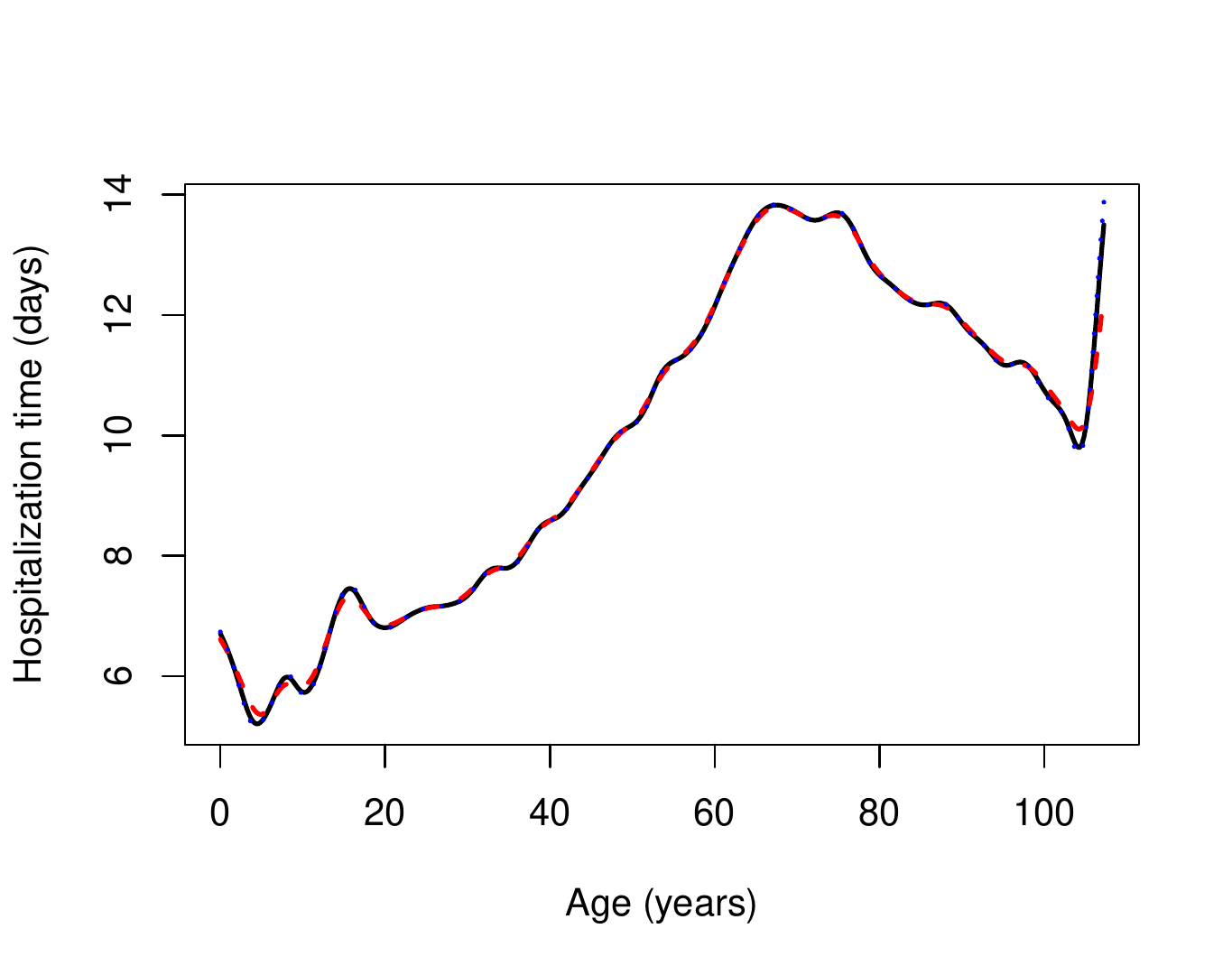}
	\par\end{centering}
	\caption{Kernel regression {estimates} for the COVID-19 data. The Nadaraya--Watson estimator with {standard (dashed line) and bagged (solid line) cross-validation bandwidths are shown. Additionally, the local linear estimator with plug-in bandwidth (dotted line) is also presented}.\label{fig:covid_est}}
\end{figure}

{Figure \ref{fig:covid_est} shows that} the Nadaraya--Watson estimator with standard cross-validation bandwidth produces a slightly smoother estimate than the one {obtained} with the bagged bandwidth, the latter being almost indistinguishable from the local linear {estimate computed} with direct plug-in bandwidth. {One} can conclude that the expected time {that} a person infected with COVID-19 will remain {in hospital} increases non-linearly with age for people under approximately $70$ years. This trend is reversed for people aged between $70$ and $100$ years. {This} could be due to the fact that patients in this age group are more likely to die and, therefore, end the hospitalization period prematurely. Finally, the expected hospitalization time grows again very rapidly with age for people over $100$ years of age, although this could be {caused by some boundary effect}, since the number of observations for people over $100$ years old is very small, specifically $155$, which corresponds to roughly $0.15\%$ of the total number of observations. In order to avoid this {possible} boundary effect, the estimators were also fitted to a modified version of the sample in which the explanatory variable was transformed {using its own} empirical distribution function. The resulting estimators are shown in Figure \ref{fig:est_nw_ll_tiempo_hosp_transfemp}, where the explanatory variable was returned to its original scale by means of {its} empirical quantile function.

\begin{figure}[htb]
\begin{centering}
	\includegraphics[scale=0.75]{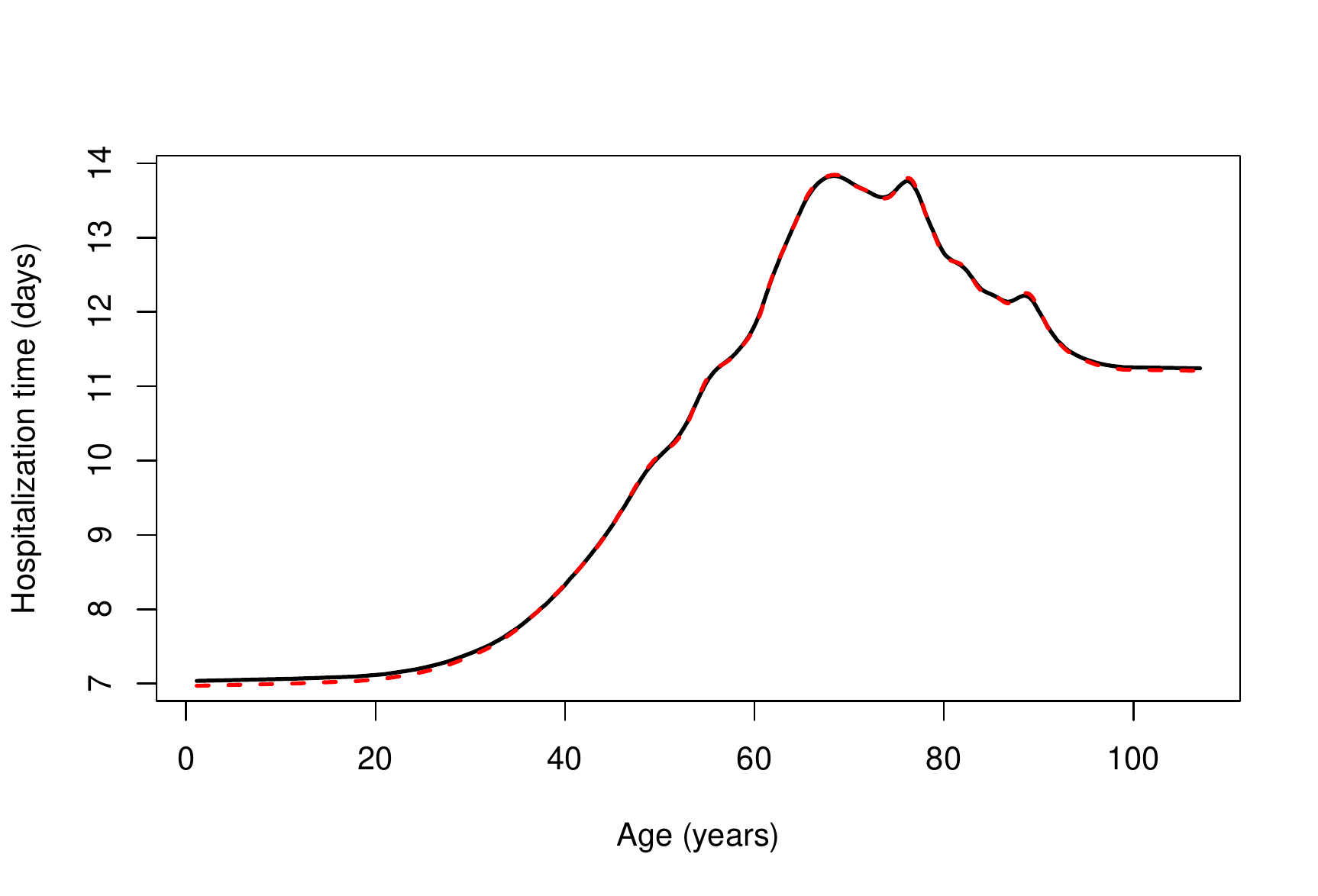}
	\par\end{centering}
	\caption{Kernel regression estimators for the COVID-19 data{, removing boundary effects}.  The Nadaraya--Watson estimator with  {standard (dashed line) and bagged (solid line) cross-validation bandwidths are shown.}\label{fig:est_nw_ll_tiempo_hosp_transfemp}}
\end{figure}

Finally, the same procedure was followed  to estimate the expected {time in hospital but splitting} the patients by {gender}, as shown in Figure \ref{fig:tiempo_hosp_por_sexos_ll}. {This figure shows that} the expected {time in hospital} is generally shorter for women, except for ages less than $30$ years or between $65$ and $85$ years. {Anyhow}, the difference in {mean time in hospital for men and women} never seems to exceed one day. {In Figure \ref{fig:tiempo_hosp_por_sexos_ll}, only the {Nadaraya--Watson} estimates computed with the bagged cross-validation bandwidths ($h = 0.03$ for men and $h = 0.028$ for women) are shown. Both the {Nadaraya--Watson} estimates with standard cross-validation bandwidths ($h = 0.028$ for men and $h = 0.023$ for women) and the local linear estimates with direct plug-in bandwidths produced very similar and graphically indistinguishable results from those shown in Figure \ref{fig:tiempo_hosp_por_sexos_ll}.}

\begin{figure}[htb]
\begin{centering}
	\includegraphics[scale=0.75]{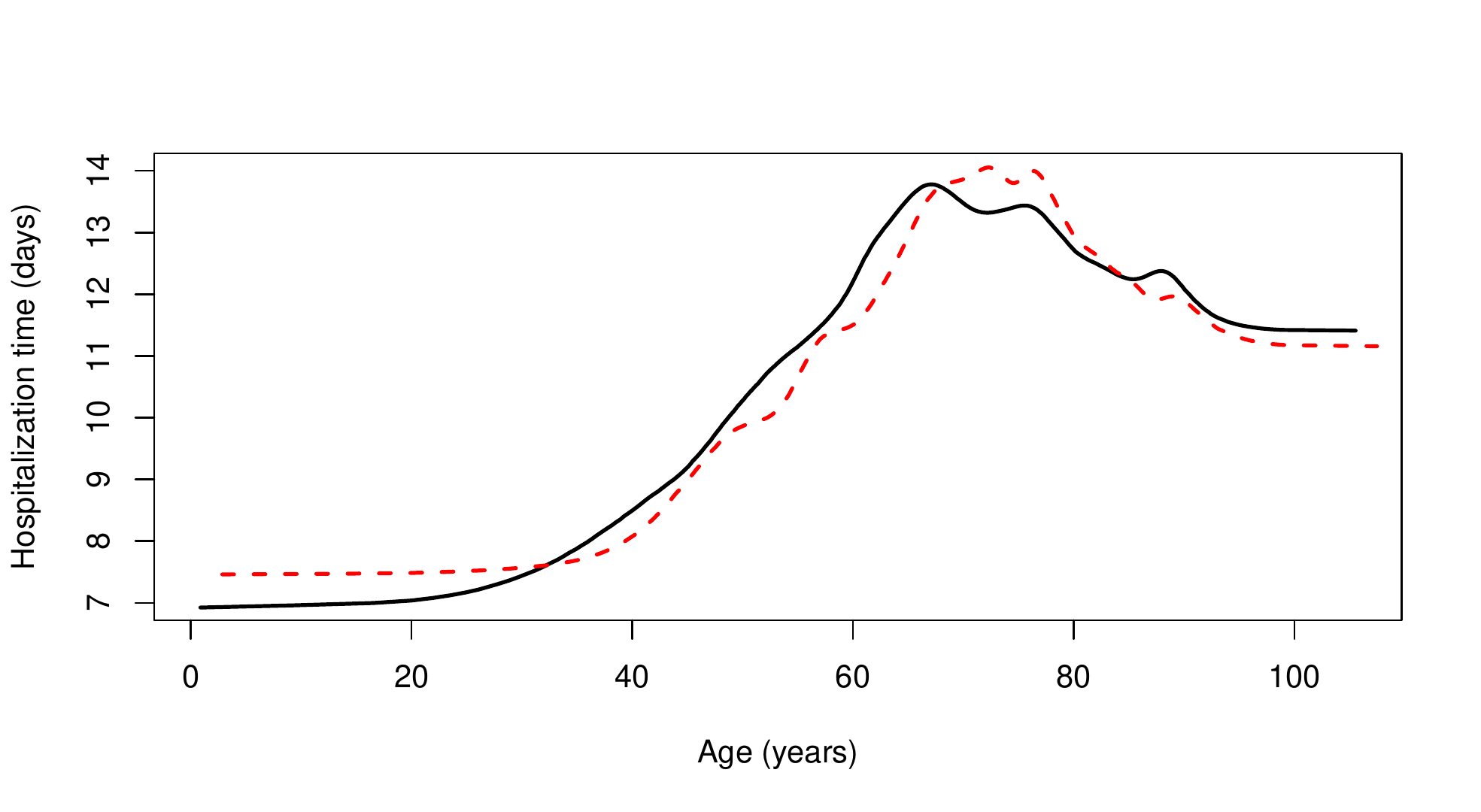}
	\par\end{centering}
	\caption{{Kernel regression estimators for the COVID-19 data by gender, removing boundary effects. The Nadaraya--Watson estimators with bagged cross-validation bandwidths are shown for male (solid line) and female (dashed line) patients.}\label{fig:tiempo_hosp_por_sexos_ll}}
\end{figure}

\section{Discussion}\label{sec:discussion}
{The} asymptotic properties of {the leave-one-out cross-validation bandwidth for the Nadaraya--Watson estimator considering a regression model with random design {have been studied in this paper}. Additionally,} {a bagged cross-validation selector have been also analyzed (theoretically and empirically)} as an alternative to standard leave-one-out cross-validation. The {advantage} of this bandwidth selector is twofold: (i) to gain computational {efficiency} with respect to standard leave-one-out cross-validation by applying the cross-validation algorithm to several subsamples of size $r < n$ rather than a single sample of size $n$, and (ii) to reduce the variability of the leave-one-out cross-validation bandwidth. {Although} the {new} bandwidth selector studied in the present paper can {outperform} the behavior of the standard cross-validation selector even for moderate sample sizes, improvements in computation time become truly significant only for large-{sized} samples. 

{The methodology presented} in this paper {can be applied} to {other} bandwidth selection techniques{, apart  from} cross-validation, {as mentioned} in \cite{PaperBiometrika}. {Extensions to bootstrap bandwidth selectors is an interesting topic for {a} future research. The bootstrap resampling plans proposed by {\cite{Cao1993}} can be used to derive {a} closed form for the bootstrap criterion function in nonparametric regression estimation, along the lines presented by {\cite{BarbeitoCaoSperlich2021}} who have dealt with matching and prediction}. 

{Another interesting} future {research} {topic} is the extension of the results presented in this paper to the case of the local linear estimator, whose behavior is known to be superior to that of the Nadaraya--Watson estimator{, especially in the boundary {regions}.} 

\section*{Supplementary Materials}

Supplementary materials include proofs and some plots completing the simulation study.

\section*{Acknowledgments}

The authors would like to thank the  Spanish Center for Coordinating Sanitary Alerts and Emergencies for kindly providing the COVID-19 hospitalization dataset.

\section*{Funding}

This research has been supported by MINECO grant  MTM2017-82724-R, and by Xunta de Galicia (Grupos de Referencia Competitiva ED431C-2020-14 and Centro de Investigaci\'on del Sistema Universitario de Galicia ED431G 2019/01), all of them through the ERDF.

\bibliography{paper}
\bibliographystyle{apalike}

\end{document}

\def\spacingset#1{\renewcommand{\baselinestretch}%
		{#1}\small\normalsize} \spacingset{1}

%\title{Bagging cross-validated bandwidths in regression with application to Big Data}
%\author{D. Barreiro-Ures \and R. Cao \and M. Francisco-Fern\'andez}
%\date{}
%
%\maketitle

\if1\blind
{
	\title{\bf Supplementary material for ``Bagging cross-validated {bandwidth selection in nonparametric regression estimation} with applications to {large-sized samples}"}
	\author{D. Barreiro-Ures
%	\thanks{
%			{\textit{This research has been supported by MINECO grant MTM2017-82724-R, and by the Xunta de Galicia (Grupos de Referencia Competitiva ED431C-2020-14 and Centro de Investigaci—n del Sistema universitario de Galicia ED431G 2019/01), all of them through the ERDF.}}}\hspace{.2cm}
			\\
		Department of Mathematics, {CITIC,} University of A Coru\~na\\
		and \\
		R. Cao \\
		Department of Mathematics, {CITIC,} University of A Coru\~na and {ITMATI}\\
		and \\
		M. Francisco-Fern\'andez \\
		Department of Mathematics, {CITIC,} University of A Coru\~na and {ITMATI}}
		
	\maketitle
} \fi

\if0\blind
{
	\bigskip
	\bigskip
	\bigskip
	\begin{center}
		{\LARGE\bf Supplementary material for ``Bagging cross-validated bandwidth selection in nonparametric regression estimation with applications to large-sized samples"}
	\end{center}
	\medskip
} \fi

\bigskip
\begin{abstract}
This supplementary material for ``Bagging cross-validated {bandwidth selection in nonparametric regression estimation} with applications to {large-sized samples}'' contains the proofs of the theoretical results included in the
main paper.  In addition, some plots completing the simulation study presented in the main paper {are} also provided. Specifically, a figure showing empirically the closeness between the MISE bandwidths when considering the Nadaraya--Watson estimator and when using its modified version, given in equation (8) of the main paper, in different scenarios{, is included}. Moreover, a figure presenting the relationship between the standard cross-validation bandwidths and the corresponding  modified cross-validation selectors (using the standard and modified version of the Nadaraya--Watson estimator, respectively) is also added.
\end{abstract}

%\noindent%
%{\it Keywords:}  cross-validation, subsampling, regression, Nadaraya--Watson, bagging%3 to 6 keywords, that do not appear in the title
%\vfill

\newpage
\spacingset{1.5} % DON'T change the spacing!

\section{Theoretical results}
\label{sec:teo}
This section includes the proofs of Lemmas 3.1 and 3.2, Theorems 3.1 and 4.1, and Corollaries 3.1 and 4.1  of the main paper. {The following assumptions are needed:
\begin{itemize}
	\item[A1.] $K$ is a symmetric and differentiable kernel function.
	\item[A2.] For every {$j = 0, \dots, 6$} the integrals $\mu_j(K)$, $\mu_j(K')$ and $\mu_j(K^2)$ exist and are finite.
	\item[A3.] The functions $m$ and $f$ are {eight times differentiable}.
	\item[A4.] The function $\sigma^2$ is {four times differentiable}.
	\item[A5.] As $r,n \to \infty$, $r = o(n)$ and $N$ tends to a
        positive constant or $\infty$.
\end{itemize}}

\begin{mlema}{3.1}
\label{th:bias_var_mh}
	{ Under assumptions {\rm A1}--{\rm A4},} the bias and the variance of $S = A+B+C+D+E$, where the terms $A$, $B$, $C$, $D$ and $E$ are defined just after equation {\rm (12)} of the main paper, satisfy:
	\beqn
	{\rm E}(S)-m(x) &=& \mu_2(K)\left[\frac{1}{2}m''(x)+\frac{m'(x)f'(x)}{f(x)}\right]h^2\\\nonumber
	&+&\left\{\mu_4(K)\left[\frac{1}{24}m^{4)}(x)+\frac{1}{6}\frac{m'''(x)f'(x)}{f(x)}+\frac{1}{4}\frac{m''(x)f''(x)}{f(x)}\right.\right.\\\nonumber
	&+&\left.\left.\frac{1}{6}\frac{m'(x)f'''(x)}{f(x)}\right]-\mu_2(K)^2\frac{f''(x)}{f(x)}\left[\frac{1}{4}m''(x)+\frac{m'(x)f'(x)}{f(x)}\right]\right\}h^4\\\nonumber
	&+&O\left(h^6+n^{-1}h\right),
	\eeqn
	\beqn
	{\rm var}(S) &=& R(K)\sigma^2(x)f(x)^{-1}n^{-1}h^{-1}\\
	&+&\left\{\mu_2(K^2)f(x)^{-2}\left[\varphi_3(x)+\frac{1}{2}m(x)^2f''(x)-2\varphi_1(x)m(x)f(x)\right]\right.\\
	&-&\left.R(K)\mu_2(K)\sigma^2(x)f(x)^{-2}f''(x)\right\}n^{-1}h+O(n^{-1}h^2).
	\eeqn
\end{mlema}

\begin{proof}[Proof of Lemma \ref{th:bias_var_mh}]
Let us start by defining
\beqn
\varphi_1(x) &=& f(x)^{-1}\left[\frac{1}{2}m''(x)f(x)+m'(x)f'(x)+\frac{1}{2}m(x)f''(x)\right],\\
\varphi_2(x) &=& f(x)^{-1}\left[\frac{1}{24}m^{4)}(x)f(x)+\frac{1}{6}m'''(x)f'(x)+\frac{1}{4}m''(x)f''(x)+\frac{1}{6}m'(x)f'''(x)\right.\\
&+&\left.\frac{1}{24}m(x)f^{4)}(x)\right],\\
\varphi_3(x) &=& \frac{1}{2}f''(x)\left[m(x)^2+\sigma^2(x)\right]+f(x)\left[m(x)m''(x)+m'(x)^2+\frac{1}{2}{\sigma^2}''(x)\right]\\
&+&f'(x)\left[2m(x)m'(x)+{\sigma^2}'(x)\right].
\eeqn

\beqn
\E\left[\frac{1}{n}\sum\limits_{i=1}^n K_h(x-X_i)Y_i\right] &=& \E\left[K_h(x-X_1)Y_1\right] = \E\left[K_h(x-X_1)\E(Y_1 \cond X_1)\right] \\
&=& E\left[K_h(x-X_1)m(X_1)\right]
= \int K_h(x-x_1)m(x_1)f(x_1)\,dx_1\\ &=& \int K(u)m(x-hu)f(x-hu)\,du\\
&=& m(x)f(x)+h^2\mu_2(K)f(x)\varphi_1(x)\\ &+&h^4\mu_4(K)f(x)\varphi_2(x)+O(h^6).
\eeqn

Therefore,
\beq\label{eq:espA}
\E(A) = m(x)+h^2\mu_2(K)\varphi_1(x)+h^4\mu_4(K){\varphi_2(x)}+O(h^6).
\eeq

\beqn
\E\left[\frac{1}{n}\sum\limits_{i=1}^n K_h(x-X_i)\right] &=& \E[K_h(x-X_1)] = \int K_h(x-x_1)f(x_1)\,dx_1\\ & = &\int K(u)f(x-hu)\,du
= f(x)+\frac{1}{2}h^2\mu_2(K)f''(x)\\ &+& \frac{1}{24}h^4\mu_4(K)f^{4)}(x)+O(h^6)
\eeqn
and, hence,
\beq\label{eq:espB}
\E(B) = -\frac{m(x)}{f(x)}\left[\frac{1}{2}h^2\mu_2(K)f''(x)+\frac{1}{24}h^4\mu_4(K)f^{4)}(x)\right]+O(h^6).
\eeq

\beqn
\E\left[Y_1K_h(x-X_1)^2\right] &=& \E\left[m(X_1)K_h(x-X_1)^2\right] = \int K_h(x-x_1)^2m(x_1)f(x_1)\,dx_1\\
&=& h^{-1}\int K(u)^2 m(x-hu)f(x-hu)\,du = R(K)m(x)f(x)h^{-1}\\
&+&h\mu_2(K^2)f(x)\varphi_1(x)+h^3\mu_4(K^2)f(x)\varphi_2(x)+O(h^5).
\eeqn

\beqn
\E(\hat a \hat e) &=& \E\left[n^{-2}\sum\limits_{i=1}^n\sum\limits_{j=1}^n Y_iK_h(x-X_1)K_h(x-X_j)\right] = n^{-2}\left\{n\E\left[Y_1K_h(x-X_1)^2\right]\right.\\
&+&\left.n(n-1)\E\left[Y_1K_h(x-X_1)K_h(x-X_2)\right]\right\} = n^{-1}\E\left[Y_1K_h(x-X_1)^2\right]\\
&+&\E\left[Y_1K_h(x-X_1)\right]\E\left[K_h(x-X_1)\right] = R(K)m(x)f(x)n^{-1}h^{-1}+m(x)f(x)^2\\
&+&h^2\mu_2(K)\left[\frac{1}{2}f''(x)m(x)f(x)+f(x)^2\varphi_1(x)\right]+h^4\left[\frac{1}{24}\mu_4(K)f^{4)}(x)m(x)f(x)\right.\\
&+&\left.\mu_4(K)f(x)^2\varphi_2(x)+\frac{1}{2}\mu_2(K)^2f''(x)f(x)\varphi_1(x)\right]+O(h^6+n^{-1}h).
\eeqn

Therefore,
\beq\label{eq:espC}
\E(C) = -R(K)m(x)f(x)^{-1}n^{-1}h^{-1}-\frac{1}{2}h^4\mu_2(K)^2f''(x)f(x)^{-1}\varphi_1(x)+O(h^6+n^{-1}h).
\eeq

\beqn
\E\left[K_h(x-X_1)^2\right] &=& \int K_h(x-x_1)f(x_1)\,dx_1 = h^{-1}\int K(u)^2f(x-hu)\,du\\
&=& R(K)f(x)h^{-1}+\frac{1}{2}h\mu_2(K^2)f''(x)+\frac{1}{24}h^3\mu_4(K^2)f^{4)}(x)+O(h^5),
\eeqn

\beqn
\E\left(\hat e^2\right) &=& \E\left[n^{-2}\sum\limits_{i=1}^n\sum\limits_{j=1}^n K_h(x-X_i)K_h(x-X_j)\right] = n^{-2}\left\{n\E\left[K_h(x-X_1)^2\right]\right.\\
&+&\left.n(n-1)\E\left[K_h(x-X_1)K_h(x-X_2)\right]\right\} \\ &=& n^{-1}\E\left[K_h(x-X_1)^2\right]+\E\left[K_h(x-X_1)\right]^2\\
&=& R(K)f(x)n^{-1}h^{-1}+f(x)^2+h^2\mu_2(K)f''(x)f(x)\\
&+& h^4\left[\frac{1}{12}\mu_4(K)f^{4)}(x)f(x) + \frac{1}{4}\mu_2(K)^2f''(x)^2\right]+O(h^6),
\eeqn
and, hence,
\beq\label{eq:espD}
\E(D) = R(K)m(x)f(x)^{-1}n^{-1}h^{-1}+\frac{1}{4}h^4\mu_2(K)^2f''(x)^2m(x)f(x)^{-2}+O(h^6+n^{-1}h).
\eeq

\beqn
\E\left[Y_1K_h(x-X_1)^3\right] = O(h^{-2}),
\eeqn

%\beqn
%\E[\hat a \hat e^2] &=& \E\left[n^{-3}\suma[i=1][n]\suma[j=1][n]\suma[k=1][n]Y_iK_h(x-X_i)K_h(x-X_j)K_h(x-X_k)\right]\\
%&=& n^{-3}\left(n\E\left[Y_1K_h(x-X_1)^3\right]+n(n-1)\E\left[Y_1K_h(x-X_1)K_h(x-X_2)^2\right]\right.\\
%&+&\left.2n(n-1)\E\left[Y_1K_h(x-X_1)^2K_h(x-X_2)\right]\right.\\
%&+&\left.n(n-1)(n-2)\E\left[Y_1K_h(x-X_1)K_h(x-X_2)K_h(x-X_3)\right]\right)\\
%&=& 3R(K)m(x)f(x)^2n^{-1}h^{-1}+\left(\frac{1}{2}\mu_2(K^2)m(x)f(x)f''(x)+R(K)\mu_2(K)\varphi_1(x)f(x)^2\right.\\
%&+&\left.2\mu_2(K^2)\varphi_1(x)f(x)^2+R(K)\mu_2(K)m(x)f(x)f''(x)\right)n^{-1}h+m(x)f(x)^3\\
%&+&\left(\mu_2(K)m(x)f(x)^2f''(x)+\mu_2(K)\varphi_1(x)f(x)^3\right)h^2+(REVISAR)h^4+O(h^6+n^{-1}h)
%\eeqn
\beqn
\E(\hat a \hat e^2) &=& \E\left[n^{-3}\suma[i=1][n]\suma[j=1][n]\suma[k=1][n]Y_iK_h(x-X_i)K_h(x-X_j)K_h(x-X_k)\right]\\
&=& n^{-3}\left\{n\E\left[Y_1K_h(x-X_1)^3\right]+n(n-1)\E\left[Y_1K_h(x-X_1)K_h(x-X_2)^2\right]\right.\\
&+&\left.2n(n-1)\E\left[Y_1K_h(x-X_1)^2K_h(x-X_2)\right]\right.\\
&+&\left.n(n-1)(n-2)\E\left[Y_1K_h(x-X_1)K_h(x-X_2)K_h(x-X_3)\right]\right\}\\
&=& 3R(K)m(x)f(x)^2n^{-1}h^{-1}+m(x)f(x)^3\\ 
&+& \left[\mu_2(K)m(x)f(x)^2f''(x)+\mu_2(K)\varphi_1(x)f(x)^3\right]h^2\\
&+&\left[\frac{1}{4}\mu_2(K)^2m(x)f(x)f''(x)^2+\mu_4(K)f(x)^3\varphi_2(x)+\mu_2(K)^2f(x)^2\varphi_1(x)f''(x)\right]h^4\\
&+&O(h^6+n^{-1}h)
\eeqn
and, so,
\beq\label{eq:espE}
\E(E) = O(h^6+n^{-1}h).
\eeq

Adding \eqref{eq:espA}, \eqref{eq:espB}, \eqref{eq:espC}, \eqref{eq:espD} and \eqref{eq:espE} we get that
\beqn
\E(S)-m(x) &=& \mu_2(K)\left[\frac{1}{2}m''(x)+\frac{m'(x)f'(x)}{f(x)}\right]h^2\\
&+&\left\{\mu_4(K)\left[\frac{1}{24}m^{4)}(x)+\frac{1}{6}\frac{m'''(x)f'(x)}{f(x)}+\frac{1}{4}\frac{m''(x)f''(x)}{f(x)}\right.\right.\\
&+&\left.\left.\frac{1}{6}\frac{m'(x)f'''(x)}{f(x)}\right]-\mu_2(K)^2\frac{f''(x)}{f(x)}\left[\frac{1}{4}m''(x)+\frac{m'(x)f'(x)}{f(x)}\right]\right\}h^4\\
&+&O\left(h^6+n^{-1}h\right).
\eeqn

{Regarding the variance of $S$, we have that}
\beq\label{eq:varS}
\var(S) &=& \var(A)+\var(B)+\var(C)+2\left[\cov(A,B)+\cov(A,C)\right.\nonumber\\
&+&\left.\cov(B,C)+\cov(A,D)+\cov(B,D)\right].
\eeq

\beqn
\E\left[Y_1^2K_h(x-X_1)^2\right] &=& \E\left[K_h(x-X_1)^2\E\left(Y_1^2 \cond X_1\right)\right] \\ & = & \E\left\{K_h(x-X_1)^2\E\left[\left(m(X_1)+\varepsilon_1\right)^2 \cond X_1\right]\right\}\\
&=& \E\left\{\left[m(X_1)^2+\sigma^2(X_1)\right]K_h(x-X_1)^2\right\} \\ 
&=& \int K_h(x-x_1)^2 \left[m(x_1)^2+\sigma^2(x_1)\right]f(x_1)\,dx_1\\
&=& h^{-1}\int K(u)^2\left[m(x-hu)^2+\sigma^2(x-hu)\right]f(x-hu)\,du\\
&=& R(K)\left[m(x)^2+\sigma^2(x)\right]f(x)h^{-1}+\mu_2(K^2)\varphi_3(x)h+O(h^3).
\eeqn

Therefore,

\beqn
\var(\hat a) &=& n^{-2}\var\left[\suma[i=1][n]Y_iK_h(x-X_i)\right] = n^{-1}\var\left[Y_1K_h(x-X_1)\right]\\
&=& n^{-1}\left\{\E\left[Y_1^2K_h(x-X_1)^2\right]-\E\left[Y_1K_h(x-X_1)\right]^2\right\}\\
&=& R(K)\left[m(x)^2+\sigma^2(x)\right]f(x)n^{-1}h^{-1}-m(x)^2f(x)^2n^{-1}\\ 
&+& \mu_2(K^2)\varphi_3(x)n^{-1}h+O(n^{-1}h^2)
\eeqn
and
\beq\label{eq:varA}
\var(A) & =& R(K)\left[m(x)^2+\sigma^2(x)\right]f(x)^{-1}n^{-1}h^{-1}-m(x)^2n^{-1}\nonumber\\
&+&\mu_2(K^2)\varphi_3(x)f(x)^{-2}n^{-1}h+O(n^{-1}h^2).
\eeq

\beqn
\var(\hat e) &=& n^{-2}\var\left[\suma[i=1][n]K_h(x-X_i)\right] = n^{-1}\var\left[K_h(x-X_1)\right]\\
&=& n^{-1}\left\{\E\left[K_h(x-X_1)^2\right]-\E\left[K_h(x-X_1)\right]^2\right\}\\
&=& R(K)f(x)n^{-1}h^{-1}-f(x)^2n^{-1}+\frac{1}{2}\mu_2(K^2)f''(x)n^{-1}h+O(n^{-1}h^2)
\eeqn
and so
\beq\label{eq:varB}
\var(B) &=& R(K)m(x)^2f(x)^{-1}n^{-1}h^{-1}-m(x)^2n^{-1}\nonumber\\
&+& \frac{1}{2}\mu_2(K^2)m(x)^2f(x)^{-2}f''(x)n^{-1}h+O(n^{-1}h^2).
\eeq

\beqn
\cov(\hat a, \hat e) &=& n^{-2}\suma[i=1][n]\suma[j=1][n]\cov\left[Y_iK_h(x-X_i),K_h(x-X_j)\right]\\
& = &n^{-1}\cov\left[Y_1K_h(x-X_1),K_h(x-X_1)\right]\\
&=& n^{-1}\left\{\E\left[Y_1K_h(x-X_1)^2\right]-\E\left[Y_1K_h(x-X_1)\right]\E\left[K_h(x-X_1)\right]\right\}\\
&=& R(K)m(x)f(x)n^{-1}h^{-1}-m(x)f(x)^2n^{-1}\\
&+&\mu_2(K^2)\varphi_1(x)f(x)n^{-1}h+O(n^{-1}h^2)
\eeqn
and hence
\beq\label{eq:covAB}
\cov(A,B) &=& -R(K)m(x)^2f(x)^{-1}n^{-1}h^{-1}+m(x)^2n^{-1}\nonumber\\
&-&\mu_2(K^2)\varphi_1(x)m(x)f(x)^{-1}n^{-1}h+O(n^{-1}h^2).
\eeq

\beqn
& & \cov\left[Y_1K_h(x-X_1),Y_1K_h(x-X_1)K_h(x-X_2)\right] \\
&=& \var\left[Y_1K_h(x-X_1)\right]\E\left[K_h(x-X_1)\right]\\
&=& R(K)\left[m(x)^2+\sigma^2(x)\right]f(x)^2h^{-1}-m(x)^2f(x)^3\\
&+&\left\{\mu_2(K^2)\varphi_3(x)f(x)\right.\\
&+&\left.\frac{1}{2}R(K)\mu_2(K)\left[m(x)^2+\sigma^2(x)\right]f(x)f''(x)\right\}h+O(h^2),
\eeqn
\beqn
& &\cov\left[Y_1K_h(x-X_1),Y_2K_h(x-X_2)K_h(x-X_1)\right] \\
&=& \cov\left[Y_1K_h(x-X_1),K_h(x-X_1)\right]\E\left[Y_1K_h(x-X_1)\right]\\
&=& R(K)m(x)^2f(x)^2h^{-1}-m(x)^2f(x)^3\\
&+&\left[\mu_2(K^2)\varphi_1(x)m(x)f(x)^2\right.\\
&+&\left.R(K)\mu_2(K)\varphi_1(x)m(x)f(x)^2\right]+O\left(h^2\right).
\eeqn

Therefore,
\beqn
\cov(\hat a, \hat a \hat e) &=& n^{-3}\suma[i=1][n]\suma[j=1][n]\suma[k=1][n]\cov\left[Y_iK_h(x-X_i),Y_jK_h(x-X_j)K_h(x-X_k)\right]\\
&=& n^{-1}\left\{\cov\left[Y_1K_h(x-X_1),Y_1K_h(x-X_1)K_h(x-X_2)\right]\right.\\
&+&\left.\cov\left[Y_1K_h(x-X_1),Y_2K_h(x-X_2)K_h(x-X_1)\right]\right\}+O(n^{-1}h^2)\\
&=& R(K)\left[2m(x)^2+\sigma^2(x)\right]f(x)^2n^{-1}h^{-1}-2m(x)^2f(x)^3n^{-1}\\
&+&\left\{\mu_2(K^2)\varphi_3(x)f(x)+\frac{1}{2}R(K)\mu_2(K)\left[m(x)^2+\sigma^2(x)\right]f(x)f''(x)\right.\\
&+&\left.\mu_2(K^2)\varphi_1(x)m(x)f(x)^2+R(K)\mu_2(K)\varphi_1(x)m(x)f(x)^2\right\}n^{-1}h+O(n^{-1}h^2)
\eeqn
and
\beq\label{eq:covAC}
\cov(A,C) &=& -R(K)\mu_2(K)\left\{\frac{1}{2}\left[m(x)^2+\sigma^2(x)\right]f(x)^{-2}f''(x)\right.\nonumber\\
&+&\left.\varphi_1(x)m(x)f(x)^{-1}\right\}n^{-1}h+O(n^{-1}h^2).
\eeq

\beqn
& & \cov\left[K_h(x-X_1),Y_1K_h(x-X_1)K_h(x-X_2)\right] \\ &=& \cov\left[K_h(x-X_1),Y_1K_h(x-X_1)\right]\E\left[K_h(x-X_1)\right]\\
&=& R(K)m(x)f(x)^2h^{-1}-m(x)f(x)^3+\left[\mu_2(K^2)\varphi_1(x)f(x)^2\right.\\
&+&\left.\frac{1}{2}R(K)\mu_2(K)m(x)f(x)f''(x)\right]h+O(h^2),
\eeqn
\beqn
& & \cov\left[K_h(x-X_1),Y_2K_h(x-X_2)K_h(x-X_1)\right] \\ &=& \var\left[K_h(x-X_1)\right]\E\left[Y_1K_h(x-X_1)\right]\\
&=& R(K)m(x)f(x)^2h^{-1}-m(x)f(x)^3\\
&+&\left[\frac{1}{2}\mu_2(K^2)m(x)f(x)f''(x)+R(K)\mu_2(K)\varphi_1(x)f(x)^2\right]h\\
&+&O(h^2).
\eeqn

Hence,
\beqn
\cov(\hat e, \hat a \hat e) &=& n^{-3}\suma[i=1][n]\suma[j=1][n]\suma[k=1][n]\cov\left[K_h(x-X_i),Y_jK_h(x-X_j)K_h(x-X_k)\right]\\
&=& n^{-1}\left\{\cov\left[K_h(x-X_1),Y_1K_h(x-X_1)K_h(x-X_2)\right]\right.\\
&+&\left.\cov\left[K_h(x-X_1),Y_2K_h(x-X_2)K_h(x-X_1)\right]\right\}\\
&=& 2R(K)m(x)f(x)^2n^{-1}h^{-1}-2m(x)f(x)^3n^{-1}\\
&+& \left[\mu_2(K^2)\varphi_1(x)f(x)^2+\frac{1}{2}\mu_2(K^2)m(x)f(x)f''(x)\right.\\
&+&\left.\frac{1}{2}R(K)\mu_2(K)m(x)f(x)f''(x)+R(K)\mu_2(K)\varphi_1(x)f(x)^2\right]n^{-1}h\\
&+& O(n^{-1}h^2)
\eeqn
and
\beq\label{eq:covBC}
\cov(B,C) &=& R(K)\mu_2(K)\left[\frac{1}{2}m(x)^2f(x)^{-2}f''(x)+\varphi_1(x)m(x)f(x)^{-1}\right]n^{-1}h\nonumber\\
&+&O(n^{-1}h^2).
\eeq

\beqn
& &\cov\left[Y_1K_h(x-X_1),K_h(x-X_1)K_h(x-X_2)\right] \\ &=& \cov\left[Y_1K_h(x-X_1),K_h(x-X_1)\right]\E\left[K_h(x-X_1)\right]\\
&=& R(K)m(x)f(x)^2h^{-1}-m(x)f(x)^3+\left[\mu_2(K^2)\varphi_1(x)f(x)^2\right.\\
&+&\left.\frac{1}{2}R(K)\mu_2(K)m(x)f(x)f''(x)\right]h+O(h^2).
\eeqn

Therefore,
\beqn
\cov(\hat a, \hat e^2) &=& n^{-3}\suma[i=1][n]\suma[j=1][n]\suma[k=1][n]\cov\left[Y_iK_h(x-X_i),K_h(x-X_j)K_h(x-X_k)\right]\\
&=& 2n^{-1}\cov\left[Y_1K_h(x-X_1),K_h(X-X_1)K_h(x-X_2)\right]+O(n^{-1}h^2)\\
&=& 2R(K)m(x)f(x)^2n^{-1}h^{-1}-2m(x)f(x)^3n^{-1}+\left[2\mu_2(K^2)\varphi_1(x)f(x)^2\right.\\
&+&\left.R(K)\mu_2(K)m(x)f(x)f''(x)\right]n^{-1}h+O(n^{-1}h^2)
\eeqn
and
\beq\label{eq:covAD}
\cov(A,D) = R(K)\mu_2(K)m(x)^2f(x)^{-2}f''(x)n^{-1}h+O(n^{-1}h^2).
\eeq

\beqn
\cov\left[K_h(x-X_1),K_h(x-X_1)K_h(x-X_2)\right] &=& \var\left[K_h(x-X_1)\right]\E\left[K_h(x-X_1)\right]\\
&=& R(K)f(x)^2h^{-1}-f(x)^3\\&+&\left[\frac{1}{2}\mu_2(K^2)f(x)f''(x)\right.\\
&+&\left.\frac{1}{2}R(K)\mu_2(K)f(x)f''(x)\right]h+O(h^2).
\eeqn

Hence,
\beqn
\cov(\hat e, \hat e^2) &=& n^{-3}\suma[i=1][n]\suma[j=1][n]\suma[k=1][n]\cov\left[K_h(x-X_i),K_h(x-X_j)K_h(x-X_k)\right]\\
&=& 2n^{-1}\cov\left[K_h(x-X_1),K_h(x-X_1)K_h(x-X_2)\right]+O(n^{-1}h^2)\\
&=& 2R(K)f(x)^2n^{-1}h^{-1}-2f(x)^3n^{-1}\\
&+&\left[\mu_2(K^2)f(x)f''(x)+R(K)\mu_2(K)f(x)f''(x)\right]n^{-1}h+ O(n^{-1}h^2)
\eeqn
and
\beq\label{eq:covBD}
\cov(B,D) = -R(K)\mu_2(K)m(x)^2f(x)^{-2}f''(x)n^{-1}h+O(n^{-1}h^2).
\eeq

Hence,
\beqn
\var(\hat a \hat e) &=& R(K)\left[4m(x)^2+\sigma^2(x)\right]f(x)^3n^{-1}h^{-1}-4m(x)^2f(x)^4n^{-1}\\
&+&\left\{\mu_2(K^2)\varphi_3(x)f(x)^2+R(K)\mu_2(K)\left[2m(x)^2+\sigma^2(x)\right]f(x)^2f''(x)\right.\\
&+&\left.4R(K)\mu_2(K)\varphi_1(x)m(x)f(x)^3+2\mu_2(K^2)\varphi_1(x)m(x)f(x)^3\right.\\
&+&\left.\frac{1}{2}\mu_2(K^2)m(x)^2f(x)^2f''(x)\right\}n^{-1}h+O\left(n^{-1}h^2\right)
\eeqn
and
\beq\label{eq:varC}
\var(C) = O\left(n^{-1}h^2\right).
\eeq

The remaining variances and covariances were not explicitly calculated because they are clearly negligible with respect to $n^{-1}h$. Namely,
\beqn
\var(D) = \var(E) = \cov(C,D) = \cov(A,E) = \cov(B,E) = \cov(C,E) = O(n^{-1}h^2).
\eeqn

Therefore, plugging \eqref{eq:varA}--\eqref{eq:varC} into \eqref{eq:varS} yields
\beqn
\var(S) &=& R(K)\sigma^2(x)f(x)^{-1}n^{-1}h^{-1}\\&+&\left\{\mu_2(K^2)f(x)^{-2}\left[\varphi_3(x)+\frac{1}{2}m(x)^2f''(x)-2\varphi_1(x)m(x)f(x)\right]\right.\\
&-&\left.R(K)\mu_2(K)\sigma^2(x)f(x)^{-2}f''(x)\right\}n^{-1}h+O(n^{-1}h^2).
\eeqn
%\beqn
%\var(S) &=& R(K)\sigma^2(x)f(x)^{-1}n^{-1}h^{-1}+\left[\mu_2(K^2)f(x)^{-2}\left(\frac{1}{2}f''(x)\sigma^2(x)+m'(x)^2f(x)+\frac{1}{2}f(x){\sigma^2}''(x)\right.\right.\\
%&+&\left.\left.f'(x){\sigma^2}'(x)\right)-R(K)\mu_2(K)\sigma^2(x)f(x)^{-2}f''(x)\right]n^{-1}h+O(n^{-1}h^2).
%\eeqn
\end{proof}

Lemma \ref{th:bias_var_cv1} provides expressions for the first and second order terms of both the expectation and variance of $CV_n'(h)${, where $CV_n(h)$ is given in equation (9) of the main paper (with notation $\widetilde{CV}_n(h)$)}.

\begin{mlema}{3.2}
\label{th:bias_var_cv1}
Let us define
\beqn
A_1 &=& 12\mu_2(K)\mu_4(K)\int f(x)^{-1}\left[\frac{1}{2}m''(x)f(x)+m'(x)f'(x)\right]\left[\frac{1}{24}m^{4)}(x)f(x)\right.\\
&+&\left.\frac{1}{6}m'''(x)f'(x)+\frac{1}{4}m''(x)f''(x)\right]\,dx,\\
A_2 &=& \mu_2(K^2)\int f(x)^{-1}\left[\frac{1}{2}{\sigma^2}''(x)f(x)+{\sigma^2}'(x)f'(x)+\frac{1}{2}\sigma^2(x)f''(x)+m'(x)^2f(x)\right]\,dx,\\
R_1 &=& 32R(K)^2\mu_2(K)^2\int \sigma^2(x)f(x)^{-1}\left[\frac{1}{4}m''(x)^2f(x)^2+m'(x)m''(x)f(x)f'(x)\right.\\
&+&\left.m'(x)^2f'(x)^2\right]\,dx,\\
R_2 &=& 4\mu_2\left[(K')^2\right]\int \sigma^2(x)^2\,dx.
\eeqn
Then, {under assumptions {\rm A1}--{\rm A4},}
\beq
{\rm E}\left[CV_n'(h)\right] &=& 4B_1 h^3-V_1 n^{-1}h^{-2}+A_1h^5+A_2n^{-1}+O\left(h^7+n^{-1}h^2\right),\label{eq:espCV1_th}\\
%\var\left(CV_n'(h)\right) &=& 32R(K)^2\mu_2(K)^2\int \sigma^2(x)f(x)^{-1}\left(\frac{1}{4}m''(x)^2f(x)^2+m'(x)m''(x)f(x)f'(x)\right.\\\nonumber
%&+&\left.m'(x)^2f'(x)^2\right)\,dx\,n^{-1}h^2
%+4\mu_2\left((K')^2\right)\int \sigma^2(x)^2\,dx\,n^{-2}h^{-3}+O\left(n^{-1}h^4+n^{-2}h\right).\label{eq:varCV1_th}
{\rm var}\left[CV_n'(h)\right] &=& R_1n^{-1}h^2+R_2n^{-2}h^{-3}+O\left(n^{-1}h^4+n^{-2}h^{-1}\right),\label{eq:varCV1_th}
\eeq
{where $B_1$ and $V_1$ are the main terms of the bias and the variance of the MISE of the {Nadaraya--Watson} estimator, given by:
\beqn
B_1 &=& \frac{1}{4}\mu_2(K)^2\int \left[m''(x)+2\frac{m'(x)f'(x)}{f(x)}\right]^2 f(x)\,dx,\\
V_1 &=& R(K)\int \sigma^2(x)\, dx,
\eeqn
}
\end{mlema}

\begin{proof}[Proof of Lemma \ref{th:bias_var_cv1}]
If we define
\beqn
\alpha_1(u) &=& K(u)+uK'(u),\\
\beta_1(u,v) &=& K(u)K(v)+K(u)K'(v)v,\\
\alpha_{1h}(u) &=& h^{-1}\alpha_1(u/h),\\
\beta_{1h}(u,v) &=& h^{-1}\beta_1(u/h,v/h),\\
\tilde \beta_{1h}(u) &=& \beta_{1h}(u,u),
\eeqn
then $CV_n'(h)$ can be expressed as follows:
\beq\label{eq:cvprimatram1}
CV_n'(h) &=& \frac{2}{n(n-1)^2h}\suma[i=1][n]\suma[\substack{j=1\\j \neq i}][n]\suma[\substack{k=1\\k \neq i}][n]\left\{\frac{Y_i-m(X_i)}{f(X_i)}[Y_j-m(X_i)]\alpha_{1h}(X_i-X_j)\right.\nonumber\\
&-&\left.\frac{1}{f(X_i)^2h}[Y_j-m(X_i)][Y_k-m(X_i)]\beta_{1h}(X_i-X_j,X_i-X_k)\right\}.
\eeq

Since
\beqn
& & \E\left\{\frac{Y_i-m(X_i)}{f(X_i)}[Y_j-m(X_i)]\alpha_{1h}(X_i-X_j)\right\} \\&=& \E\left\{\frac{Y_j-m(X_i)}{f(X_i)}\alpha_{1h}(X_i-X_j)\E\left[Y_i-m(X_i)\cond X_i\right]\right\} = 0,
\eeqn
we have that
\beq\label{eq:espCV1}
\E\left[CV_n'(h)\right] = -\frac{2}{(n-1)h^2}\left[E_1+(n-2)E_2\right],
\eeq
where
\beqn
E_1 &=& \E\left\{f(X_1)^{-2}\tilde \beta_{1h}(X_1-X_2)\left[Y_2-m(X_1)\right]^2\right\},\\
E_2 &=& \E\left\{f(X_1)^{-2}\beta_{1h}(X_1-X_2,X_1-X_3)[Y_2-m(X_1)][Y_3-m(X_1)]\right\}.
\eeqn

We have that
\beq\label{eq:E1}
E_1 &=& \E\left(f(X_1)^{-2}\tilde \beta_{1h}(X_1-X_2)\E\left\{[Y_2-m(X_1)]^2\cond X_1,X_2\right\}\right)\nonumber\\\nonumber
&=& \E\left(f(X_1)^{-2}\tilde\beta_{1,h}(X_1-X_2)\left\{\sigma^2(X_2)+[m(X_2)-m(X_1)]^2\right\}\right)\\\nonumber
&=& \iint f(x_1)^{-2}\tilde\beta_{1h}(x_1-x_2)\left\{\sigma^2(x_2)+[m(x_2)-m(x_1)]^2\right\}f(x_1)f(x_2)\,dx_1dx_2\\\nonumber
&=& \iint f(x_1)^{-1}\tilde\beta_1(u)\left\{\sigma^2(x_1-hu)+[m(x_1-hu)-m(x_1)]^2\right\}f(x_1-hu)\,dx_1du\\\nonumber
&=& \frac{1}{2}R(K)\int \sigma^2(x)\,dx-\frac{1}{2}\mu_2(K^2)h^2 \int f(x)^{-1}\left[\frac{1}{2}{\sigma^2}''(x)f(x)+{\sigma^2}'(x)f'(x) \right.\\
&+ &\left.\frac{1}{2}\sigma^2(x)f''(x)+m'(x)^2f(x)\right]\,dx + O\left(h^4\right),
\eeq
where we have used the fact that
\beqn
\int \tilde\beta_1(u)u^i\,du = \frac{1-i}{2}\mu_i(K^2) = 0 \iff \mbox{$i$ is an odd number}.
\eeqn

\beqn
E_2 &=& \E\left\{f(X_1)^{-2}\beta_{1h}(X_1-X_2,X_1-X_3)\E\left[Y_2-m(X_1)\cond X_1,X_2\right]\right.\\
& & \left.\E\left[Y_3-m(X_1)\cond X_1,X_3\right]\right\}\\
&=& \E\left\{f(X_1)^{-2}\beta_{1h}(X_1-X_2,X_1-X_3)[m(X_2)-m(X_1)][m(X_3)-m(X_1)]\right\}\\
&=& \iiint f(x_1)^{-2}\beta_{1h}(x_1-x_2,x_1-x_3)[m(x_2)-m(x_1)]\\
& & [m(x_3)-m(x_1)]f(x_1)f(x_2)f(x_3)\,dx_1dx_2dx_3\\
&=& h\iiint f(x_1)^{-1}\beta_1(u,v)[m(x_1-hu)-m(x_1)][m(x_1-hv)-m(x_1)]\\
& &f(x_1-hu)f(x_1-hv)\,dx_1dudv.\\
\eeqn

After some Taylor developments, we obtain that
\beq\label{eq:E2}
E_2&=& -2\mu_2(K)^2h^5\int f(x)^{-1}\left[\frac{1}{4}m''(x)^2f(x)^2+m'(x)^2f'(x)^2+m'(x)m''(x)f(x)f'(x)\right]\,dx\nonumber\\
&-&6\mu_2(K)\mu_4(K)h^7\int f(x)^{-1}\left[\frac{1}{2}m''(x)f(x)+m'(x)f'(x)\right]\nonumber\\
& &\left[\frac{1}{24}m^{4)}(x)f(x)+\frac{1}{6}m'''(x)f'(x)+\frac{1}{4}m''(x)f''(x)\right]\,dx+O\left(h^9\right),
\eeq
where we have used the fact that
\beqn
\iint \beta_1(u,v)u^iv^j\,dudv = -j\mu_i(K)\mu_j(K) = 0 \iff \mbox{$j = 0$ or ($i$ or $j$ is an odd number)}.
\eeqn

Plugging \eqref{eq:E1} and \eqref{eq:E2} into \eqref{eq:espCV1} we get the {main} order terms in \eqref{eq:espCV1_th}. {However, the modified version of the {Nadaraya--Watson} estimator, given in equation (8) of the main paper,} is a first order approximation of $\hat m_h$ and so the second order terms of \eqref{eq:cvprimatram1} do not match those of its ``classical'' counterpart. Accordingly, we will use the following, more precise, theoretical approximation of the Nadaraya--Watson estimator,
\beqn
\overline{m}_h(x) &=& m(x)+\frac{1}{n^2f(x)}\suma[i=1][n]\suma[j=1][n]\left\{K_h(x-X_i)[Y_i-m(x)]\right.\\
&+&\left.f(x)^{-1}\left[K_h(x-X_i)Y_i-m(x)f(x)\right]\left[f(x)-K_h(x-X_j)\right]\right\},
\eeqn
and we will denote by $\overline{CV}(h)$ the corresponding cross-validation function, that is,
\beqn
\overline{CV}(h) = \frac{1}{n}\suma[i=1][n]\left[\overline{m}_h^{-i}(X_i)-Y_i\right]^2.
\eeqn

Therefore, we need to study the second order terms of
\beq\label{eq:cvbarprima}
\E\left[\overline{CV}'(h)\right] = \frac{2}{(n-1)^4h}\suma[j=2][n]\suma[k=2][n]\suma[l=2][n]\suma[r=2][n]\left(\Omega_1^j+\Omega_2^{jk}\right)\left(\Omega_3^l+\Omega_4^{lr}\right),
\eeq
where
\beqn
\Omega_1^j &=& f(X_1)^{-2}K_h(X_1-X_j),\\
\Omega_2^{jk} &=& f(X_1)^{-3}\left[K_h(X_1-X_j)Y_j-m(X_1)f(X_1)\right]\left[f(X_1)-K_h(X_1-X_k)\right],\\
\Omega_3^l &=& -f(X_1)^{-2}\alpha_{1h}(X_1-X_l),\\
\Omega_4^{lr} &=& f(X_1)^{-3}\left\{\alpha_{1h}(X_1-X_r)\left[K_h(X_1-X_l)Y_l-m(X_1)f(X_1)\right]\right.\\
&-&\left.\alpha_{1h}(X_1-X_l)Y_l\left[f(X_1)-K_h(X_1-X_r)\right]\right\}.
\eeqn

For the sake of simplicity, we will denote by ``$Z(h,n) \overset{2}{=}$" the second order terms of a function $Z(h,n)$. 

Let us also define
\beqn
\varphi_4(x) &=& f(x)^{-1}\left\{\frac{1}{2}f''(x)\sigma^2(x)+f'(x){\sigma^2}'(x)+f(x)\left[\frac{1}{2}{\sigma^2}''(x)+m'(x)^2\right]\right\},\\
\varphi_6(x) &=& f(x)^{-1}\left[\frac{1}{24}m^{4)}(x)f(x)+\frac{1}{6}m'''(x)f'(x)+\frac{1}{4}m''(x)f''(x)+\frac{1}{6}m'(x)f'''(x)\right],\\
\varphi_7(x) &=& f(x)^{-1}\left[\frac{1}{2}m''(x)f(x)+m'(x)f'(x)\right],\\
\varphi_8(x) &=& f(x)^{-1}\left[\frac{1}{6!}m^{6)}(x)f(x)+\frac{1}{5!}m^{5)}(x)f'(x)+\frac{1}{48}m^{4)}(x)f''(x)+\frac{1}{36}m'''(x)f'''(x)\right.\\
&+&\left.\frac{1}{48}m''(x)f^{4)}(x)+\frac{1}{5!}m'(x)f^{5)}(x)\right],\\
\varphi_9(x) &=& f(x)^{-1}\left\{\frac{1}{2}f''(x)\sigma^2(x)+f'(x)\left[{\sigma^2}'(x)+m(x)m'(x)\right]+f(x)\left[\frac{1}{2}{\sigma^2}''(x)+m'(x)^2\right.\right.\\
&+&\left.\left.\frac{1}{2}m(x)m''(x)\right]\right\},\\
\varphi_{10}(x) &=& f(x)^{-1}\left[\frac{1}{6!}m^{6)}(x)f(x)+\frac{1}{5!}m^{5)}(x)f'(x)+\frac{1}{48}m^{4)}(x)f''(x)+\frac{1}{36}m'''(x)f'''(x)\right.\\
&+&\left.\frac{1}{48}m''(x)f^{4)}(x)+\frac{1}{5!}m'(x)f^{5)}(x)+\frac{1}{6!}m(x)f^{6)}(x)\right].
\eeqn

\beq\label{eq:omega13}
\E\left(\suma[j=2][n]\suma[k=2][n]\suma[l=2][n]\suma[r=2][n]\Omega_1^j\Omega_3^l\right) = (n-1)^2\left[(n-1)\E\left(\Omega_1^2\Omega_3^2\right)+(n-1)(n-2)\E\left(\Omega_1^2\Omega_3^3\right)\right].
\eeq

\beq\label{eq:omega1232}
\E\left(\Omega_1^2\Omega_3^2\right) &=& -\E\left(f(X_1)^{-2}K_h(X_1-X_2)\alpha_{1h}(X_1-X_2)\left\{[m(X_2)-m(X_1)]^2+\sigma^2(X_2)\right\}\right)\nonumber\\
&=& -h^{-1}\iint f(x_1)^{-1}K(u)\alpha_1(u)\left\{[m(x_1-hu)-m(x_1)]^2+\sigma^2(x_1-hu)\right\}\nonumber\\
&& f(x_1-hu)\,dx_1du \overset{2}{=} \frac{1}{2}\mu_2(K^2)h\int \varphi_4,
\eeq
where we have used the fact that
\beqn
\int K(u)\alpha_1(u)u^i\,du = \frac{1-i}{2}\mu_i(K^2) = 0 \iff \mbox{$i$ is odd}.
\eeqn

\beq\label{eq:omega1233}
\E\left(\Omega_1^2\Omega_3^3\right) &=& -\E\left\{f(X_1)^{-2}K_h(X_1-X_2)\alpha_{1h}(X_1-X_3)\right.\nonumber\\\nonumber
&&\left.[m(X_2)-m(X_1)][m(X_3)-m(X_1)]\right\}\\\nonumber
&=& -\iiint f(x_1)^{-1}K(u)\alpha_1(v)[m(x_1-hu)-m(x_1)][m(x_1-hv)-m(x_1)]\\\nonumber
&&f(x_1-hu)f(x_1-hv)\,dx_1dudv\\
&\overset{2}{=}& 6\mu_2(K)\mu_4(K)h^6\int \varphi_6\varphi_7f,
\eeq
where we have used the fact that
\beqn
\iint K(u)\alpha_1(v)u^iv^j\,dudv = -j\mu_i(K)\mu_j(K) = 0 \iff \mbox{$j = 0$ or ($i$ or $j$ are odd)}.
\eeqn

Plugging \eqref{eq:omega1232} and \eqref{eq:omega1233} into \eqref{eq:omega13} we get
\beq\label{eq:omega13_2}
\E\left(\suma[j=2][n]\suma[k=2][n]\suma[l=2][n]\suma[r=2][n]\Omega_1^j\Omega_3^l\right) \overset{2}{=} \frac{1}{2}\mu_2(K^2)n^3h\int \varphi_4+6\mu_2(K)\mu_4(K)n^4h^6\int \varphi_6\varphi_7f.
\eeq

\beq\label{eq:omega23}
\E\left(\suma[j=2][n]\suma[k=2][n]\suma[l=2][n]\suma[r=2][n]\Omega_2^{jk}\Omega_3^l\right) &=& (n-1)\left\{n^3\E\left(\Omega_2^{23}\Omega_3^4\right)+n^2\left[\E\left(\Omega_2^{22}\Omega_3^3\right)\right.\right.\nonumber\\
&+&\left.\left.\E\left(\Omega_2^{23}\Omega_3^2\right)+\E\left(\Omega_2^{32}\Omega_3^3\right)\right]\right\}+o\left(n^4h^6+n^3h\right).
\eeq

\beq\label{eq:omega22334}
\E\left(\Omega_2^{23}\Omega_3^4\right) = I_1+I_2+I_3+I_4,
\eeq
where
\beqn
I_1 &=& -\E\left\{f(X_1)^{-2}K_h(X_1-X_2)Y_2\alpha_{1h}(X_1-X_4)[Y_4-m(X_1)]\right\},\\
I_2 &=& \E\left\{f(X_1)^{-3}K_h(X_1-X_2)Y_2K_h(X_1-X_3)\alpha_{1h}(X_1-X_4)[Y_4-m(X_1)]\right\},\\
I_3 &=& \E\left\{f(X_1)^{-1}m(X_1)\alpha_{1h}(X_1-X_4)[Y_4-m(X_1)]\right\},\\
I_4 &=& -\E\left\{f(X_1)^{-2}m(X_1)K_h(X_1-X_3)\alpha_{1h}(X_1-X_4)[Y_4-m(X_1)]\right\}.
\eeqn

\beq\label{eq:I1}
I_1 &=& -\iiint f(x_1)^{-1}K(u)\alpha_1(v)m(x_1-hu)\left[m(x_1-hv)-m(x_1)\right]\nonumber\\
&&f(x_1-hu)f(x_1-hv)\,dx_1dudv\nonumber\\
&\overset{2}{=}& h^6\left[6\mu_6(K)\int mf\varphi_8+4\mu_2(K)\mu_4(K)\int f\varphi_1\varphi_6+2\mu_2(K)\mu_4(K)\int f\varphi_2\varphi_7\right],
\eeq
\beq\label{eq:I2}
I_2 &=& \iiiint f(x_1)^{-2}K(u)K(v)\alpha_1(w)m(x_1-hu)\left[m(x_1-hw)\right.\nonumber\\
&-&\left.m(x_1)\right]f(x_1-hu)f(x_1-hv)f(x_1-hw)\,dx_1dudvdw\nonumber\\
&\overset{2}{=}& -h^6\left[6\mu_6(K)\int mf\varphi_8+4\mu_2(K)\mu_4(K)\int f\varphi_1\varphi_6+2\mu_2(K)\mu_4(K)\int mf''\varphi_6\right.\nonumber\\
&+&\left.2\mu_2(K)\mu_4(K)\int f\varphi_2\varphi_7+\frac{1}{12}\mu_2(K)\mu_4(K)\int mf^{4)}\varphi_7+\mu_2(K)^3\int f''\varphi_1\varphi_7\right],
\eeq
\beq\label{eq:I3}
I_3 &=& \iint m(x_1)\alpha_1(u)\left[m(x_1-hu)-m(x_1)\right]f(x_1-hu)\,dx_1du\nonumber\\
&\overset{2}{=}& -6\mu_6(K)h^6\int mf\varphi_8,
\eeq
\beq\label{eq:I4}
I_4 &=& -\iiint f(x_1)^{-1}m(x_1)K(u)\alpha_1(v)\left[m(x_1-hv)-m(x_1)\right]\nonumber\\
&&f(x_1-hu)f(x_1-hv)\,dx_1dudv\nonumber\\
&\overset{2}{=}& h^6\left[6\mu_6(K)\int mf\varphi_8+2\mu_2(K)\mu_4(K)\int mf''\varphi_6+\frac{1}{12}\mu_2(K)\mu_4(K)\int mf^{4)}\varphi_7\right].
\eeq

Plugging \eqref{eq:I1}, \eqref{eq:I2}, \eqref{eq:I3} and \eqref{eq:I4} into \eqref{eq:omega22334} we get
\beq\label{eq:omega22334_2}
\E\left(\Omega_2^{23}\Omega_3^4\right) = -\mu_2(K)^3h^6\int f''\varphi_1\varphi_7.
\eeq

\beq\label{eq:omega22233}
\E\left(\Omega_2^{22}\Omega_3^3\right) = I_1+I_5+I_3+I_4,
\eeq
where
\beqn
I_5 = \E\left\{f(X_1)^{-3}K_h(X_1-X_2)^2Y_2\alpha_{1h}(X_1-X_3)\left[Y_3-m(X_1)\right]\right\}.
\eeqn

\beq\label{eq:I5}
I_5 &=& h^{-1}\iiint f(x_1)^{-2}K(u)^2\alpha_1(v)m(x_1-hu)\left[m(x_1-hv)\right.\nonumber\\
&-&\left.m(x_1)\right]f(x_1-hu)f(x_1-hv)\,dx_1dudv\nonumber\\
&\overset{2}{=}& -2R(K)\mu_2(K)h\int m\varphi_7,
\eeq
where we have used the fact that
\beqn
\iint K(u)^2\alpha_1(v)u^iv^j\,dudv = -j\mu_i(K^2)\mu_j(K) = 0 \iff \mbox{$j = 0$ or ($i$ or $j$ are odd)}.
\eeqn

Plugging \eqref{eq:I5} into \eqref{eq:omega22233} yields
\beq\label{eq:omega22233_2}
\E\left(\Omega_2^{22}\Omega_3^3\right) = -2R(K)\mu_2(K)\int m\varphi_7.
\eeq

\beq\label{eq:omega22332}
\E\left(\Omega_2^{23}\Omega_3^2\right) = I_6+I_7+I_3+I_4,
\eeq
where
\beqn
I_6 &=& -\E\left\{f(X_1)^{-2}K_h(X_1-X_2)Y_2\alpha_{1h}(X_1-X_2)\left[Y_2-m(X_1)\right]\right\},\\
I_7 &=& \E\left\{f(X_1)^{-3}K_h(X_1-X_2)Y_2K_h(X_1-X_3)\alpha_{1h}(X_1-X_2)\left[Y_2-m(X_1)\right]\right\}.
\eeqn

\beq\label{eq:I6}
I_6 &=& -h^{-1}\iint f(x_1)^{-1}K(u)\alpha_1(u)\left\{\sigma^2(x_1-hu)+m(x_1-hu)\right.\nonumber\\
&&\left.\left[m(x_1-hu)-m(x_1)\right]\right\}f(x_1-hu)f(x_1-hv)\,dx_1dudv\nonumber\\
&\overset{2}{=}& \frac{1}{2}\mu_2(K^2)h\int \varphi_9,
\eeq
\beq\label{eq:I7}
I_7 &=& h^{-1}\iiint f(x_1)^{-2}K(u)K(v)\alpha_1(u)\left\{\sigma^2(x_1-hu)+m(x_1-hu)\right.\nonumber\\
&&\left.\left[m(x_1-hu)-m(x_1)\right]\right\}f(x_1-hu)f(x_1-hv)\,dx_1dudv\nonumber\\
&\overset{2}{=}& h\left[\frac{1}{4}R(K)\mu_2(K)\int f^{-1}f''\sigma^2-\frac{1}{2}\mu_2(K^2)\int \varphi_9\right].
\eeq

Plugging \eqref{eq:I6} and \eqref{eq:I7} into \eqref{eq:omega22332} we get
\beq\label{eq:omega22332_2}
\E\left(\Omega_2^{23}\Omega_3^2\right) = \frac{1}{4}R(K)\mu_2(K)h\int f^{-1}f''\sigma^2.
\eeq

\beq\label{eq:omega23233}
\E\left(\Omega_2^{32}\Omega_3^3\right) = I_6+I_7+I_3+I_4 = \E\left(\Omega_2^{23}\Omega_3^2\right).
\eeq

Then, plugging \eqref{eq:omega22334_2}, \eqref{eq:omega22233_2}, \eqref{eq:omega22332_2} and \eqref{eq:omega23233} into \eqref{eq:omega23} yields
\beq\label{eq:omega23_2}
\E\left(\suma[j=2][n]\suma[k=2][n]\suma[l=2][n]\suma[r=2][n]\Omega_2^{jk}\Omega_3^l\right) &\overset{2}{=}& -\mu_2(K)^3n^4h^6\int f''\varphi_1\varphi_7+n^3h\left[\frac{1}{2}R(K)\mu_2(K)\right.\nonumber\\
&&\left.\int f^{-1}f''\sigma^2-2R(K)\mu_2(K)\int m\varphi_7\right].
\eeq

\beq\label{eq:omega14}
\E\left(\suma[j=2][n]\suma[k=2][n]\suma[l=2][n]\suma[r=2][n]\Omega_1^j\Omega_4^{lr}\right) &=& n^4\E\left(\Omega_1^2\Omega_4^{34}\right)+n^3\left[\E\left[\Omega_1^2\Omega_4^{23}\right)\right.\nonumber\\
&+&\left.\E\left(\Omega_1^2\Omega_4^{32}\right)+\E\left(\Omega_1^3\Omega_4^{22}\right)\right]+o\left(n^4h^6+n^3h\right).
\eeq

\beq\label{eq:omega12434}
\E\left(\Omega_1^2\Omega_4^{34}\right) = J_1+J_2+J_3+J_4,
\eeq
where
\beqn
J_1 &=& \E\left\{f(X_1)^{-3}K_h(X_1-X_2)\left[Y_2-m(X_1)\right]\alpha_{1h}(X_1-X_4)K_h(X_1-X_3)Y_3\right\},\\
J_2 &=& -\E\left\{f(X_1)^{-2}K_h(X_1-X_2)\left[Y_2-m(X_1)\right]\alpha_{1h}(X_1-X_4)m(X_1)\right\},\\
J_3 &=& -\E\left\{f(X_1)^{-2}K_h(X_1-X_2)\left[Y_2-m(X_1)\right]\alpha_{1h}(X_1-X_3)Y_3\right\},\\
J_4 &=& \E\left\{f(X_1)^{-3}K_h(X_1-X_2)\left[Y_2-m(X_1)\right]\alpha_{1h}(X_1-X_3)Y_3K_h(X_1-X_4)\right\}.
\eeqn

\beq\label{eq:J1}
J_1 &=& \iiiint f(x_1)^{-2}K(u)K(v)\alpha_1(w)\left[m(x_1-hu)-m(x_1)\right]\nonumber\\
&&m(x_1-hv)f(x_1-hu)f(x_1-hv)f(x_1-hw)\,dx_1dudvdw\overset{2}{=}-h^6\left[\frac{1}{6}\mu_2(K)\mu_4(K)\right.\nonumber\\
&&\left. \int mf^{4)}\varphi_7+\mu_2(K)\mu_4(K)\int mf''\varphi_6+\mu_2(K)^3\int f''\varphi_1\varphi_7\right].
\eeq

\beq\label{eq:J2}
J_2 &=& -\iiint f(x_1)^{-1}m(x_1)K(u)\alpha_1(v)\left[m(x_1-hu)-m(x_1)\right]\nonumber\\
&&f(x_1-hu)f(x_1-hv)\,dx_1dudv\nonumber\\
&\overset{2}{=}&h^6\left[\frac{1}{6}\mu_2(K)\mu_4(K)\int mf^{4)}\varphi_7+\mu_2(K)\mu_4(K)\int mf''\varphi_6\right].
\eeq

\beq\label{eq:J3}
J_3 &=& -\iiint f(x_1)^{-1}K(u)\alpha_1(v)\left[m(x_1-hu)-m(x_1)\right]m(x_1-hv)\nonumber\\
&&f(x_1-hu)f(x_1-hv)\,dx_1dudv\nonumber\\
&\overset{2}{=}& h^6\left[4\mu_2(K)\mu_4(K)\int f\varphi_2\varphi_7+2\mu_2(K)\mu_4(K)\int f\varphi_1\varphi_6\right].
\eeq

\beq\label{eq:J4}
J_4 &=& \iiiint f(x_1)^{-2}K(u)K(v)\alpha_1(w)\left[m(x_1-hu)-m(x_1)\right]\nonumber\\
&&m(x_1-hw)f(x_1-hu)f(x_1-hv)f(x_1-hw)\,dx_1dudvdw\overset{2}{=}-h^6\left[4\mu_2(K)\mu_4(K)\right.\nonumber\\
&& \left.\int f\varphi_2\varphi_7+2\mu_2(K)\mu_4(K)\int f\varphi_1\varphi_6+\mu_2(K)^3\int f''\varphi_1\varphi_7\right].
\eeq

Plugging \eqref{eq:J1}, \eqref{eq:J2}, \eqref{eq:J3} and \eqref{eq:J4} into \eqref{eq:omega12434} we get
\beq\label{eq:omega12434_2}
\E\left(\Omega_1^2\Omega_4^{34}\right) \overset{2}{=} -2\mu_2(K)^3\int f''\varphi_1\varphi_7.
\eeq

\beq\label{eq:omega12423}
\E\left(\Omega_1^2\Omega_4^{23}\right) = J_5+J_2+J_6+J_7,
\eeq
where
\beqn
J_5 &=& \E\left\{f(X_1)^{-3}K_h(X_1-X_2)^2\left[Y_2-m(X_1)\right]\alpha_{1h}(X_1-X_3)Y_2\right\},\\
J_6 &=& -\E\left\{f(X_1)^{-2}K_h(X_1-X_2)\left[Y_2-m(X_1)\right]\alpha_{1h}(X_1-X_2)Y_2\right\},\\
J_7 &=& \E\left\{f(X_1)^{-3}K_h(X_1-X_2)\left[Y_2-m(X_1)\right]\alpha_{1h}(X_1-X_2)Y_2K_h(X_1-X_3)\right\}.
\eeqn

\beq\label{eq:J5}
J_5 &=& h^{-1}\iiint f(x_1)^{-2}K(u)^2\alpha_1(v)\left\{\sigma^2(x_1-hu)+m(x_1-hu)\right.\nonumber\\
&&\left.\left[m(x_1-hu)-m(x_1)\right]\right\}f(x_1-hu)f(x_1-hv)\,dx_1dudv\nonumber\\
&\overset{2}{=}& -R(K)\mu_2(K)h\int f^{-1}f''\sigma^2.
\eeq

\beq\label{eq:J6}
J_6 &=& -h^{-1}\iint f(x_1)^{-1}K(u)\alpha_1(u)\left\{\sigma^2(x_1-hu)+m(x_1-hu)\right.\nonumber\\
&&\left.\left[m(x_1-hu)-m(x_1)\right]\right\}f(x_1-hu)\,dx_1du\nonumber\\
&\overset{2}{=}& \frac{1}{2}\mu_2(K^2)h\int \varphi_9.
\eeq

\beq\label{eq:J7}
J_7 &=& h^{-1}\iiint f(x_1)^{-2}K(u)\alpha_1(u)K(v)\left\{\sigma^2(x_1-hu)+m(x_1-hu)\right.\nonumber\\
&&\left.\left[m(x_1-hu)-m(x_1)\right]\right\}f(x_1-hu)f(x_1-hv)\,dx_1dudv\nonumber\\
&\overset{2}{=}& h\left[\frac{1}{4}R(K)\mu_2(K)\int f^{-1}f''\sigma^2-\frac{1}{2}\mu_2(K^2)\int \varphi_9\right].
\eeq

Plugging \eqref{eq:J5}, \eqref{eq:J6} and \eqref{eq:J7} into \eqref{eq:omega12423} we get
\beq\label{eq:omega12423_2}
\E\left(\Omega_1^2\Omega_4^{23}\right) \overset{2}{=} -\frac{3}{4}R(K)\mu_2(K)\int f^{-1}f''\sigma^2.
\eeq

\beq\label{eq:omega12432}
\E\left(\Omega_1^2\Omega_4^{32}\right) = J_8+J_9+J_3+J_{10},
\eeq
where
\beqn
J_8 &=& \E\left\{f(X_1)^{-3}K_h(X_1-X_2)\left[Y_2-m(X_1)\right]\alpha_{1h}(X_1-X_2)K_h(X_1-X_3)Y_3\right\},\\
J_9 &=& -\E\left\{f(X_1)^{-2}K_h(X_1-X_2)\left[Y_2-m(X_1)\right]\alpha_{1h}(X_1-X_2)m(X_1)\right\},\\
J_{10} &=& \E\left\{f(X_1)^{-3}K_h(X_1-X_2)^2\left[Y_2-m(X_1)\right]\alpha_{1h}(X_1-X_3)Y_3\right\}.
\eeqn

\beq\label{eq:J8}
J_8 &=& h^{-1}\iiint f(x_1)^{-2}K(u)\alpha_1(u)K(v)\left[m(x_1-hu)-m(x_1)\right]\nonumber\\
&&m(x_1-hv)f(x_1-hu)f(x_1-hv)\,dx_1dudv\nonumber\\
&\overset{2}{=}& -\frac{1}{2}\mu_2(K^2)h\int m\varphi_7.
\eeq

\beq\label{eq:J9}
J_9 &=& -h^{-1}\iint f(x_1)^{-1}m(x_1)K(u)\alpha_1(u)\left[m(x_1-hu)-m(x_1)\right]\nonumber\\
&&f(x_1-hu)\,dx_1du\nonumber\\
&\overset{2}{=}& \frac{1}{2}\mu_2(K^2)h\int m\varphi_7.
\eeq

\beq\label{eq:J10}
J_{10} &=& h^{-1}\iiint f(x_1)^{-2}K(u)^2\alpha_1(v)\left[m(x_1-hu)-m(x_1)\right]\nonumber\\
&&m(x_1-hv)f(x_1-hu)f(x_1-hv)\,dx_1dudv = O\left(h^3\right).
\eeq

Plugging \eqref{eq:J8}, \eqref{eq:J9} and \eqref{eq:J10} into \eqref{eq:omega12432} yields
\beq\label{eq:omega12432_2}
\E\left(\Omega_1^2\Omega_4^{32}\right) = O\left(h^3\right).
\eeq

\beq\label{eq:omega13422}
\E\left(\Omega_1^3\Omega_4^{22}\right) = 2J_{11}+J_2+J_3,
\eeq
where
\beqn
J_{11} = \E\left\{f(X_1)^{-3}K_h(X_1-X_3)\left[Y_3-m(X_1)\right]\alpha_{1h}(X_1-X_2)K_h(X_1-X_2)Y_2\right\}.
\eeqn

\beq\label{eq:J11}
J_{11} &=& h^{-1}\iiint f(x_1)^{-2}K(u)\alpha_1(v)K(v)\left[m(x_1-hv)-m(x_1)\right]\nonumber\\
&&m(x_1-hu)f(x_1-hu)f(x_1-hv)\,dx_1dudv\nonumber\\
&\overset{2}{=}& \frac{1}{2}R(K)\mu_2(K)\int m\varphi_7.
\eeq

Plugging \eqref{eq:J11} into \eqref{eq:omega13422} we get
\beq\label{eq:omega13422_2}
\E\left(\Omega_1^3\Omega_4^{22}\right) \overset{2}{=} R(K)\mu_2(K)\int m\varphi_7.
\eeq

Plugging \eqref{eq:omega12434_2}, \eqref{eq:omega12423_2}, \eqref{eq:omega12432_2} and \eqref{eq:omega13422_2} into \eqref{eq:omega14} yields
\beq\label{eq:omega14_2}
\E\left(\suma[j=2][n]\suma[k=2][n]\suma[l=2][n]\suma[r=2][n]\Omega_1^j\Omega_4^{lr}\right) &\overset{2}{=}& -2\mu_2(K)^3n^4h^6\int f''\varphi_1\varphi_7 +n^3h\left[R(K)\mu_2(K)\right. \nonumber\\
&&\left.\int m\varphi_7-\frac{3}{4}R(K)\mu_2(K)\int f^{-1}f''\sigma^2\right].
\eeq

\beq\label{eq:omega24}
\E\left(\suma[j=2][n]\suma[k=2][n]\suma[l=2][n]\suma[r=2][n]\Omega_2^{jk}\Omega_4^{lr}\right) &=& n^4\E\left(\Omega_2^{23}\Omega_4^{45}\right)+n^3\left[\E\left(\Omega_2^{22}\Omega_4^{34}\right)\right.\nonumber\\
&+&\left.\E\left(\Omega_2^{23}\Omega_4^{24}\right)+\E\left(\Omega_2^{23}\Omega_4^{42}\right)+\E\left(\Omega_2^{32}\Omega_4^{24}\right)+\E\left(\Omega_2^{32}\Omega_4^{42}\right)\right.\nonumber\\
&+&\left.\E\left(\Omega_2^{34}\Omega_4^{22}\right)\right]+o\left(n^4h^6+n^3h\right).
\eeq

\beq\label{eq:omega223445}
\E\left(\Omega_2^{23}\Omega_4^{45}\right) = H_1+H_2+\dots+H_{16},
\eeq
where
\beqn
H_1 &=& \E\left[f(X_1)^{-3}K_h(X_1-X_2)Y_2\alpha_{1h}(X_1-X_5)K_h(X_1-X_4)Y_4\right],\\
H_2 &=& -\E\left[f(X_1)^{-2}K_h(X_1-X_2)Y_2\alpha_{1h}(X_1-X_5)m(X_1)\right],\\
H_3 &=& -\E\left[f(X_1)^{-2}K_h(X_1-X_2)Y_2\alpha_{1h}(X_1-X_4)Y_4\right],\\
H_4 &=& \E\left[f(X_1)^{-3}K_h(X_1-X_2)Y_2\alpha_{1h}(X_1-X_4)K_h(X_1-X_5)Y_4\right],\\
H_5 &=& -\E\left[f(X_1)^{-4}K_h(X_1-X_2)Y_2\alpha_{1h}(X_1-X_5)K_h(X_1-X_4)Y_4K_h(X_1-X_3)\right],\\
H_6 &=& \E\left[f(X_1)^{-3}K_h(X_1-X_2)Y_2\alpha_{1h}(X_1-X_5)K_h(X_1-X_3)m(X_1)\right],\\
H_7 &=& \E\left[f(X_1)^{-3}K_h(X_1-X_2)Y_2\alpha_{1h}(X_1-X_4)Y_4K_h(X_1-X_3)\right],\\
H_8 &=& -\E\left[f(X_1)^{-4}K_h(X_1-X_2)Y_2\alpha_{1h}(X_1-X_4)K_h(X_1-X_5)K_h(X_1-X_3)Y_4\right],\\
H_9 &=& -\E\left[f(X_1)^{-2}m(X_1)\alpha_{1h}(X_1-X_5)K_h(X_1-X_4)Y_4\right],\\
H_{10} &=& \E\left[f(X_1)^{-1}m(X_1)^2\alpha_{1h}(X_1-X_5)\right],\\
H_{11} &=& \E\left[f(X_1)^{-1}m(X_1)\alpha_{1h}(X_1-X_4)Y_4\right],\\
H_{12} &=& -\E\left[f(X_1)^{-2}m(X_1)\alpha_{1h}(X_1-X_4)Y_4K_h(X_1-X_5)\right],\\
H_{13} &=& \E\left[f(X_1)^{-3}m(X_1)K_h(X_1-X_3)\alpha_{1h}(X_1-X_5)K_h(X_1-X_4)Y_4\right],\\
H_{14} &=& -\E\left[f(X_1)^{-2}m(X_1)^2K_h(X_1-X_3)\alpha_{1h}(X_1-X_5)\right],\\
H_{15} &=& -\E\left[f(X_1)^{-2}m(X_1)K_h(X_1-X_3)\alpha_{1h}(X_1-X_4)Y_4\right],\\
H_{16} &=& \E\left[f(X_1)^{-3}m(X_1)K_h(X_1-X_3)\alpha_{1h}(X_1-X_4)Y_4K_h(X_1-X_5)\right].
\eeqn

\beq\label{eq:H1}
H_1 &=& \iiiint f(x_1)^{-2}K(u)K(v)\alpha_1(w)m(x_1-hu)m(x_1-hv)\nonumber\\
&&f(x_1-hu)f(x_1-hv)f(x_1-hw)\,dx_1dudvdw\nonumber\\
&\overset{2}{=}& -h^6\left[\frac{1}{120}\mu_6(K)\int m^2f^{6)}+\frac{1}{3}\mu_2(K)\mu_4(K)\int mf^{4)}\varphi_1\right.\nonumber\\
&+&\left.2\mu_2(K)\mu_4(K)\int mf''\varphi_2+\mu_2(K)^3\int f''\varphi_1^2\right].
\eeq

\beq\label{eq:H2}
H_2 &=& -\iiint f(x_1)^{-1}m(x_1)K(u)\alpha_1(v)m(x_1-hu)f(x_1-hu)\nonumber\\
&&f(x_1-hv)\,dx_1dudv\nonumber\\
&\overset{2}{=}& h^6\left[\frac{1}{120}\mu_6(K)\int m^2f^{6)}+\frac{1}{6}\mu_2(K)\mu_4(K)\int mf^{4)}\varphi_1\right.\nonumber\\
&+&\left.\mu_2(K)\mu_4(K)\int mf''\varphi_2\right].
\eeq

\beq\label{eq:H3}
H_3 &=& -\iiint f(x_1)^{-1}K(u)\alpha_1(v)m(x_1-hu)m(x_1-hv)f(x_1-hu)\nonumber\\
&&f(x_1-hv)\,dx_1dudv\nonumber\\
&\overset{2}{=}& h^6\left[6\mu_6(K)\int mf\varphi_{10}+6\mu_2(K)\mu_4(K)\int f\varphi_1\varphi_2\right].
\eeq

\beq\label{eq:H4}
H_4 &=& \iiiint f(x_1)^{-2}K(u)K(v)\alpha_1(w)m(x_1-hu)m(x_1-hw)\nonumber\\
&&f(x_1-hu)f(x_1-hv)f(x_1-hw)\,dx_1dudvdw\nonumber\\
&\overset{2}{=}& -h^6\left[6\mu_6(K)\int mf\varphi_{10}+6\mu_2(K)\mu_4(K)\int f\varphi_1\varphi_2+2\mu_2(K)\mu_4(K)\int mf''\varphi_2\right.\nonumber\\
&+&\left.\frac{1}{12}\mu_2(K)\mu_4(K)\int mf^{4)}\varphi_1+\mu_2(K)^3\int f''\varphi_1^2\right].
\eeq

\beq\label{eq:H5}
H_5 &=& -\int\dots\int f(x_1)^{-3}K(u)K(v)K(w)\alpha_1(z)m(x_1-hu)\nonumber\\
&&m(x_1-hw)f(x_1-hu)f(x_1-hv)f(x_1-hw)f(x_1-hz)\,dx_1dudvdwdz\nonumber\\
&\overset{2}{=}& h^6\left[\frac{1}{120}\mu_6(K)\int m^2f^{6)}+\frac{1}{3}\mu_2(K)\mu_4(K)\int mf^{4)}\varphi_1\right.\nonumber\\
&+&\left.\frac{1}{12}\mu_2(K)\mu_4(K)\int m^2f^{-1}f''f^{4)}+2\mu_2(K)\mu_4(K)\int mf''\varphi_2\right.\nonumber\\
&+&\left.\frac{1}{24}\mu_2(K)\mu_4(K)\int m^2f^{-1}f''f^{4)}+\mu_2(K)^3\int mf^{-1}(f'')^2\varphi_1\right.\nonumber\\
&+&\left.\mu_2(K)^3\int f''\varphi_1^2\right].
\eeq

\beq\label{eq:H6}
H_6 &=& \iiint f(x_1)^{-2}m(x_1)K(u)K(v)\alpha_1(w)m(x_1-hu)f(x_1-hu)\nonumber\\
&&f(x_1-hv)f(x_1-hw)\,dx_1dudvdw\\\nonumber
&\overset{2}{=}& -h^6\left[\frac{1}{120}\mu_6(K)\int m^2f^{6)}+\frac{1}{6}\mu_2(K)\mu_4(K)\int mf^{4)}\varphi_1\right.\nonumber\\
&+&\left.\frac{1}{8}\mu_2(K)\mu_4(K)\int m^2f^{-1}f''f^{4)}\right.\nonumber\\
&+&\left.\mu_2(K)\mu_4(K)\int mf''\varphi_2+\frac{1}{2}\mu_2(K)^3\int mf^{-1}(f'')^2\varphi_1\right].
\eeq

\beq\label{eq:H7}
H_7 &=& \iiiint f(x_1)^{-2}K(u)K(v)\alpha_1(w)m(x_1-hu)m(x_1-hw)\nonumber\\
&&f(x_1-hu)f(x_1-hv)f(x_1-hw)\,dx_1dudvdw\nonumber\\
&\overset{2}{=}& -h^6\left[6\mu_6(K)\int mf\varphi_{10}+6\mu_2(K)\mu_4(K)\int f\varphi_1\varphi_2+2\mu_2(K)\mu_4(K)\int mf''\varphi_2\right.\nonumber\\
&+&\left.\frac{1}{12}\mu_2(K)\mu_4(K)\int mf^{4)}\varphi_1+\mu_2(K)^3\int f''\varphi_1^2\right].
\eeq

\beq\label{eq:H8}
H_8 &=& -\int\dots\int f(x_1)^{-3}K(u)K(v)K(w)\alpha_1(z)m(x_1-hu)\nonumber\\
&&m(x_1-hz)f(x_1-hu)f(x_1-hv)f(x_1-hw)f(x_1-hz)\,dx_1dudvdwdz\nonumber\\
&\overset{2}{=}& h^6\left[6\mu_6(K)\int mf\varphi_{10}+6\mu_2(K)\mu_4(K)\int f\varphi_1\varphi_2+4\mu_2(K)\mu_4(K)\int mf''\varphi_2\right.\nonumber\\
&+&\left.\frac{1}{6}\mu_2(K)\mu_4(K)\int mf^{4)}\varphi_1+2\mu_2(K)^3\int f''\varphi_1^2+\frac{1}{2}\mu_2(K)^3\int mf^{-1}(f'')^2\varphi_1\right].
\eeq

\beq\label{eq:H9}
H_9 &=& -\iiint f(x_1)^{-1}m(x_1)K(u)\alpha_1(v)m(x_1-hu)f(x_1-hu)f(x_1-hv)\,dx_1dudv\nonumber\\
&\overset{2}{=}& h^6\left[\frac{1}{120}\mu_6(K)\int m^2f^{6)}+\frac{1}{6}\mu_2(K)\mu_4(K)\int mf^{4)}\varphi_1\right.\nonumber\\
&+& \left.\mu_2(K)\mu_4(K)\int mf''\varphi_2\right].
\eeq

\beq\label{eq:H10}
H_{10} = \iint m(x_1)^2\alpha_1(u)f(x_1-hu)\,dx_1du \overset{2}{=} -\frac{1}{120}\mu_6(K)h^6\int m^2f^{6)}.
\eeq

\beq\label{eq:H11}
H_{11} = \iint m(x_1)\alpha_1(u)m(x_1-hu)f(x_1-hu)\,dx_1du \overset{2}{=} -6\mu_6(K)h^6\int mf\varphi_{10}.
\eeq

\beq\label{eq:H12}
H_{12} &=& -\iiint f(x_1)^{-1}m(x_1)K(u)\alpha_1(v)m(x_1-hv)f(x_1-hu)f(x_1-hv)\,dx_1dudv\nonumber\\
&\overset{2}{=}& h^6\left[6\mu_6(K)\int mf\varphi_{10}+2\mu_2(K)\mu_4(K)\int mf''\varphi_2\right.\nonumber\\
&+& \left.\frac{1}{12}\mu_2(K)\mu_4(K)\int mf^{4)}\varphi_1\right].
\eeq

\beq\label{eq:H13}
H_{13} &=& \iiiint f(x_1)^{-2}m(x_1)K(u)K(v)\alpha_1(w)m(x_1-hv)\nonumber\\
&&f(x_1-hu)f(x_1-hv)f(x_1-hw)\,dx_1dudvdw\nonumber\\
&\overset{2}{=}& -h^6\left[\frac{1}{120}\mu_6(K)\int m^2f^{6)}+\frac{1}{8}\mu_2(K)\mu_4(K)\int m^2f^{-1}f''f^{4)}\right.\nonumber\\
&+&\left.\frac{1}{6}\mu_2(K)\mu_4(K)\int mf^{4)}\varphi_1\right.\nonumber\\
&+&\left.\mu_2(K)\mu_4(K)\int mf''\varphi_2+\frac{1}{2}\mu_2(K)^3\int mf^{-1}(f'')^2\varphi_1\right].
\eeq

\beq\label{eq:H14}
H_{14} &=& -\iiint f(x_1)^{-1}m(x_1)^2K(u)\alpha_1(v)f(x_1-hu)f(x_1-hv)\,dx_1dudv\nonumber\\
&\overset{2}{=}& h^6\left[\frac{1}{120}\mu_6(K)\int m^2f^{6)}+\frac{1}{8}\mu_2(K)\mu_4(K)\int m^2f^{-1}f''f^{4)}\right].
\eeq

\beq\label{eq:H15}
H_{15} &=& -\iiint f(x_1)^{-1}m(x_1)K(u)\alpha_1(v)m(x_1-hv)f(x_1-hu)f(x_1-hv)\,dx_1dudv\nonumber\\
&\overset{2}{=}& h^6\left[6\mu_6(K)\int mf\varphi_{10}+2\mu_2(K)\mu_4(K)\int mf''\varphi_2\right.\nonumber\\
&+&\left.\frac{1}{12}\mu_2(K)\mu_4(K)\int mf^{4)}\varphi_1\right].
\eeq

\beq\label{eq:H16}
H_{16} &=& \iiiint f(x_1)^{-2}m(x_1)K(u)K(v)\alpha_1(w)m(x_1-hw)f(x_1-hu)\nonumber\\
&&f(x_1-hv)f(x_1-hw)\,dx_1dudvdw\nonumber\\
&\overset{2}{=}& -h^6\left[6\mu_6(K)\int mf\varphi_{10}+4\mu_2(K)\mu_4(K)\int mf''\varphi_2+\frac{1}{6}\mu_2(K)\mu_4(K)\int mf^{4)}\varphi_1\right.\nonumber\\
&+&\left.\frac{1}{2}\mu_2(K)^3\int mf^{-1}(f'')^2\varphi_1\right].
\eeq

Plugging \eqref{eq:H1}--\eqref{eq:H16} into \eqref{eq:omega223445} we get
\beq\label{eq:omega223445_2}
\E\left(\Omega_2^{23}\Omega_4^{45}\right) \overset{2}{=} \frac{1}{6}\mu_2(K)\mu_4(K)h^6\int mf^{4)}\varphi_1.
\eeq

\beq\label{eq:omega222434}
\E\left(\Omega_2^{22}\Omega_4^{34}\right) = H_1+\dots+H_4+F_1+F_2+F_3+F_4+H_9+\dots+H_{16},
\eeq
where
\beqn
F_1 &=& -\E\left[f(X_1)^{-4}K_h(X_1-X_2)^2Y_2K_h(X_1-X_4)\alpha_{1h}(X_1-X_5)Y_4\right],\\
F_2 &=& \E\left[f(X_1)^{-3}K_h(X_1-X_2)^2Y_2\alpha_{1h}(X_1-X_5)m(X_1)\right],\\
F_3 &=& \E\left[f(X_1)^{-3}K_h(X_1-X_2)^2Y_2\alpha_{1h}(X_1-X-4)Y_4\right],\\
F_4 &=& -\E\left[f(X_1)^{-4}K_h(X_1-X_2)^2Y_2K_h(X_1-X_5)\alpha_{1h}(X_1-X_4)Y_4\right].
\eeqn

\beq\label{eq:F1}
F_1 &=& -h^{-1}\iiiint f(x_1)^{-3}K(u)^2K(v)\alpha_1(w)m(x_1-hu)m(x_1-hv)\nonumber\\
&&f(x_1-hu)f(x_1-hv)f(x_1-hw)\,dx_1dudvdw\nonumber\\
&\overset{2}{=}& R(K)\mu_2(K)h\int m^2f^{-1}f''.
\eeq

\beq\label{eq:F2}
F_2 &=& h^{-1}\iiint f(x_1)^{-2}m(x_1)K(u)^2\alpha_1(v)m(x_1-hu)\nonumber\\
&&f(x_1-hu)f(x_1-hv)\,dx_1dudv\nonumber\\
&\overset{2}{=}& -R(K)\mu_2(K)h\int m^2f^{-1}f''.
\eeq

\beq\label{eq:F3}
F_3 &=& h^{-1}\iiint f(x_1)^{-2}K(u)^2\alpha_1(v)m(x_1-hu)m(x_1-hv)\nonumber\\
&&f(x_1-hu)f(x_1-hv)\,dx_1dudv\nonumber\\
&\overset{2}{=}& -2R(K)\mu_2(K)h\int m\varphi_1.
\eeq

\beq\label{eq:F4}
F_4 &=& -h^{-1}\iiiint f(x_1)^{-3}K(u)^2K(v)\alpha_1(w)m(x_1-hu)\nonumber\\
&&m(x_1-hw)f(x_1-hu)f(x_1-hv)f(x_1-hw)\,dx_1dudvdw\nonumber\\
&\overset{2}{=}& 2R(K)\mu_2(K)h\int m\varphi_1.
\eeq

Plugging \eqref{eq:F1}, \eqref{eq:F2}, \eqref{eq:F3} and \eqref{eq:F4} into \eqref{eq:omega222434} yields
\beq\label{eq:omega222434_2}
\E\left(\Omega_2^{22}\Omega_4^{34}\right) = O\left(h^3\right).
\eeq

\beq\label{eq:omega223424}
\E\left(\Omega_2^{23}\Omega_4^{24}\right) = G_1+G_2+2G_3+G_4+G_5,
\eeq
where
\beqn
G_1 &=& \E\left[f(X_1)^{-3}K_h(X_1-X_2)^2Y_2^2\alpha_{1h}(X_1-X_5)\right],\\
G_2 &=& -\E\left[F(X_1)^{-2}K_h(X_1-X_2)Y_2^2\alpha_{1h}(X_1-X_2)\right],\\
G_3 &=& \E\left[f(X_1)^{-3}K_h(X_1-X_2)Y_2^2\alpha_{1h}(X_1-X_2)K_h(X_1-X_5)\right],\\
G_4 &=& -\E\left[f(X_1)^{-4}K_h(X_1-X_2)^2Y_2^2K_h(X_1-X_3)\alpha_{1h}(X_1-X_5)\right],\\
G_5 &=& -\E\left[f(X_1)^{-4}K_h(X_1-X_2)Y_2^2\alpha_{1h}(X_1-X_2)K_h(X_1-X_3)K_h(X_1-X_5)\right].
\eeqn

\beq\label{eq:G1}
G_1 &=& h^{-1}\iiint f(x_1)^{-2}K(u)^2\alpha_1(v)\left[m(x_1-hu)^2+\sigma^2(x_1-hu)\right]\nonumber\\
&&f(x_1-hu)f(x_1-hv)\,dx_1dudv\nonumber\\
&\overset{2}{=}& -R(K)\mu_2(K)h\int \left(m^2+\sigma^2\right)f^{-1}f''.
\eeq

\beq\label{eq:G2}
G_2 &=& -h^{-1}\iint f(x_1)^{-1}K(u)\alpha_1(u)\left[m(x_1-hu)^2+\sigma^2(x_1-hu)\right]f(x_1-hu)\,dx_1du\nonumber\\
&\overset{2}{=}& \frac{1}{2}\mu_2(K^2)\int f^{-1}\varphi_3.
\eeq

\beq\label{eq:G3}
G_3 &=& h^{-1}\iiint f(x_1)^{-2}K(u)\alpha_1(u)K(v)\left[m(x_1-hu)^2\right.\nonumber\\
&+&\left.\sigma^2(x_1-hu)\right]f(x_1-hu)f(x_1-hv)\,dx_1dudv\nonumber\\
&\overset{2}{=}& h\left(\frac{1}{4}R(K)\mu_2(K)\int \left(m^2+\sigma^2\right)f^{-1}f''-\frac{1}{2}\mu_2(K^2)\int f^{-1}\varphi_3\right).
\eeq

\beq\label{eq:G4}
G_4 &=& -h^{-1}\iiiint f(x_1)^{-3}K(u)^2K(v)\alpha_1(w)\left[m(x_1-hu)^2\right.\nonumber\\
&+&\left.\sigma^2(x_1-hu)\right]f(x_1-hu)f(x_1-hv)f(x_1-hw)\,dx_1dudvdw\nonumber\\
&\overset{2}{=}& \frac{1}{2}R(K)\mu_2(K)h\int f^{-1}\varphi_3.
\eeq

\beq\label{eq:G5}
G_5 &=& -h^{-1}\iiiint f(x_1)^{-3}K(u)\alpha_1(u)K(v)K(w)\left[m(x_1-hu)^2\right.\nonumber\\
&+&\left.\sigma^2(x_1-hu)\right]f(x_1-hu)f(x_1-hv)f(x_1-hw)\,dx_1dudvdw\nonumber\\
&\overset{2}{=}& h\left[\frac{1}{2}R(K)\mu_2(K)\int \left(m^2+\sigma^2\right)f^{-1}f''-\frac{1}{2}\mu_2(K^2)\int f^{-1}\varphi_3\right].
\eeq

Plugging \eqref{eq:G1}, \eqref{eq:G2}, \eqref{eq:G3}, \eqref{eq:G4} and \eqref{eq:G5} into \eqref{eq:omega223424} we get
\beq\label{eq:omega223424_2}
\E\left(\Omega_2^{23}\Omega_4^{24}\right) \overset{2}{=} -\frac{1}{2}\mu_2(K^2)\int f^{-1}\varphi_3.
\eeq

\beq\label{eq:omega223442}
\E\left(\Omega_2^{23}\Omega_4^{42}\right) = L_1+L_2+L_3+L_4+L_5+F_4,
\eeq
where
\beqn
L_1 &=& \E\left[f(X_1)^{-3}K_h(X_1-X_2)Y_2\alpha_{1h}(X_1-X_2)K_h(X_1-X_4)Y_4\right],\\
L_2 &=& -\E\left[f(X_1)^{-2}K_h(X_1-X_2)Y_2\alpha_{1h}(X_1-X_2)m(X_1)\right],\\
L_3 &=& \E\left[f(X_1)^{-3}K_h(X_1-X_2)^2Y_2\alpha_{1h}(X_1-X_4)Y_4\right],\\
L_4 &=& -\E\left[f(X_1)^{-4}K_h(X_1-X_2)Y_2\alpha_{1h}(X_1-X_2)K_h(X_1-X_3)K_h(X_1-X_4)Y_4\right],\\
L_5 &=& \E\left[f(X_1)^{-3}K_h(X_1-X_2)Y_2\alpha_{1h}(X_1-X_2)K_h(X_1-X_3)m(X_1)\right].
\eeqn

\beq\label{eq:L1}
L_1 &=& h^{-1}\iiint f(x_1)^{-2}K(u)\alpha_1(u)K(v)m(x_1-hu)m(x_1-hv)\nonumber\\
&&f(x_1-hu)f(x_1-hv)\,dx_1dudv\nonumber\\
&\overset{2}{=}& h\left[\frac{1}{2}R(K)\mu_2(K)\int m\varphi_1-\frac{1}{2}\mu_2(K^2)\int m\varphi_1\right].
\eeq

\beq\label{eq:L2}
L_2 &=& -h^{-1}\iint f(x_1)^{-1}m(x_1)K(u)\alpha_1(u)m(x_1-hu)\nonumber\\
&&f(x_1-hu)\,dx_1du \overset{2}{=} \frac{1}{2}\mu_2(K^2)h\int m\varphi_1.
\eeq

\beq\label{eq:L3}
L_3 &=& h^{-1}\iiint f(x_1)^{-2}K(u)^2\alpha_1(v)m(x_1-hu)m(x_1-hv)\nonumber\\
&&f(x_1-hu)f(x_1-hv)\,dx_1dudv \overset{2}{=} -2R(K)\mu_2(K)h\int m\varphi_1.
\eeq

\beq\label{eq:L4}
L_4 &=& -h^{-1}\iiiint f(x_1)^{-3}K(u)\alpha_1(u)K(v)K(w)m(x_1-hu)\nonumber\\
&&m(x_1-hw)f(x_1-hu)f(x_1-hv)f(x_1-hw)\,dx_1dudvdw\nonumber\\
&\overset{2}{=}& -h\left[\frac{1}{4}R(K)\mu_2(K)\int m^2f^{-1}f''+\frac{1}{2}R(K)\mu_2(K)\int m\varphi_1\right.\nonumber\\
&-&\left.\frac{1}{2}\mu_2(K^2)\int m\varphi_1\right].
\eeq

\beq\label{eq:L5}
L_5 &=& h^{-1}\iiint f(x_1)^{-2}m(x_1)K(u)\alpha_1(u)K(v)m(x_1-hu)\nonumber\\
&&f(x_1-hu)f(x_1-hv)\,dx_1dudv\nonumber\\
&\overset{2}{=}& h\left[\frac{1}{4}R(K)\mu_2(K)\int m^2f^{-1}f''-\frac{1}{2}\mu_2(K^2)\int m\varphi_1\right].
\eeq

Plugging \eqref{eq:F4} and \eqref{eq:L1}--\eqref{eq:L5} into \eqref{eq:omega223442} yields
\beq\label{eq:omega223442_2}
\E\left(\Omega_2^{23}\Omega_4^{42}\right) \overset{2}{=} h\left(R(K)\mu_2(K)-\mu_2(K^2)\right)\int m\varphi_1.
\eeq

\beq\label{eq:omega232424}
\E\left(\Omega_2^{32}\Omega_4^{24}\right) = F_1+L_1+L_4+F_2+L_2+L_5 \overset{2}{=} O\left(h^3\right).
\eeq

\beq\label{eq:omega232442}
\E\left(\Omega_2^{32}\Omega_4^{42}\right) = N_1+N_2+\dots+N_6,
\eeq
where
\beqn
N_1 &=& -\E\left[f(X_1)^{-4}K_h(X_1-X_2)Y_2K_h(X_1-X_3)\alpha_{1h}(X_1-X_3)K_h(X_1-X_4)Y_4\right],\\
N_2 &=& \E\left[f(X_1)^{-3}m(X_1)K_h(X_1-X_2)Y_2K_h(X_1-X_3)\alpha_{1h}(X_1-X_3)\right],\\
N_3 &=& -\E\left[f(X_1)^{-4}K_h(X_1-X_2)Y_2K_h(X_1-X_3)^2\alpha_{1h}(X_1-X_4)Y_4\right],\\
N_4 &=& \E\left[f(X_1)^{-3}m(X_1)K_h(X_1-X_3)\alpha_{1h}(X_1-X_3)K_h(X_1-X_4)Y_4\right],\\
N_5 &=& -\E\left[f(X_1)^{-2}m(X_1)^2K_h(X_1-X_3)\alpha_{1h}(X_1-X_3)\right],\\
N_6 &=& \E\left[f(X_1)^{-3}m(X_1)K_h(X_1-X_3)^2\alpha_{1h}(X_1-X_4)Y_4\right].
\eeqn

\beq\label{eq:N1}
N_1 &=& -h^{-1}\iiiint f(x_1)^{-3}K(u)\alpha_1(u)K(v)K(w)m(x_1-hv)\nonumber\\
&&m(x_1-hw)f(x_1-hu)f(x_1-hv)f(x_1-hw)\,dx_1dudvdw\nonumber\\
&\overset{2}{=}& h\left[\frac{1}{4}\mu_2(K^2)\int m^2f^{-1}f''-R(K)\mu_2(K)\int m\varphi_1\right].
\eeq

\beq\label{eq:N2}
N_2 &=& h^{-1}\iiint f(x_1)^{-2}m(x_1)K(u)\alpha_1(u)K(v)m(x_1-hv)\nonumber\\
&&f(x_1-hu)f(x_1-hv)\,dx_1dudv\nonumber\\
&\overset{2}{=}& h\left[\frac{1}{2}R(K)\mu_2(K)\int m\varphi_1-\frac{1}{4}\mu_2(K^2)\int m^2f^{-1}f''\right].
\eeq

\beq\label{eq:N3}
N_3 &=& -h^{-1}\iiiint f(x_1)^{-3}K(u)^2K(v)\alpha_1(w)m(x_1-hv)\nonumber\\
&&m(x_1-hw)f(x_1-hu)f(x_1-hv)f(x_1-hw)\,dx_1dudvdw\nonumber\\
&\overset{2}{=}& 2\mu_2(K^2)h\int m\varphi_1.
\eeq

\beq\label{eq:N4}
N_4 &=& h^{-1}\iiint f(x_1)^{-2}m(x_1)K(u)\alpha_1(u)K(v)m(x_1-hv)\nonumber\\
&&f(x_1-hu)f(x_1-hv)\,dx_1dudv\nonumber\\
&\overset{2}{=}& h\left[\frac{1}{2}R(K)\mu_2(K)\int m\varphi_1-\frac{1}{4}\mu_2(K^2)\int m^2f^{-1}f''\right].
\eeq

\beq\label{eq:N5}
N_5 &=& -h^{-1}\iint f(x_1)^{-1}m(x_1)^2K(u)\alpha_1(u)f(x_1-hu)\,dx_1du\nonumber\\
&\overset{2}{=}& \frac{1}{4}\mu_2(K^2)h\int m^2f^{-1}f''.
\eeq

\beq\label{eq:N6}
N_6 &=& h^{-1}\iiint f(x_1)^{-2}m(x_1)K(u)^2\alpha_1(v)m(x_1-hv)\nonumber\\
&&f(x_1-hu)f(x_1-hv)\,dx_1dudv\nonumber\\
&\overset{2}{=}& -2\mu_2(K^2)h\int m\varphi_1.
\eeq

Plugging \eqref{eq:N1}--\eqref{eq:N6} into \eqref{eq:omega232442} yields
\beq\label{eq:omega232442_2}
\E\left(\Omega_2^{32}\Omega_4^{42}\right) \overset{2}{=} O\left(h^3\right).
\eeq

\beq\label{eq:omega234422}
\E\left(\Omega_2^{34}\Omega_4^{22}\right) = 2\left(R_1+R_2+R_3+R_4\right),
\eeq
where
\beqn
R_1 &=& \E\left[f(X_1)^{-3}K_h(X_1-X_2)Y_2K_h(X_1-X_4)\alpha_{1h}(X_1-X_4)Y_4\right],\\
R_2 &=& -\E\left[f(X_1)^{-4}K_h(X_1-X_2)Y_2K_h(X_1-X_3)K_h(X_1-X_4)\alpha_{1h}(X_1-X_4)Y_4\right],\\
R_3 &=& -\E\left[f(X_1)^{-2}m(X_1)K_h(X_1-X_4)\alpha_{1h}(X_1-X_4)Y_4\right],\\
R_4 &=& \E\left[f(X_1)^{-3}m(X_1)K_h(X_1-X_3)K_h(X_1-X_4)\alpha_{1h}(X_1-X_4)Y_4\right].
\eeqn

\beq\label{eq:R1}
R_1 &=& h^{-1}\iiint f(x_1)^{-2}K(u)\alpha_1(u)K(v)m(x_1-hu)m(x_1-hv)\nonumber\\
&&f(x_1-hu)f(x_1-hv)\,dx_1dudv\nonumber\\
&\overset{2}{=}& h\left[\frac{1}{2}R(K)\mu_2(K)\int m\varphi_1-\frac{1}{2}\mu_2(K^2)\int m\varphi_1\right].
\eeq

\beq\label{eq:R2}
R_2 &=& -h^{-1}\iiiint f(x_1)^{-3}K(u)\alpha_1(u)K(v)K(w)m(x_1-hv)\nonumber\\
&&m(x_1-hw)f(x_1-hu)f(x_1-hv)f(x_1-hw)\,dx_1dudvdw\nonumber\\
&\overset{2}{=}& -h\left[\frac{1}{4}R(K)\mu_2(K)\int m^2f^{-1}f''+\frac{1}{2}R(K)\mu_2(K)\int m\varphi_1-\frac{1}{2}\mu_2(K^2)\int m\varphi_1\right].
\eeq

\beq\label{eq:R3}
R_3 &=& -h^{-1}\iint f(x_1)^{-1}m(x_1)K(u)\alpha_1(u)m(x_1-hu)f(x_1-hu)\nonumber\\
&&\,dx_1du \overset{2}{=} \frac{1}{2}\mu_2(K^2)h\int m\varphi_1.
\eeq

\beq\label{eq:R4}
R_4 &=& h^{-1}\iiint f(x_1)^{-2}m(x_1)K(u)\alpha_1(u)K(v)m(x_1-hu)\nonumber\\
&&f(x_1-hu)f(x_1-hv)\,dx_1dudv\nonumber\\
&\overset{2}{=}& h\left[\frac{1}{4}R(K)\mu_2(K)\int m^2f^{-1}f''-\frac{1}{2}\mu_2(K^2)\int m\varphi_1\right].
\eeq

Plugging \eqref{eq:R1}--\eqref{eq:R4} into \eqref{eq:omega234422} we get
\beq\label{eq:omega234422_2}
\E\left(\Omega_2^{34}\Omega_4^{22}\right) \overset{2}{=} O\left(h^3\right).
\eeq

Plugging \eqref{eq:omega223445_2}, \eqref{eq:omega222434_2}, \eqref{eq:omega223424_2}, \eqref{eq:omega223442_2}, \eqref{eq:omega232424}, \eqref{eq:omega232442_2} and \eqref{eq:omega234422_2} into \eqref{eq:omega24} yields
\beq\label{eq:omega24_2}
\E\left(\suma[j=2][n]\suma[k=2][n]\suma[l=2][n]\suma[r=2][n]\Omega_2^{jk}\Omega_4^{lr}\right) &\overset{2}{=}& \frac{1}{6}\mu_2(K)\mu_4(K)n^4h^6\int mf^{4)}\varphi_1\nonumber\\
&+&n^3h\left[R(K)\mu_2(K)\int m\varphi_1-\mu_2(K^2)\int m\varphi_1\right.\nonumber\\
&-&\left.\frac{1}{2}\mu_2(K^2)\int f^{-1}\varphi_3\right].
\eeq

Finally, plugging \eqref{eq:omega13_2}, \eqref{eq:omega23_2}, \eqref{eq:omega14_2} and \eqref{eq:omega24_2} into \eqref{eq:cvbarprima} we get
\beqn
\E\left[\overline{CV}'(h)\right] \overset{2}{=}A_1h^5+A_2n^{-1}.
\eeqn

%%%%%%%%%%%%%%
%%% Fin de la demostración para la parte del sesgo %%%
%%%%%%%%%%%%%%

Let us now define
\beqn
P_{ij} &=& \frac{Y_i-m(X_i)}{f(X_i)}[Y_j-m(X_i)]\alpha_{1h}(X_i-X_j),\\
Q_{ijk} &=& f(X_i)^{-2}[Y_j-m(X_i)][Y_k-m(X_i)]\beta_{1h}(X_i-X_j,X_i-X_k).
\eeqn

Then,
\beqn
\var\left[CV_n'(h)\right] = \frac{4}{n^2(n-1)^4h^2}\suma[i=1][n]\suma[\substack{j=1\\j \neq i}][n]\suma[\substack{k=1\\k \neq i}][n]\suma[l=1][n]\suma[\substack{r=1\\r \neq l}][n]\suma[\substack{s=1\\s \neq l}][n]C_{ijklrs},
\eeqn
where
\beqn
C_{ijklrs} = \cov\left(P_{ij},P_{lr}\right)-h^{-1}\cov\left(P_{ij},Q_{lrs}\right)-h^{-1}\cov\left(P_{lr},Q_{ijk}\right)+h^{-2}\cov\left(Q_{ijk},Q_{lrs}\right).
\eeqn

By counting the possible cases we get
\beqn
\var\left[CV_n'(h)\right] &=& \frac{4}{n^2(n-1)^4h^2}\left[n(n-1)(n-2)(n-3)(n-4)(n-5)C_{123456}\right.\\
&+&\left.n(n-1)(n-2)(n-3)(n-4)\left(C_{123145}+2C_{123415}+2C_{123451}+2C_{123455}\right.\right.\\
&+&\left.\left.C_{123425}+2C_{123452}+C_{123453}\right)+n(n-1)(n-2)(n-3)\left(2C_{122134}+C_{123124}\right.\right.\\
&+&\left.\left.2C_{123142}+C_{123143}+2C_{122314}+C_{123214}+2C_{123412}+2C_{123314}+2C_{123413}\right.\right.\\
&+&\left.\left.2C_{122341}+2C_{123421}+C_{123341}+2C_{123431}+C_{122344}+C_{123423}+C_{123432}+\right.\right.\\
&+&\left.\left.2C_{123411}+2C_{122324}+2C_{122342}\right)+n(n-1)(n-2)\left(C_{122322}+C_{122133}\right.\right.\\
&+&\left.\left.C_{123123}+C_{123132}+C_{123213}+2C_{123312}+C_{123321}+2C_{122311}+2C_{123211}\right.\right.\\
&+&\left.\left.2C_{123311}+2C_{123122}+2C_{123322}+2C_{122132}+2C_{122312}\right)\right.\\
&+&\left.n(n-1)\left(C_{122122}+C_{122211}\right)\right].
\eeqn

Among the previous covariances, it can be argued that the only ones that contribute to the dominant term of $\var\left[CV_n'(h)\right]$ are $C_{123145}$, $C_{123245}$, $C_{123425}$ and $C_{123124}$. Before we continue and with the intention of facilitating the calculations of the four $C_{ijklrs}$ that we need, let us obtain general expressions for each of the summands that make up $C_{ijklrs}$. Since
\beqn
\E\left(P_{ij}\cond X_i,X_j,X_l,X_r,Y_j,Y_r\right) = 0
\eeqn
and
\beqn
\cov\left(Y_i-m(X_i),Y_l-m(X_l)\cond X_i, X_l\right) = \delta_{il}\sigma^2(X_i)
\eeqn
then
\beqn
\cov\left(P_{ij},P_{lr}\right) &=& \E\left[\cov\left(P_{ij},P_{lr}\cond X_i,X_j,X_l,X_r,Y_j,Y_r\right)\right]\\
&=& \E\left\{f(X_i)^{-1}f(X_l)^{-1}\alpha_{1h}(X_i-X_j)\alpha_{1h}(X_l-X_r)[Y_j-m(X_i)][Y_r-m(X_l)]\right.\\
&&\left.\cov\left(Y_i-m(X_i),Y_l-m(X_l)\cond X_i, X_l\right)\right\}\\
&=& \delta_{il}\E\left\{f(X_i)^{-2}\alpha_{1h}(X_i-X_j)\alpha_{1h}(X_i-X_r)[Y_j-m(X_i)][Y_r-m(X_i)]\sigma^2(X_i)\right\}.
\eeqn

\beqn
\cov\left(P_{ij}, Q_{lrs}\right) &=& \E\left\{f(X_i)^{-1}f(X_l)^{-2}\alpha_{1h}(X_i-X_j)\beta_{1h}(X_l-X_r,X_l-X_s)\right.\\
&&\left.[Y_i-m(X_i)][Y_j-m(X_i)][Y_r-m(X_l)][Y_s-m(X_l)]\right\}
\eeqn

If $r, s \neq i$ it is clear that $\cov\left(P_{ij}, Q_{lrs}\right) = 0$. Now, for the cases $r = i$ and $s = i$ (both cases imply $i \neq l$), let us define
\beqn
t = \begin{cases}
	s,\,\mbox{if } r = i\\
	r,\,\mbox{if } s = i
\end{cases}
\eeqn
and note that
\beqn
\cov\left[Y_i-m(X_i), Y_i-m(X_l)\right] = \sigma^2(X_i).
\eeqn

Then,
\beqn
\cov\left(P_{ij},Q_{lrs}\right) &=& \E\left\{f(X_i)^{-1}f(X_l)^{-2}\alpha_{1h}(X_i-X_j)\beta_{1h}(X_l-X_r,X_l-X_s)\right.\\
&&\left.[Y_j-m(X_i)][Y_t-m(X_l)]\cov\left[Y_i-m(X_i),Y_i-m(X_l)\cond X_i, X_l\right]\right\}\\
&=& \E\left\{f(X_i)^{-1}f(X_l)^{-2}\alpha_{1h}(X_i-X_j)\beta_{1h}(X_l-X_r,X_l-X_s)\right.\\
&&\left.[Y_j-m(X_i)][Y_t-m(X_l)]\sigma^2(X_i)\right\}.
\eeqn

\beqn
&&\cov\left(Q_{ijk},Q_{lrs}\right) \\ &=& \E\left[\cov\left(Q_{ijk},Q_{lrs}\cond X_i,X_j,X_k,X_l,X_r,X_s\right)\right]\\
&+&\cov\left[\E\left(Q_{ijk}\cond X_i,X_j,X_k,X_l,X_r,X_s\right),\E\left(Q_{lrs}\cond X_i,X_j,X_k,X_l,X_r,X_s\right)\right]\\
&=& \E\left(f(X_i)^{-2}f(X_l)^{-2}\beta_{1h}(X_i-X_j,X_i-X_k)\beta_{1h}(X_l-X_r,X_l-X_s)\right.\\
&&\left.\cov\left\{[Y_j-m(X_i)][Y_k-m(X_i)],[Y_r-m(X_l)][Y_s-m(X_l)]\cond X_i,X_j,X_k,X_l,X_r,X_s\right\}\right)\\
&+& \cov\left[\E\left(Q_{ijk}\cond X_i,X_j,X_k,X_l,X_r,X_s\right),\E\left(Q_{lrs}\cond X_i,X_j,X_k,X_l,X_r,X_s\right)\right].
\eeqn

Note that, if $\{j,k\} \cap \{r,s\} = \emptyset$, then
\beqn
\cov\left\{[Y_j-m(X_i)][Y_k-m(X_i)],[Y_r-m(X_l)][Y_s-m(X_l)]\cond X_i,X_j,X_k,X_l,X_r,X_s\right\} = 0.
\eeqn

Now, regarding the term $C_{123245}$, since $i \neq l$ and $r,s \neq i$, we have
\beqn
\cov\left(P_{12}, P_{24}\right) = \cov\left(P_{12},Q_{245}\right) = 0
\eeqn
and
\beqn
\cov\left(P_{24},Q_{123}\right) &=& \E\left\{f(X_2)^{-1}f(X_1)^{-2}\alpha_{1h}(X_2-X_4)\beta_{1h}(X_1-X_2,X_1-X_3)[Y_4-m(X_2)]\right.\\
&&\left.[Y_3-m(X_1)]\sigma^2(X_2)\right\} \\
& = &\E\left\{f(X_2)^{-1}f(X_1)^{-2}\alpha_{1h}(X_2-X_4)\beta_{1h}(X_1-X_2,X_1-X_3)\right.\\
&&\left.[m(X_4)-m(X_2)][m(X_3)-m(X_1)]\sigma^2(X_2)\right\}\\
&=& \iiiint f(x_2)^{-1}f(x_1)^{-2}\alpha_{1h}(x_2-x_4)\beta_{1h}(x_1-x_2,x_1-x_3)[m(x_4)-m(x_2)]\\
&&[m(x_3)-m(x_1)]f(x_1)f(x_2)f(x_3)f(x_4)\,dx_1dx_2dx_3dx_4
\eeqn

Making the following change of variables,
\beqn
\begin{cases}
	x_4 = x_2-hu_4\\
	x_3 = x_1-hu_3\\
	x_2 = x_1-hu_2
\end{cases}
\eeqn
and using the fact that
\beqn
&&\iiint \alpha_1(u_4)\beta_1(u_2,u_3)u_4^iu_2^ju_3^k\,du_4du_2du_3 = ik\mu_i(K)\mu_j(K)\mu_k(K) = 0\\
&&\iff \mbox{$i = 0$ or $k = 0$ or ($i$, $j$ or $k$ is an odd number)},
\eeqn
we obtain that
\beqn
\cov\left(P_{24},Q_{123}\right)&=& h\iiiint f(x_1)^{-1}\alpha_1(u_4)\beta_1(u_2,u_3)[m(x_1-hu_2-hu_4)-m(x_1-hu_2)]\\
&&[m(x_1-hu_3)-m(x_1)]\sigma^2(x_1-hu_2)f(x_1-hu_3)\\
&& f(x_1-hu_2-hu_4)\,dx_1du_2du_3du_4\\
&=& h\iiiint f(x_1)^{-1}\alpha_1(u_4)\beta_1(u_2,u_3)u_4^2u_3^2h^4\left[\frac{1}{4}m''(x_1)^2\sigma^2(x_1)f(x_1)^2\right.\\
&+&\left.m'(x_1)^2\sigma^2(x_1)f'(x_1)^2+m'(x_1)m''(x_1)\sigma^2(x_1)f(x_1)f'(x_1)\right]\,dx_1du_2du_3du_4\\
&+&O\left(h^7\right)\\
&=& 4\mu_2(K)^2h^5\int f(x)^{-1}\sigma^2(x)\left[\frac{1}{4}m''(x)^2f(x)^2+m'(x)^2f'(x)^2\right.\\
&+&\left.m'(x)m''(x)f(x)f'(x)\right]dx+O\left(h^7\right),
\eeqn
%where we have made the following change of variables,
%\beqn
%\begin{cases}
%	x_4 = x_2-hu_4\\
%	x_3 = x_1-hu_3\\
%	x_2 = x_1-hu_2
%\end{cases}
%\eeqn
%and used the fact that
%\beqn
%&&\iiint \alpha_1(u_4)\beta_1(u_2,u_3)u_4^iu_2^ju_3^k\,du_4du_2du_3 = ik\mu_i(K)\mu_j(K)\mu_k(K) = 0\\
%&&\iff \mbox{$i = 0$ or $k = 0$ or ($i$, $j$ or $k$ is an odd number)}.
%\eeqn

Since $\{j,k\} \cap \{r,s\} = \emptyset$,
\beqn
\cov\left(Q_{123},Q_{245}\right) &=& \cov\left\{f(X_1)^{-2}\beta_{1h}(X_1-X_2,X_1-X_3)[m(X_2)-m(X_1)]\right.\\
&&\left.[m(X_3)-m(X_1)], f(X_2)^{-2}\beta_{1h}(X_2-X_4,X_2-X_5)[m(X_4)-m(X_2)]\right.\\
&&\left.[m(X_5)-m(X_2)]\right\}\\
&=& \E\left\{f(X_1)^{-2}f(X_2)^{-2}\beta_{1h}(X_1-X_2,X_1-X_3)\beta_{1h}(X_2-X_4,X_2-X_5)\right.\\
&&\left.[m(X_2)-m(X_1)][m(X_3)-m(X_1)][m(X_4)-m(X_2)]\right.\\
&&\left.[m(X_5)-m(X_2)]\right\}\\
&-&\E\left\{f(X_1)^{-2}\beta_{1h}(X_1-X_2,X_1-X_3)[m(X_2)-m(X_1)]\right.\\
&&\left.[m(X_3)-m(X_1)]\right\}^2\\
&=& O\left(h^{10}\right).
\eeqn

Therefore,
\beqn
C_{123245} &=& -4\mu_2(K)^2h^4\int f(x)^{-1}\sigma^2(x)\left[\frac{1}{4}m''(x)^2f(x)^2+m'(x)^2f'(x)^2\right.\\
&+&\left.m'(x)m''(x)f(x)f'(x)\right]\,dx+O\left(h^6\right).
\eeqn

Regarding the term $C_{123425}$, since $i \neq l$, $j,k \neq l$ and $r,s \neq i$, then
\beqn
\cov\left(P_{12},P_{42}\right) = \cov\left(P_{12},Q_{425}\right) = \cov\left(P_{42},Q_{123}\right) = 0.
\eeqn

%Let $\mathcal{X} = \{X_1,X_2,X_3,X_4,X_5\}$. Then
%\beqn
%\cov\left((Y_2-m(X_1))(Y_3-m(X_1)),(Y_2-m(X_4))(Y_5-m(X_4))\cond \mathcal{X}\right) &=& \sigma^2(X_2)(m(X_3)-m(X_1))\\
%&&(m(X_5)-m(X_4)).
%\eeqn
%
%\beqn
%&&\E\left[f(X_1)^{-2}f(X_4)^{-2}\beta_{1h}(X_1-X_2,X_1-X_3)\beta_{1h}(X_4-X_2,X_4-X_5)\sigma^2(X_2)(m(X_3)-m(X_1))\right.\\
%&&\left.(m(X_5)-mX(X_4))\right] = \int\dots\int f(x_1)^{-2}f(x_4)^{-2} \beta_{1h}(x_1-x_2,x_1-x_3)\beta_{1h}(x_4-x_2,x_4-x_5)\sigma^2(x_2)\\
%&&(m(x_3)-m(x_1))(m(x_5)-m(x_4))f(x_1)f(x_2)f(x_3)f(x_4)f(x_5)\,dx_1dx_2dx_3dx_4dx_5\\
%&&= h^2\int\dots\int f(x_1)^{-1}f(x_1-hu_2+hu_4)^{-1}\beta_1(u_2,u_3)\beta_1(u_4,u_5)\sigma^2(x_1-hu_2)(m(x_1-hu_3)-m(x_1))\\
%&&(m(x_1-hu_2+hu_4-hu_5)-m(x_1-hu_2+hu_4))f(x_1-hu_2)f(x_1-hu_3)f(x_1-hu_2+hu_4-hu_5)\\
%&&dx_1du_2du_3du_4du_5 =,
%\eeqn
We have that
\beqn
&&h^2\int\dots\int f(x_1)^{-2}f(x_4)^{-2}\beta_{1h}(x_1-x_2,x_1-x_3)\beta_{1h}(x_4-x_2,x_4-x_5)[m(x_3)-m(x_1)]\\
&& [m(x_5)-m(x_4)]\left\{\sigma^2(x_2)+[m(x_2)-m(x_1)][m(x_2)-m(x_4)]\right\}\\
&&f(x_1)f(x_2)f(x_3)f(x_4)f(x_5)dx_1dx_2dx_3dx_4dx_5\\
&&= h^2\int\dots\int f(x_1)^{-1}f(x_1-hu_2+hu_4)^{-1}\beta_1(u_2,u_3)\beta_1(u_4,u_5)[m(x_1-hu_3)-m(x_1)]\\
&&[m(x_1-hu_2+hu_4-hu_5)-m(x_1-hu_2+hu_4)]\left\{\sigma^2(x_1-hu_2)\right.\\
&+&\left.[m(x_1-hu_2)-m(x_1)][m(x_1-hu_2)-m(x_1-hu_2+hu_4)]\right\}f(x_1-hu_2)\\
&&f(x_1-hu_3)f(x_1-hu_2+hu_4-hu_5)dx_1du_2du_3du_4du_5\\
& =& 4R(K)^2\mu_2(K)^2h^6\int\sigma^2f(x)^{-1}\left[\frac{1}{4}(m'')^2f^2+m'm''ff'+(m')^2(f')^2\right]+ O\left(h^8\right)
\eeqn
where we have made the following change of variables,
\beqn
\begin{cases}
	x_2 = x_1-hu_2\\
	x_3 = x_1-hu_3\\
	x_4 = x_2+hu_4\\
	x_5 = x_2-hu_5
\end{cases}
\eeqn
and used the fact that
\beqn
&&\iiiint \beta_1(u_2,u_3)\beta_1(u_4,u_5)u_2^iu_3^ju_4^ku_5^l\,du_2du_3du_4du_5 = jl\mu_i(K)\mu_j(K)\mu_k(K)\mu_l(K) = 0\\
&&\iff \mbox{$j = 0$ or $l = 0$ or ($i$, $j$, $k$ or $l$ is an odd number)}.
\eeqn

Therefore,
\beqn
C_{123425} = 8R(K)^2\mu_2(K)^2h^4\int\sigma^2f(x)^{-1}\left[\frac{1}{4}(m'')^2f^2+m'm''ff'+(m')^2(f')^2\right]+O\left(h^6\right).
\eeqn

As for the term $C_{123124}$, since $r,s \neq i$ and $j,k \neq l$, then
\beqn
\cov\left(P_{12},Q_{124}\right) = \cov\left(P_{12},Q_{123}\right) = 0.
\eeqn

We have that $i = l$ and so
\beqn
\cov\left(P_{12},P_{12}\right) &=& \E\left\{f(X_1)^{-2}\alpha_{1h}(X_1-X_2)^2\sigma^2(X_1)\left[\left(m(X_2)-m(X_1)\right)^2+\sigma^2(X_2)\right]\right\}\\
&=& h^{-1}\iint f(x_1)^{-1}\alpha_1(u)^2\sigma^2(x_1)\left\{\sigma^2(x_1-hu)+\left[m(x_1-hu)-m(x_1)\right]^2\right\}\\
&&f(x_1-hu)\,dx_1du = \mu_2\left[(K')^2\right]h^{-1}\int (\sigma^2)^2+O\left(h\right),
\eeqn
where we have used the fact that
\beqn
\int \alpha_1(u)^2u^i\,du = -i\mu_i(K^2)+\mu_{i+2}\left[(K')^2\right] = 0 \iff \mbox{$i$ is odd}.
\eeqn

\beqn
\cov\left(Q_{123},Q_{124}\right) &=& \E\left(f(X_1)^{-4}\beta_{1h}(X_1-X_2,X_1-X_3)\beta_{1h}(X_1-X_2,X_1-X_4)\right.\\
&&\left.[m(X_3)-m(X_1)][m(X_4)-m(X_1)]\left\{\sigma^2(X_2)+[m(X_2)-m(X_1)]^2\right\}\right)\\
&-&\E\left\{f(X_1)^{-2}\beta_{1h}(X_1-X_2,X_1-X_3)[m(X_2)-m(X_1)]\right.\\
&& \left.[m(X_3)-m(X_1)]\right\}^2=O\left(h^5\right).
\eeqn

Therefore,
\beqn
C_{123124} = \mu_2\left[(K')^2\right]h^{-1}\int (\sigma^2)^2+O\left(h\right).
\eeqn

%\underline{$C_{123145}$}\\

Finally, using similar reasoning and calculations we get that
\beqn
C_{123145} = 4\mu_2(K)^2h^4\int f^{-1}\sigma^2\left[\frac{1}{4}(m'')^2f^2+(m')^2(f')^2+m'm''ff'\right]+O\left(h^6\right).
\eeqn
\end{proof}

Now, from the following decomposition (equation (11) of the main paper):
\beq\label{eq:hcvtaylor}
\hat h_{CV,n}-h_{n0} &\approx& -\frac{CV_n'(h_{n0})-M_n'(h_{n0})}{M_n''(h_{n0})}\nonumber\\
&+&\frac{[CV_n'(h_{n0})-M_n'(h_{n0})][CV_n''(h_{n0})-M_n''(h_{n0})]}{M_n''(h_{n0})^2},
\eeq
and using Lemmas \ref{th:bias_var_mh} and \ref{th:bias_var_cv1}, the asymptotic bias and variance of the cross-validation bandwidth that minimizes the modified version of the cross-validation criterion given in equation (9) of the main paper:
\beq\label{eq:cvtram}
\widetilde{CV}_n(h) = \frac{1}{n}\sum\limits_{i=1}^n \left[\tilde m_h^{(-i)}(X_i)-Y_i\right]^2,
\eeq
can be obtained. 
Theorem \ref{th:dist_asint_h_tramposo} contains this result.

\begin{mtheo}{3.1}
\label{th:dist_asint_h_tramposo}
{Under assumptions {\rm A1}--{\rm A4},} the {asymptotic} bias and the variance of the bandwidth that minimizes \eqref{eq:cvtram} {are:}
\beqn
{\rm E}\left(\hat h_{CV,n}\right)-h_{n0} &=& Bn^{-3/5}+o\left(n^{-3/5}\right),\\
{\rm var}\left(\hat h_{CV,n}\right) &=& Vn^{-3/5}+o\left(n^{-3/5}\right),
\eeqn
where
\beqn
B &=& \frac{6B_2C_0^5+V_2-A_1C_0^5-A_2}{12B_1C_0^2+2V_1C_0^{-3}},\\
V &=& \frac{R_1C_0^2+R_2C_0^{-3}}{\left(12B_1C_0^2+2V_1C_0^{-3}\right)^2}.
\eeqn
\end{mtheo}

\begin{proof}[Proof of Theorem \ref{th:dist_asint_h_tramposo}]
From equation (\ref{eq:hcvtaylor}),
it follows that, up to first order,
\beq\label{eq:dem_bias_hcv}
\E\left(\hat h_{CV,n}\right)-h_{n0} &=& \frac{M_n'(h_{n0})-\E\left[CV'_n(h_{n0})\right]}{M_n''(h_{n0})},\\\label{eq:dem_var_hcv}
\var\left(\hat h_{CV,n}\right) &=& \frac{\var\left[CV_n'(h_{n0})\right]}{M_n''(h_{n0})^2}.
\eeq

Since the first-order terms of $M_n'(h_{n0})$ and $\E\left[CV_n'(h_{n0})\right]$ coincide, we must consider the second-order terms of $M_n'(h_{n0})$ and $\E\left[CV_n'(h_{n0})\right]$  for the bias of $\hat h_{CV,n}$, while for the variance, it will suffice to consider the first-order term of $\var\left[CV_n'(h_{n0})\right]$. Therefore, to proof Theorem \ref{th:dist_asint_h_tramposo}, we only have to plug the results of Lemma \ref{th:bias_var_mh} and Lemma \ref{th:bias_var_mh} into \eqref{eq:dem_bias_hcv} and \eqref{eq:dem_var_hcv}.
\end{proof}

\begin{mcorol}{3.1}
\label{cor:limdist_hcv}
{Under assumptions of Theorem \ref{th:dist_asint_h_tramposo},} the asymptotic distribution of the bandwidth that minimizes {the modified version of the cross-validation criterion, given in equation {\rm (9)} of the main paper,} satisfies
\beqn
n^{3/10}\left(\hat h_{CV,n}-h_{n0}\right) \xrightarrow[]{d} N(0,V),
\eeqn
where the constant $V$ was defined in Theorem \ref{th:dist_asint_h_tramposo}.
\end{mcorol}

\begin{proof}[Proof of Corollary \ref{cor:limdist_hcv}]
Using the Cram\'er-Wold device \cite{CramerWold1936} and a reasoning similar to that followed in \cite{PaperBiometrika}, it is possible to derive the asymptotic normality of the statistic of interest, namely $n^{3/10}\left(\hat h_{CV,n}-h_{n0}\right)$. The mean and variance of the asymptotic distribution of this statistic are an immediate consequence of Theorem \ref{th:dist_asint_h_tramposo}.
\end{proof}

Expressions for the {asymptotic} bias and the variance of the bagged cross-validation bandwidth,
\beq\label{eq:hcvbag}
\hat h(r,N) = \frac{1}{N}\suma[i=1][N] \tilde h_{r,i}.
\eeq
are given in Theorem \ref{th:hcvbag_moments}.

\begin{mtheo}{4.1}
\label{th:hcvbag_moments}
{Under assumptions {\rm A1}--{\rm A5},} the {asymptotic bias and the variance of the} bagged cross-validation bandwidth defined in \eqref{eq:hcvbag} {are:}
\beqn
{\rm E}\left[\hat h(r,N)\right]-h_{n0} &=& (B+C_1)r^{-2/5}n^{-1/5}+o\left(r^{-2/5}n^{-1/5}\right),\\
{\rm var}\left[\hat h(r,N)\right] &=& Vr^{-1/5}n^{-2/5}\left[\frac{1}{N}+\left(\frac{r}{n}\right)^2\right]+o\left(\frac{r^{-1/5}n^{-2/5}}{N}+r^{9/5}n^{-12/5}\right),
\eeqn
where the constants $B$ and $V$ were defined in Theorem \ref{th:dist_asint_h_tramposo} and the constant $C_1$ is defined in \eqref{eq:C1}.
\end{mtheo}

\begin{proof}[Proof of Theorem \ref{th:hcvbag_moments}]
If we define
\beq\label{eq:C1}
C_1 = -\frac{6B_2C_0^5+V_2}{12B_1C_0^2+2V_1C_0^{-3}},
\eeq
then we have
\beqn
h_{r0} &=& C_0r^{-1/5}+C_1r^{-3/5}+o\left(r^{-3/5}\right),\\
\left(\frac{r}{n}\right)^{1/5}h_{r0} &=& C_0n^{-1/5}+C_1r^{-2/5}n^{-1/5}+o\left(r^{-2/5}n^{-1/5}\right)
\eeqn
and
\beqn
\left(\frac{r}{n}\right)^{1/5}h_{r0}-h_{n0} &=& C_1\left(r^{-2/5}n^{-1/5}-n^{-3/5}\right)+o\left(r^{-2/5}n^{-1/5}+n^{-3/5}\right)\\
&=& C_1r^{-2/5}n^{-1/5}+o\left(r^{-2/5}n^{-1/5}\right),
\eeqn
where we have used the fact that $r = o(n)$. Therefore,
\beqn
\E\left[\hat h(r,N)\right]-h_{n0} &=& \E\left[\left(\frac{r}{n}\right)^{1/5}\hat h_{CV,r,1}\right]-h_{n0} \\
&=& \left(\frac{r}{n}\right)^{1/5}\E\left(\hat h_{CV,r,1}-h_{r0}\right)+\left[\left(\frac{r}{n}\right)^{1/5}h_{r0}-h_{n0}\right]\\
&=& (B+C_1)r^{-2/5}n^{-1/5}+o\left(r^{-2/5}n^{-1/5}\right).
\eeqn

Regarding the variance, we have
\beq\label{eq:varhbagdescomp}
\var\left[\hat h(r,N)\right] = \frac{1}{N}\left(\frac{r}{n}\right)^{2/5}\left[\var\left(\hat h_{CV,r,1}\right)+(N-1)\cov\left(\hat h_{CV,r,1},\hat h_{CV,r,2}\right)\right]
\eeq
and
\beq\label{eq:covhbag12}
\cov\left(\hat h_{CV,r,1},\hat h_{CV,r,2}\right) \approx M_r''(h_{r0})^{-2}\cov\left[CV_1'(h_{r0}),CV_2'(h_{r0})\right],
\eeq
where
\beqn
CV_q'(h) = \frac{2}{r(r-1)^2h}\sum\limits_{\substack{i,j,k \in I_q\\j,k\neq i}} \left[A_{ij}\alpha_{1h}(X_i-X_j)-h^{-1}B_{ijk}\beta_{1h}(X_i-X_j,X_i-X_k)\right],\, q \in \{1,2\},
\eeqn
$I_1, I_2 \sim U(\mathcal{P})$ and $\mathcal{P} = \{I \in \{1,\dots,n\} \mid \#I = r\}$.

\beqn
\cov\left[CV_1'(h),CV_2'(h)\right] &=& \cov\left\{\E\left[CV_1'(h)\cond I_1,I_2\right],\E\left[CV_2'(h)\cond I_1,I_2\right]\right\}\\
&+&\E\left\{\cov\left[CV_1'(h),CV_2'(h)\cond I_1,I_2\right]\right\}\\
&=& \E\left\{\cov\left[CV_1'(h),CV_2'(h)\cond I_1,I_2\right]\right\}
\eeqn
since $\E\left[CV_q'(h)\cond I_1,I_2\right]$, for $q \in \{1,2\}$, does not depend on $I_1,I_2$ and is therefore not random.

\beq\label{eq:covcv1cv2}
\cov\left[CV_1'(h),CV_2'(h)\cond I_1,I_2\right] &=& \frac{4}{r^2(r-1)^4h^2}\sum\limits_{\substack{i,j,k \in I_1\\l,s,t \in I_2\\j,k \neq i\\ s,t \neq l}}\cov\left[A_{ij}\alpha_{1h}(X_i-X_j)\right.\\\nonumber
&-&\left.h^{-1}B_{ijk}\beta_{1h}(X_i-X_j,X_i-X_k),A_{ls}\alpha_{1h}(X_l-X_s)\right.\\\nonumber
&-&\left.h^{-1}B_{lst}\beta_{1h}(X_l-X_s,X_l-X_t)\right].
\eeq

Following the proof of Lemma \ref{th:bias_var_cv1}, we only need to count the number of cases associated with $C_{123124}$ and $C_{123425}$. If we define $M = \#\left(I_1 \cap I_2\right)$, which is a random variable, then the number of times $C_{123124}$ and $C_{123425}$ appear in \eqref{eq:covcv1cv2} is $M(M-1)(r^2-4r-M) = M^2r^2+o\left(M^2r^2\right)$ for $C_{123124}$, and $M^2r^3+o\left(M^2r^3\right)$ for $C_{123425}$.
%\beq\label{eq:conteo_casos}
%C_{123124}&:& M(M-1)(r^2-4r-M) = M^2r^2+o\left(M^2r^2\right),\\\nonumber
%C_{123425}&:& M^2r^3+o\left(M^2r^3\right).
%\eeq

Plugging these numbers into \eqref{eq:covcv1cv2}, we get
\beqn
\cov\left[CV_1'(h),CV_2'(h)\cond I_1,I_2\right] = \frac{4}{r^2(r-1)^4h^2}\left(C_{123124}M^2r^2+C_{123425}M^2r^3\right)+Z,
\eeqn
where $Z= o_p\left(C_{123124}M^2r^{-4}+C_{123425}M^2r^{-3}\right)$.

\beqn
\E\left(M^2 \cond I_1\right) &=& \E\left\{\left[\sum\limits_{i \in I_1}1_{I_2}(i)\right]^2 \cond I_1\right\} = \E\left[\sum\limits_{i \in I_1}\sum\limits_{j \in I_1}1_{I_2}(i)1_{I_2}(j) \cond I_1\right] \\
&=& \sum\limits_{i \in I_1}\sum\limits_{j \in I_1} \mathbb{P}(i,j \in I_2 \cond I_1)= r\mathbb{P}(1 \in I_2)+r(r-1)\mathbb{P}(1 \in I_2)^2 \\
&=& r\frac{r}{n}+r(r-1)\frac{r^2}{n^2}\\ 
&=& \frac{r^2[n+r(r-1)]}{n^2} \\
&=& \E\left(M^2\right),
\eeqn
where $1_{I_2}(\cdot)$ denotes the indicator function of $I_2$ and we have used the fact that $1_{I_2}(i)$ is a Bernoulli distribution with parameter $r/n$. Therefore,
\beqn
\cov\left[CV_1'(h),CV_2'(h)\right] &=& R_1(n^{-1}r^{-1}h^2+rn^{-2}h^2)\\
&+&R_2n^{-2}h^{-3}+O\left(n^{-2}h^{-1}+n^{-1}r^{-1}h^4+n^{-2}rh^4\right)
\eeqn
and
\beq\label{eq:covcv1cv2_hr0}
\cov\left[CV_1'(h_{r0}),CV_2'(h_{r0})\right] & = & R_1C_0^2(n^{-1}r^{-7/5}+n^{-2}r^{3/5})\nonumber\\ & + &R_2C_0^{-3}n^{-2}r^{3/5}+O\left(n^{-1}r^{-9/5}+n^{-2}r^{1/5}\right).
\eeq

Now, plugging \eqref{eq:covcv1cv2_hr0} into \eqref{eq:covhbag12} we get
\beq\label{eq:covhbag12_hr0}
\cov\left(\hat h_{CV,r,1},\hat h_{CV,r,2}\right) = Vn^{-2}r^{7/5}+Wn^{-1}r^{-3/5}+O\left(n^{-2}r+n^{-1}r^{-1}\right),
\eeq
where
\beqn
W = \frac{R_1C_0^2}{\left(12B_1C_0^2+2V_1C_0^{-3}\right)^2}.
\eeqn

Finally, plugging \eqref{eq:covhbag12_hr0} into \eqref{eq:varhbagdescomp} yields
\beqn
\var\left[\hat h(r,N)\right] = Vr^{-1/5}n^{-2/5}\left[\frac{1}{N}+\left(\frac{r}{n}\right)^2\right]+o\left(r^{9/5}n^{-12/5}\right).
\eeqn

\end{proof}

\begin{mcorol}{4.1}\label{cor:limdist_hcvbag}
{Under the assumptions of Thorem \ref{th:hcvbag_moments},} the asymptotic distribution of the bagged cross-validation bandwidth defined in \eqref{eq:hcvbag} satisfies:
\beqn
\frac{r^{1/10}n^{1/5}}{\sqrt{\frac{1}{N}+\left(\frac{r}{n}\right)^2}}\left[\hat h(r,N)-h_{n0}\right] \xrightarrow[]{d} N(0, V),
\eeqn
where the constant $V$ was defined in Theorem \ref{th:dist_asint_h_tramposo}. In particular, if we assume that $r = o\left(n/\sqrt{N}\right)$, then,
\beqn
r^{1/10}n^{1/5}\sqrt{N}\left[\hat h(r,N)-h_{n0}\right] \xrightarrow[]{d} N(0,V).
\eeqn
\end{mcorol}

\begin{proof}[Proof of Corollary \ref{cor:limdist_hcvbag}]
The result is obtained immediately from Corollary \ref{cor:limdist_hcv} and Theorem \ref{th:hcvbag_moments}.
\end{proof}

\section{Simulation studies}\label{sec:sims}
\begin{figure}[htb]
	\begin{subfigure}{.3\textwidth}
		\centering
	\includegraphics[width=100px]{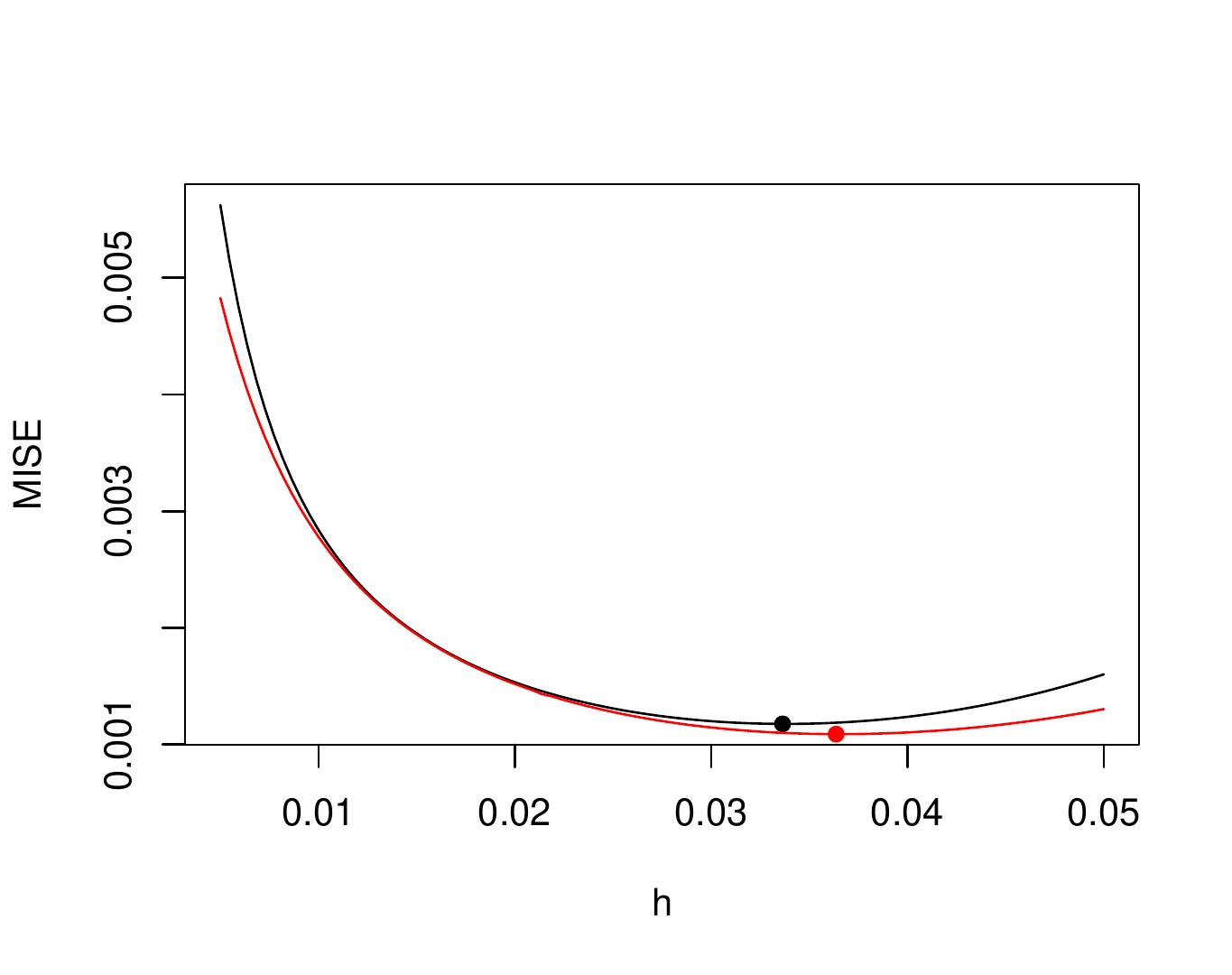}
	\caption{Model M1, $n = 100$}
	\end{subfigure}
	\begin{subfigure}{.3\textwidth}
		\centering
	\includegraphics[width=100px]{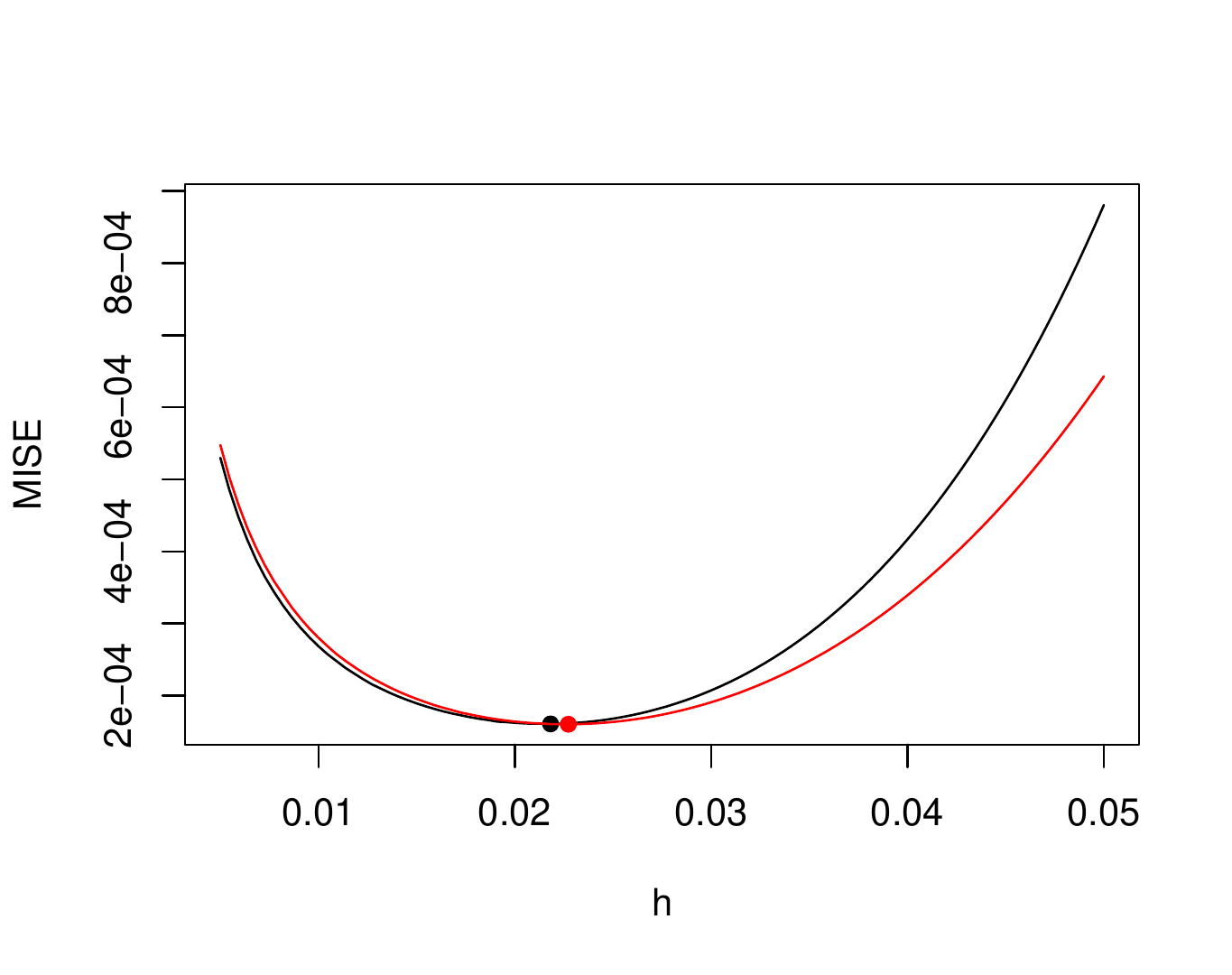}
	\caption{Model M1, $n = 1000$}
	\end{subfigure}
	\begin{subfigure}{.3\textwidth}
		\centering
	\includegraphics[width=100px]{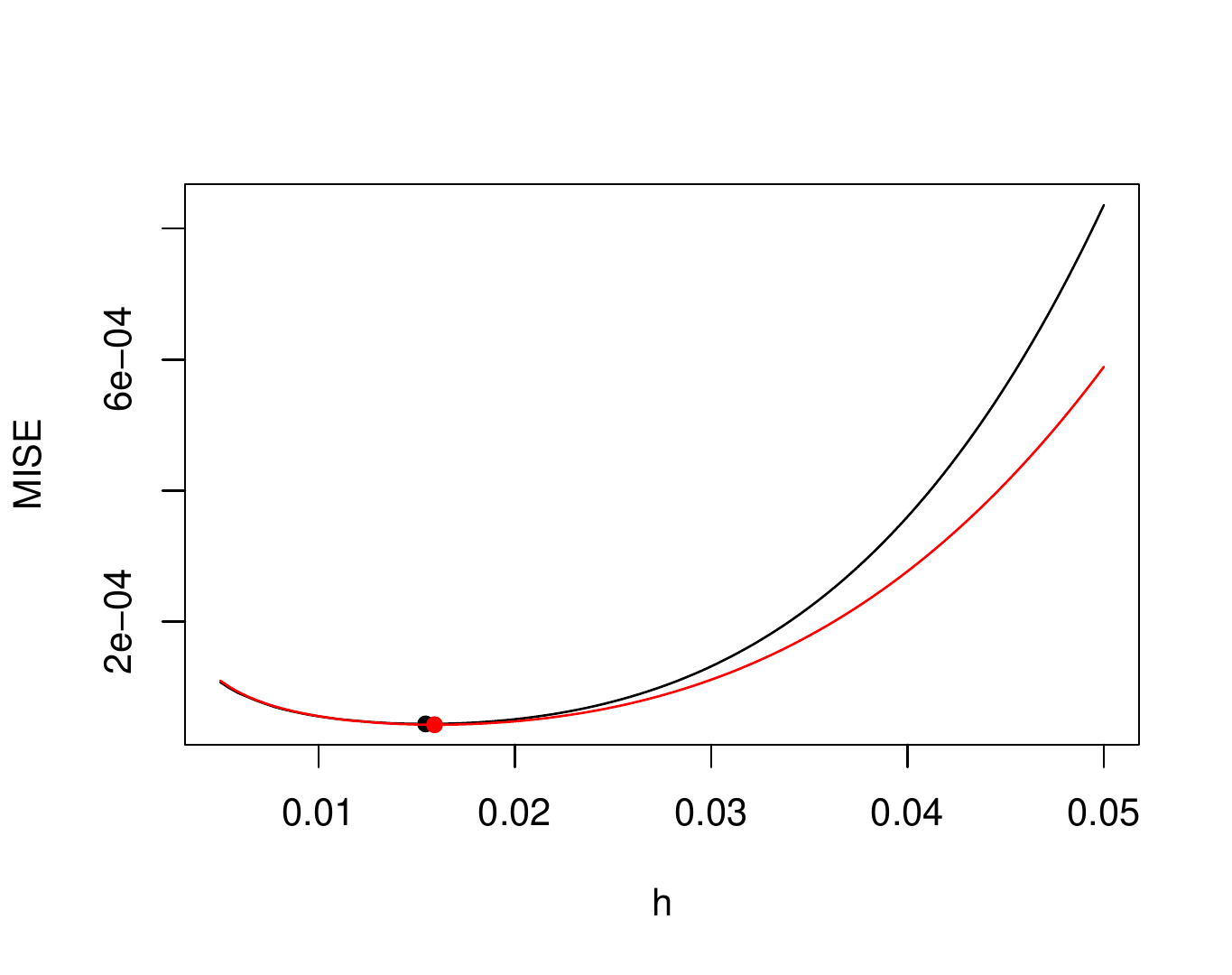}
	\caption{Model M1, $n = 5000$}
	\end{subfigure}
	\newline
	\begin{subfigure}{.3\textwidth}
		\centering
	\includegraphics[width=100px]{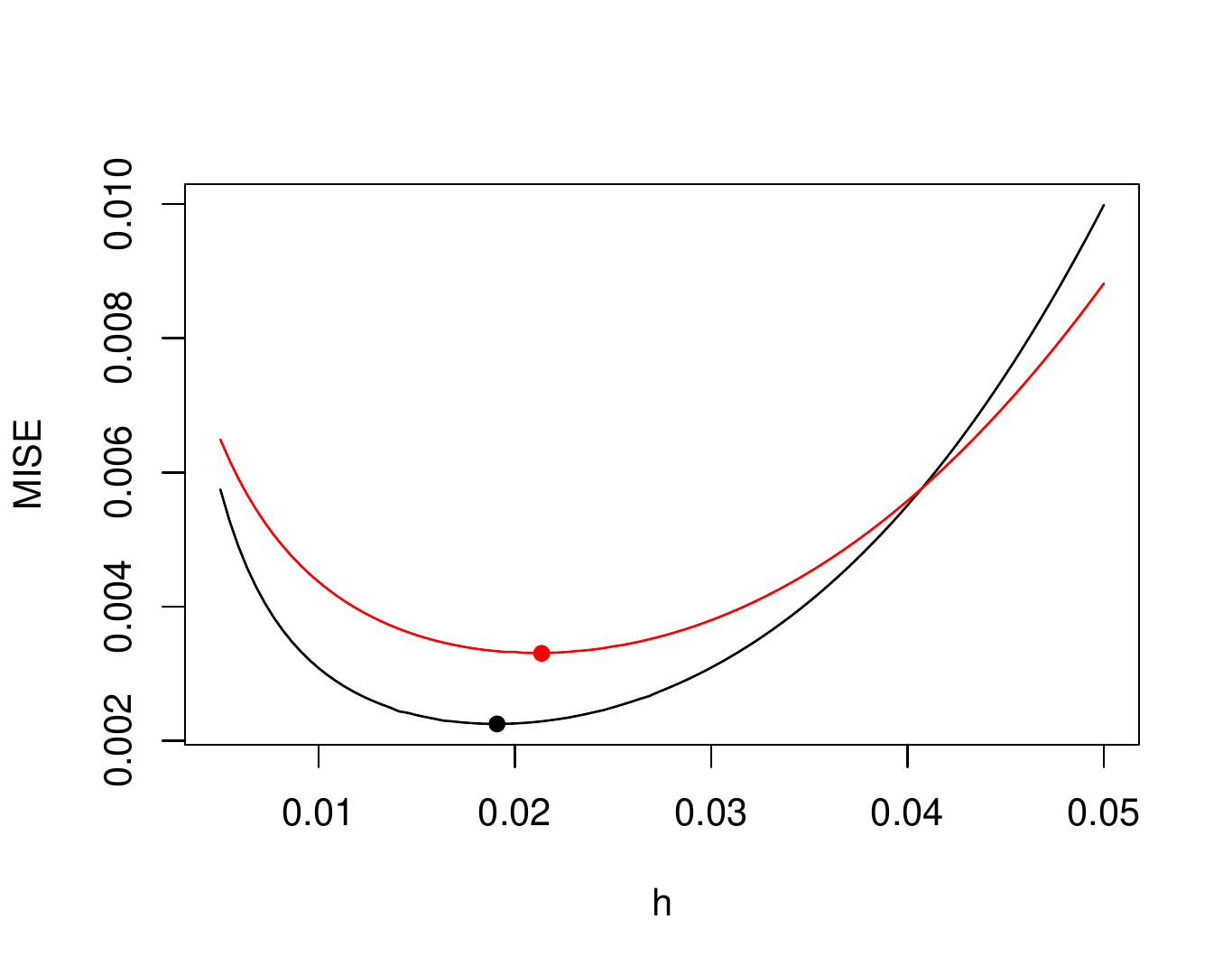}
	\caption{Model M2, $n = 100$}
	\end{subfigure}
	\begin{subfigure}{.3\textwidth}
		\centering
	\includegraphics[width=100px]{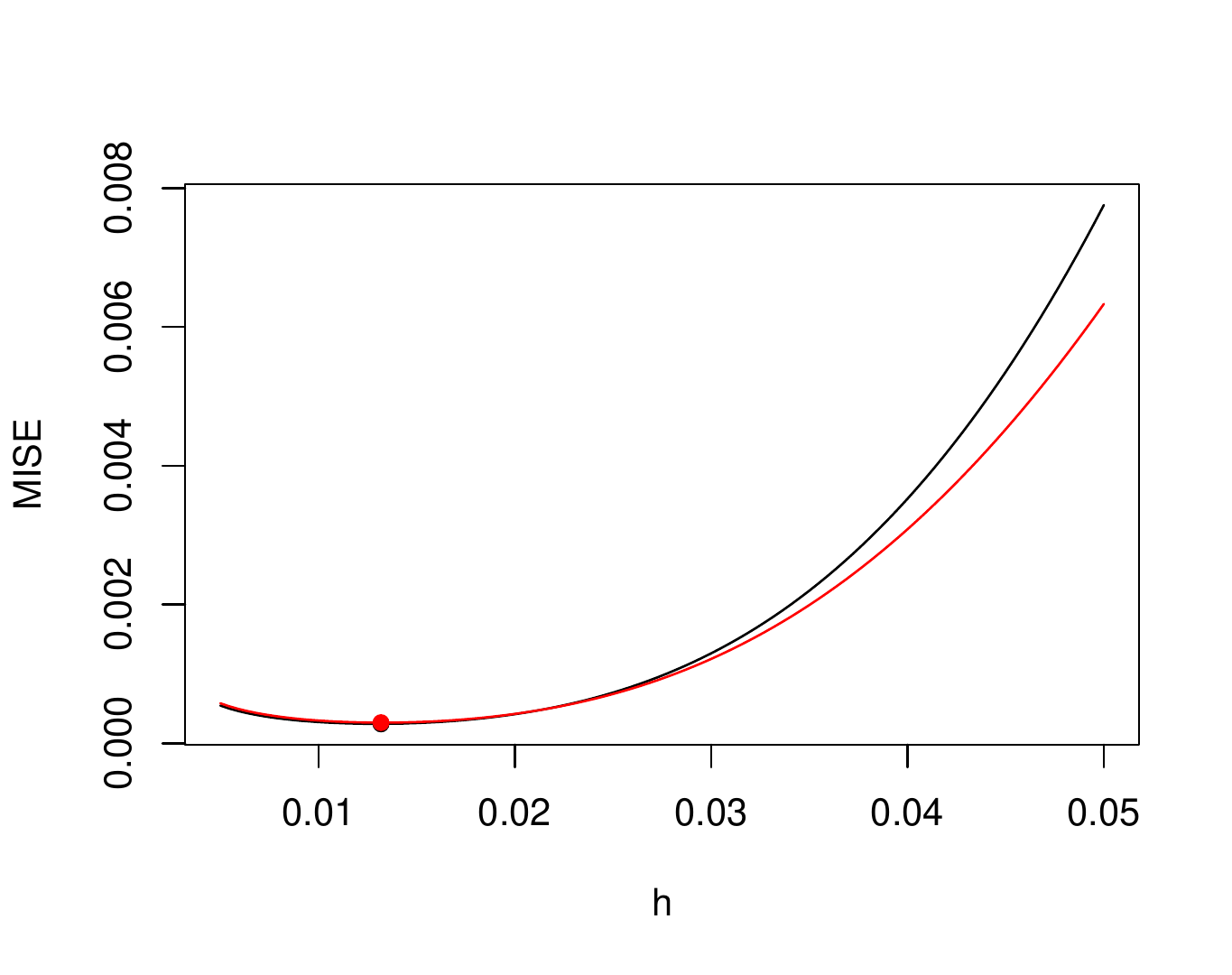}
	\caption{Model M2, $n = 1000$}
	\end{subfigure}
	\begin{subfigure}{.3\textwidth}
		\centering
	\includegraphics[width=100px]{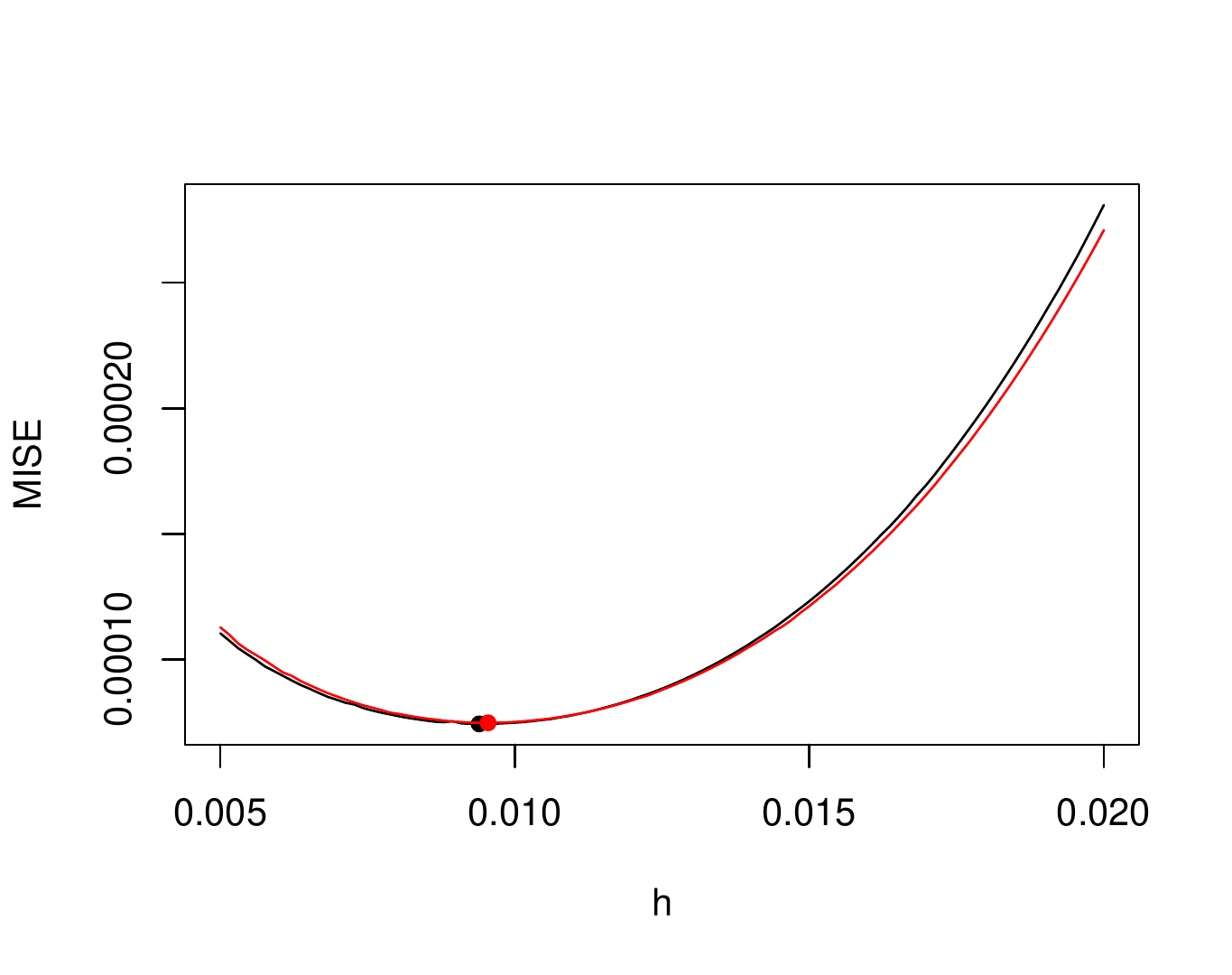}
	\caption{Model M2, $n = 5000$}
	\end{subfigure}
	\newline
	\begin{subfigure}{.3\textwidth}
		\centering
	\includegraphics[width=100px]{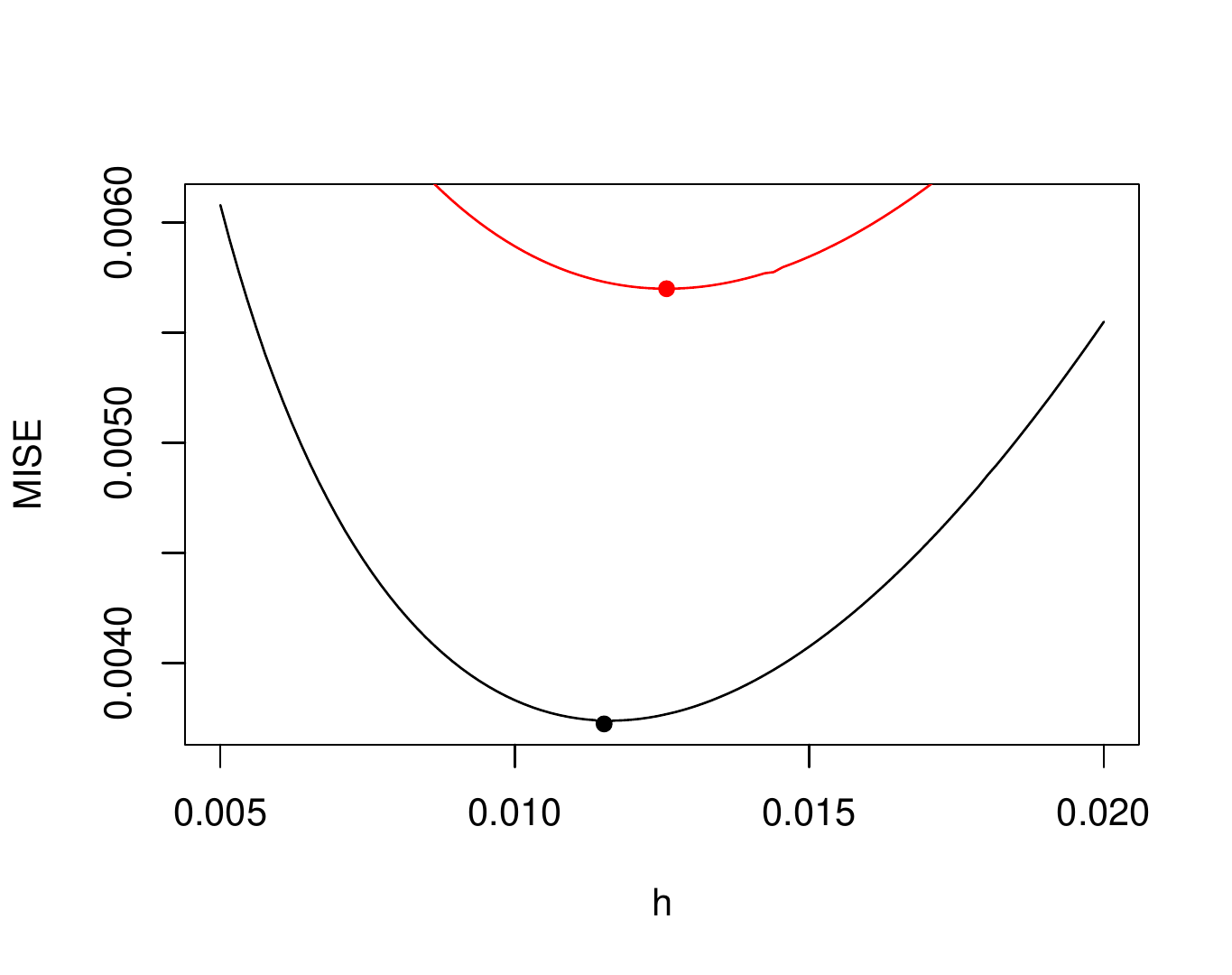}
	\caption{Model M3, $n = 100$}
	\end{subfigure}
	\begin{subfigure}{.3\textwidth}
		\centering
	\includegraphics[width=100px]{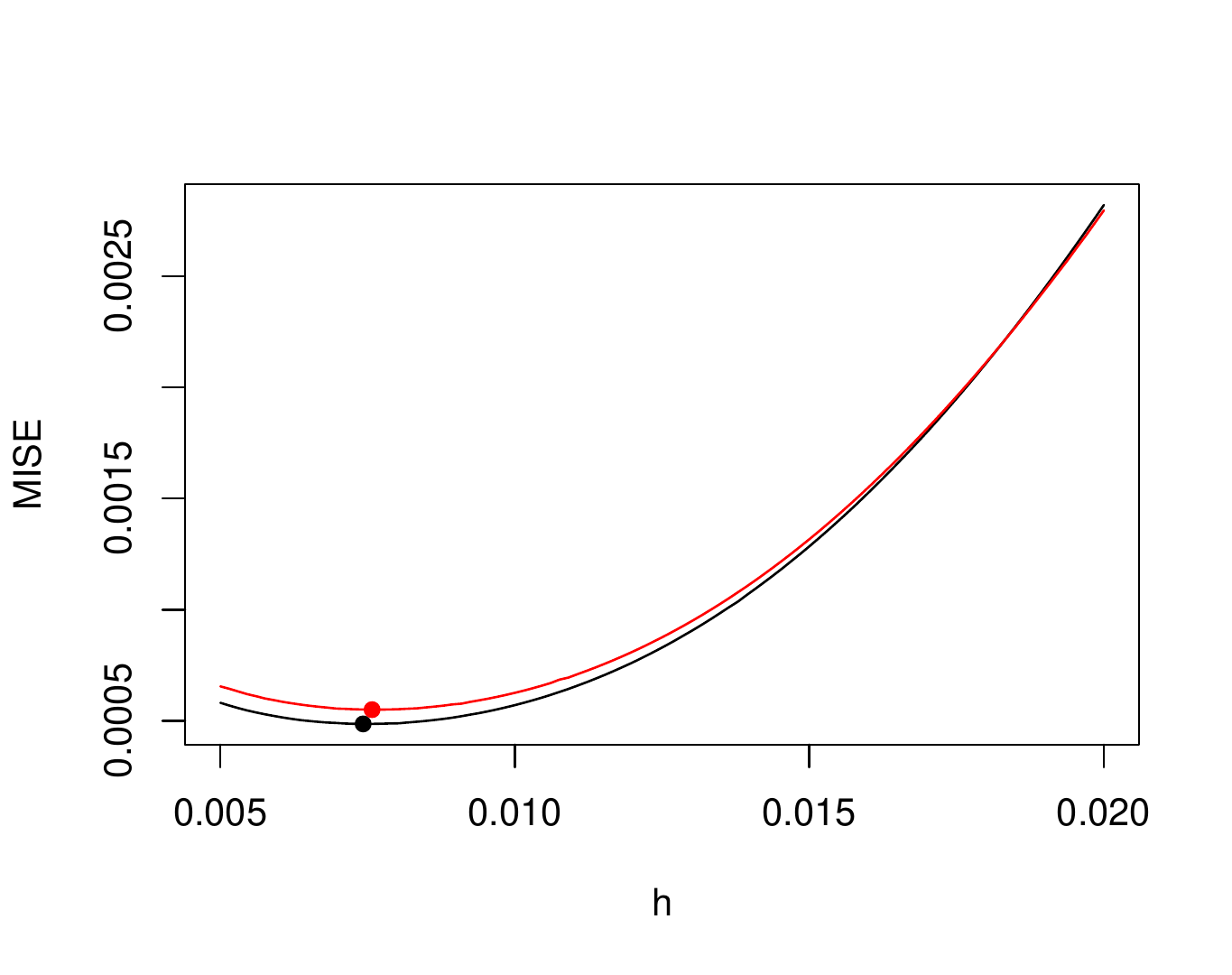}
	\caption{Model M3, $n = 1000$}
	\end{subfigure}
	\begin{subfigure}{.3\textwidth}
		\centering
	\includegraphics[width=100px]{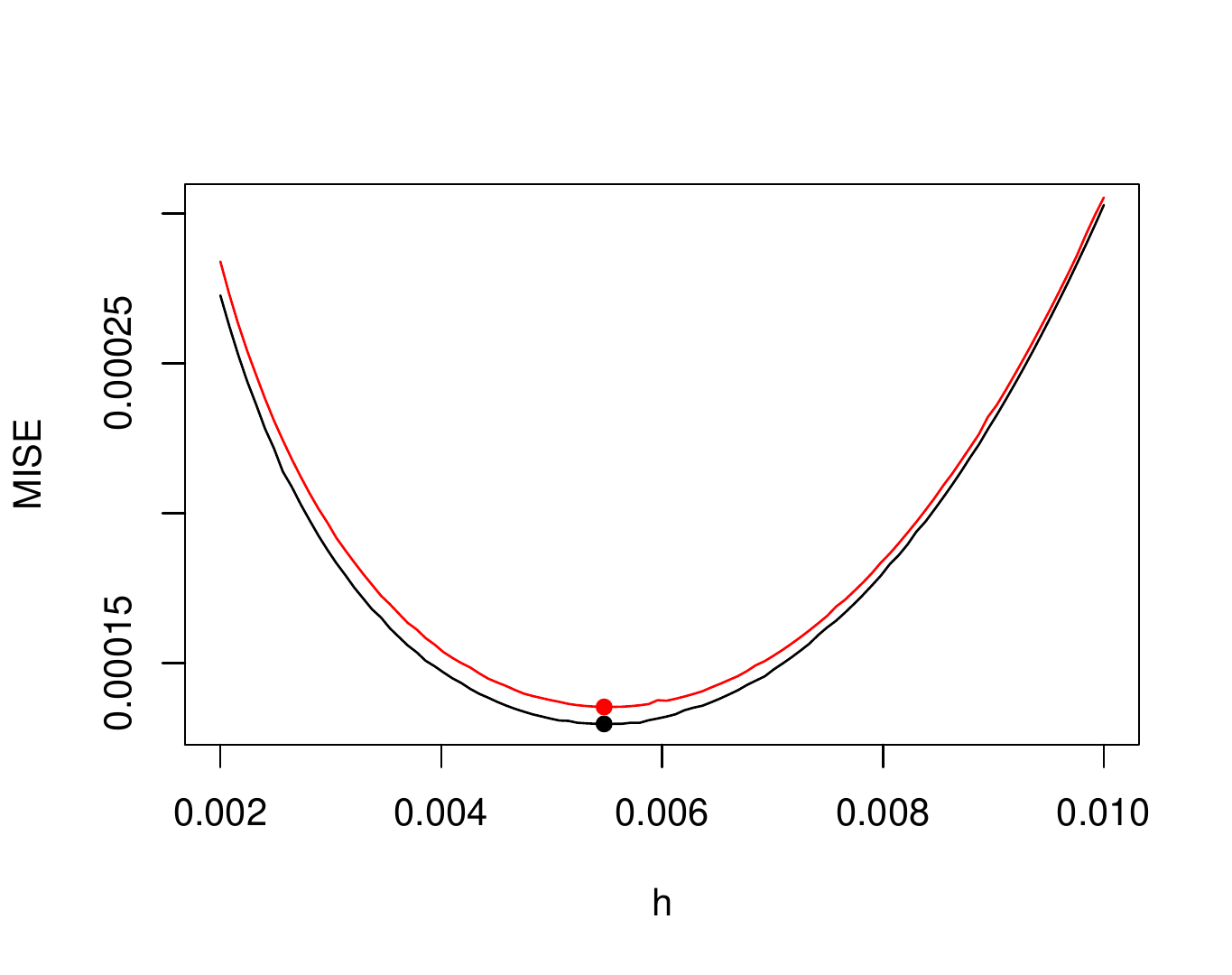}
	\caption{Model M3, $n = 5000$}
	\end{subfigure}
	\caption{MISE curve for the standard Nadaraya--Watson estimator (black line) for its modified version (red line) and  with their minima (red and black points, respectively).\label{fig:mise_tramposo}}
\end{figure}

\begin{figure}[htb]
	\includegraphics[width=\linewidth,height=450px]{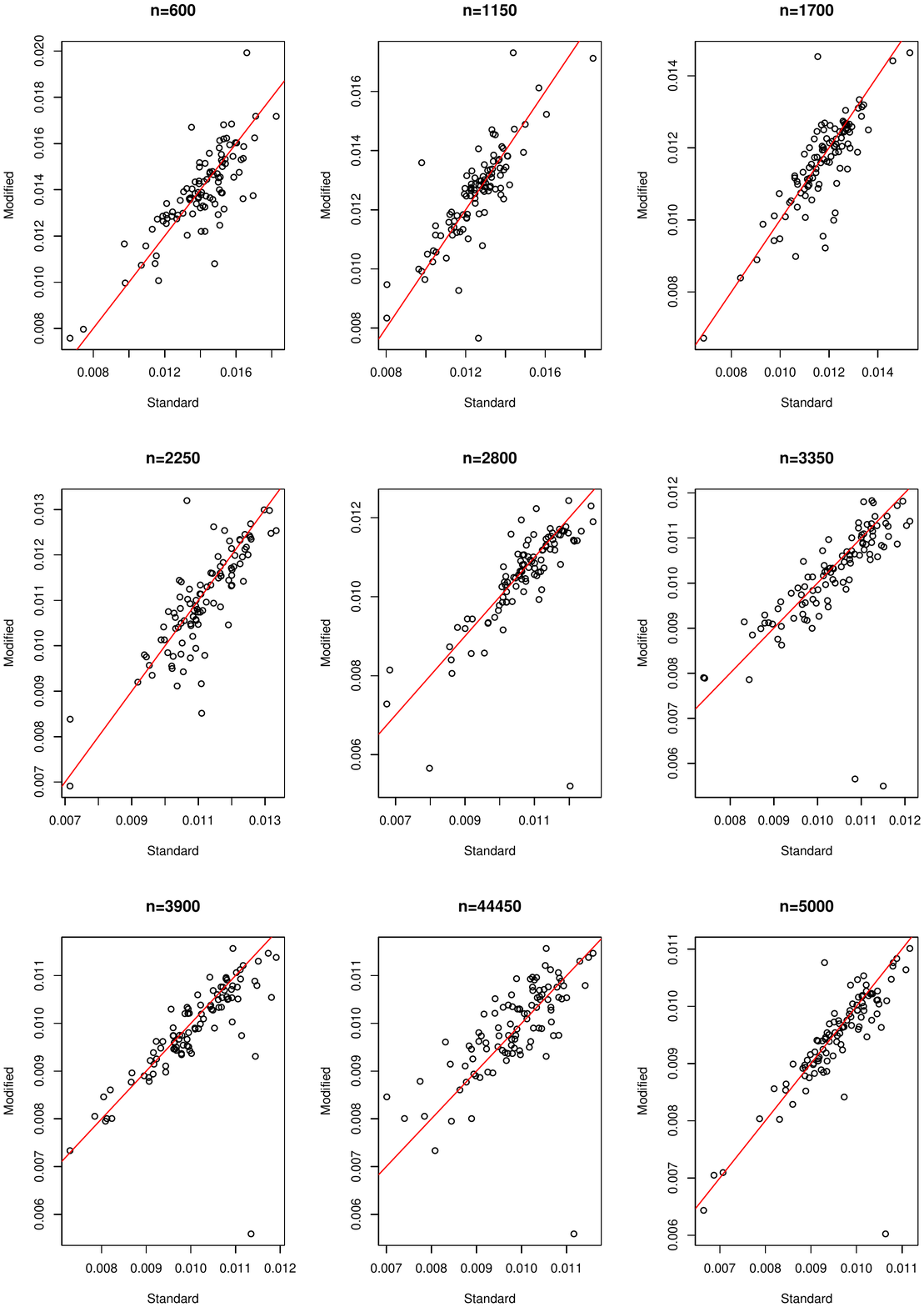}
	\caption{Cross-validation bandwidths using the standard Nadaraya--Watson estimator (x-axis) and its modified version (y-axis) for samples of sizes ranging from $600$ to $5000$ drawn from model M2.\label{fig:hcv_hcvtramp}}
\end{figure}

In this section, we complete the simulations presented in the main paper, adding two additional plots not included in the paper for reasons of space. In this
 simulation study, we considered the following regression models:
\begin{enumerate}
	\item[M1:] $Y = m(X)+\varepsilon$, $m(x) = 2x$, $X \sim \mbox{Beta}(3,3)$, $\varepsilon \sim \mbox{N}(0,0.1^2)$,
	\item[M2:] $Y = m(X)+\varepsilon$, $m(x) = \sin(2\pi x)^2$, $X \sim \mbox{Beta}(3,3)$, $\varepsilon \sim \mbox{N}(0,0.1^2)$,
	\item[M3:] $Y = m(X)+\varepsilon$, $m(x) = x+x^2\sin(8\pi x)^2$, $X \sim \mbox{Beta}(3,3)$, $\varepsilon \sim \mbox{N}(0,0.1^2)$,
\end{enumerate}
whose regression functions are plotted in Figure 1 of the main paper. In a first step, we empirically checked how close the bandwidths that minimize the MISE of the {Nadaraya--Watson} estimator and its modified version given, respectively, in equations {(2)} and (8) of the main paper are. For this, we simulated $100$ samples of different sizes ($100$, $1000$ and $5000$) from models M1, M2 and M3 and compute the corresponding MISE curves for the standard Nadaraya--Watson estimator and for its modified version. Figure \ref{fig:mise_tramposo} shows these curves. It can be observed that the bandwidth that minimizes the MISE of standard Nadaraya--Watson estimator and the MISE of its modified version appear to be quite close, even for moderately small sample sizes. Naturally, the distance between the minima of both curves tends to zero as the sample size increases. On the other hand, Figure \ref{fig:hcv_hcvtramp} shows the standard and modified cross-validation bandwidths (using the standard and modified version of the Nadaraya--Watson estimator, respectively) obtained for samples of sizes ranging from $600$ to $5000$ drawn from model M2. It can be seen that both bandwidth selectors provide similar results, which in turn get closer as $n$ increases.

\bibliography{paper}
\bibliographystyle{apalike}